\documentclass[journal, twoside]{IEEEtran}
\usepackage{cite}
\usepackage{graphicx}
\graphicspath{{./Figs/},{./ScoreMap/},{./ResMap/}}
\DeclareGraphicsExtensions{.pdf,.jpeg,.png,.tif,.jpg,.eps}
\usepackage{amsmath}
\usepackage{amssymb}
\usepackage{bm}
\usepackage{algorithm}
\usepackage{algorithmic}
\usepackage{array}
\usepackage[caption=false,font=footnotesize]{subfig}
\usepackage{dblfloatfix}
\usepackage{colortbl}  
\usepackage{xcolor}
\usepackage{url}
\usepackage[bookmarks=true,colorlinks,linkcolor=red,anchorcolor=cyan,citecolor=magenta,backref=slide]{hyperref}
\usepackage{diagbox}
\usepackage{multirow}
\usepackage{makecell}
\usepackage{utfsym}
\newcommand{\PreserveBackslash}[1]{\let\temp=\\#1\let\\=\temp}
\newcolumntype{C}[1]{>{\PreserveBackslash\centering}p{#1}}
\newcolumntype{R}[1]{>{\PreserveBackslash\raggedleft}p{#1}}
\newcolumntype{L}[1]{>{\PreserveBackslash\raggedright}p{#1}}

\newcommand{\eqrefnew}[1]{Eq.~(\ref{#1})}
\newcommand{\secref}[1]{Section~\ref{#1}}
\newcommand{\figref}[1]{\figurename~\ref{#1}}

\newcommand{\tabref}[1]{Table~\ref{#1}}

\usepackage{color}

\usepackage{todonotes}
\presetkeys{todonotes}{inline, bordercolor=white, color=blue!30}{}
\graphicspath{{./Figs/},{./ScoreMap/},{./ResMap/}}
\UseRawInputEncoding
\begin{document}
\title{You Only Train Once: Learning a General Anomaly Enhancement Network with Random Masks for Hyperspectral Anomaly Detection}
\author{Zhaoxu Li, Yingqian Wang, Chao Xiao, Qiang Ling, Zaiping Lin, and Wei An
\thanks{Manuscript received XXX; revised XXX.  
	This work was supported in part by the Foundation for Innovative Research Groups of the National Natural Science Foundation of China under Grant 61921001.
	Zhaoxu Li, Yingqian Wang, Chao Xiao, Qiang Ling, Zaiping Lin, and Wei An are with the College of Electronic Science and Technology, National University of Defense Technology, Changsha 410073, China (email: lizhaoxu@nudt.edu.cn; wangyingqian16@nudt.edu.cn; xiaochao12@nudt.edu.cn; lingqiang16@nudt.edu.cn; linzaiping@nudt.edu.cn; anwei@nudt.edu.cn)}
\thanks{Corresponding authors: Zaiping Lin; Qiang Ling.}
}

\markboth{}{}


\maketitle

\begin{abstract}

In this paper, we introduce a new approach to address the challenge of generalization in hyperspectral anomaly detection (AD). Our method eliminates the need for adjusting parameters or retraining on new test scenes as required by most existing methods. Employing an image-level training paradigm, we achieve a general anomaly enhancement network for hyperspectral AD that only needs to be trained once. Trained on a set of anomaly-free hyperspectral images with random masks, our network can learn the spatial context characteristics between anomalies and background in an unsupervised way. Additionally, a plug-and-play model selection module is proposed to search for a spatial-spectral transform domain that is more suitable for AD task than the original data. To establish a unified benchmark to comprehensively evaluate our method and existing methods, we develop a large-scale hyperspectral AD dataset (HAD100) that includes 100 real test scenes with diverse anomaly targets.
In comparison experiments, we combine our network with a parameter-free detector and achieve the optimal balance between detection accuracy and inference speed among state-of-the-art AD methods. Experimental results also show that our method still achieves competitive performance when the training and test set are captured by different sensor devices. Our code is available at \href{https://github.com/ZhaoxuLi123/AETNet}{https://github.com/ZhaoxuLi123/AETNet}.

\end{abstract}


\IEEEpeerreviewmaketitle

\section{Introduction}
\begin{figure}[t]
	\centering
	\includegraphics[width=3.5in]{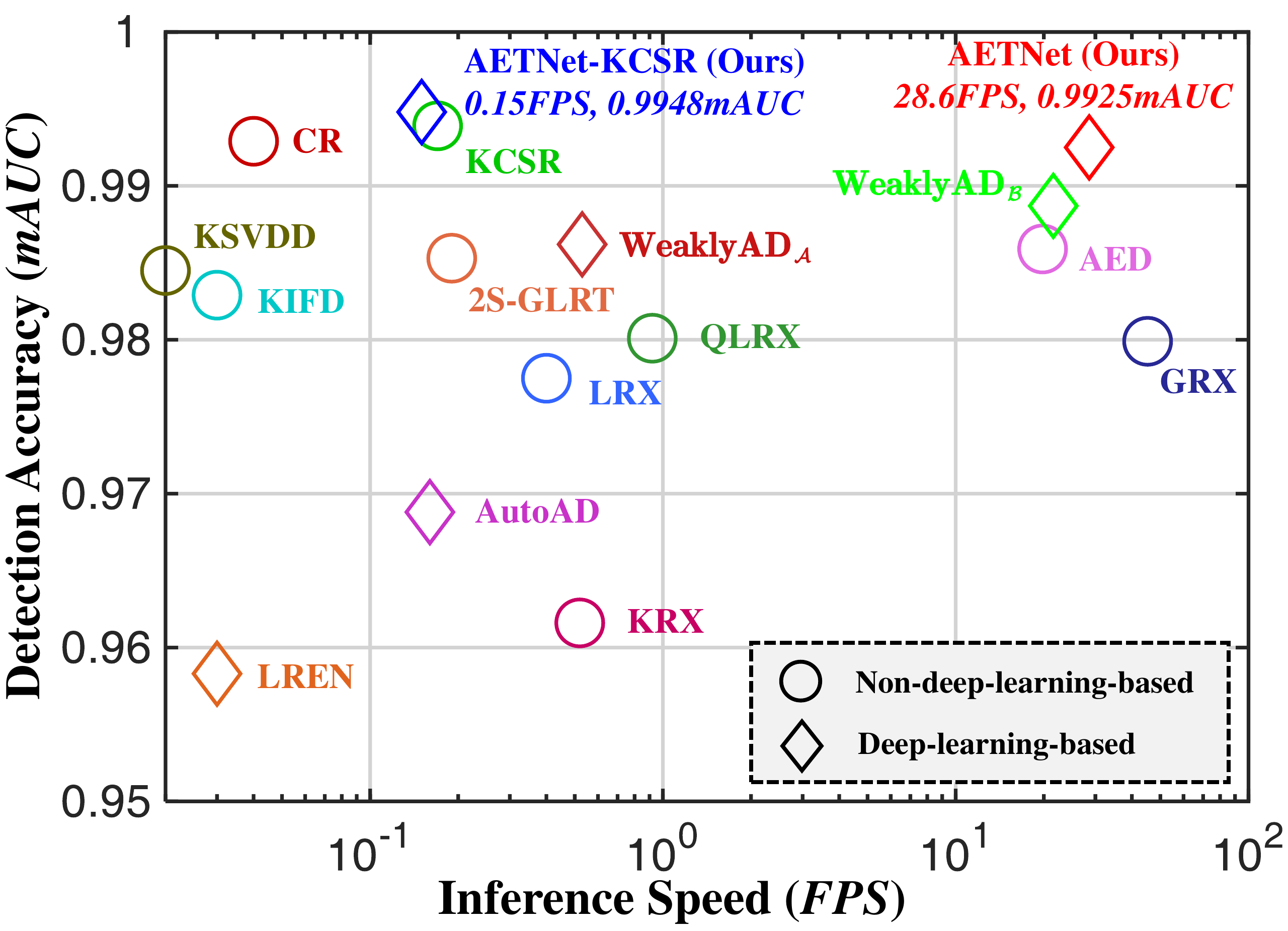}
	\caption{Comparison of the detection accuracy (mAUC) and inference speed (frames per second, FPS) of different AD methods on our HAD100 dataset. }
	\label{fig:auc_pfs}
\end{figure}   
\label{sec:Intro}
\IEEEPARstart{H}{yperspectral} imagery (HSI)  records abundant spectral  information which  can reflect the essential characteristics of different materials\cite{HSI}.
With the development of hyperspectral spectrometers,  hyperspectral imaging has been widely applied to different fields, such as agricultural estimation \cite{AgrcEst}, civilian rescue \cite{SrchRsc}, mineral exploration \cite{MnrExpl}, quality monitoring \cite{FoodSafety}, and explosive detection \cite{ExplDct}.
Among them,  hyperspectral  anomaly  detection is a critical technology to find targets from background without any  prior information.
  
During the past years, many methods have been developed for hyperspectral anomaly detection (AD). 
RX method\cite{RX-AD}, named after its proposers, is the cornerstone among AD methods. 
It supposes background spectra in the HSI obey a multi-Gaussian distribution and calculates the Mahalanobis distance between the test sample and the background samples to search anomalies. 
Depending on whether the global or local background spectrum is used, RX can be divided into global RX (GRX) and local RX (LRX)\cite{GRX-LRX-AD}. 
The later adopts a dual concentric rectangular window centering around the test pixel and selects the area between the inner and outer windows as background samples.
Inspired by LRX, many dual-window-based methods are developed, such as quasi-local-RX (QLRX)\cite{QLRX-AD},  locally adaptable iterative RX (LAIRX)\cite{LAIRX-AD}, weighted RX\cite{WghtRX-AD}, linear filter-based RX\cite{WghtRX-AD}, kernel RX (KRX)\cite{KRX-AD}, cluster KRX (CKRX)\cite{CKRX-AD}, support vector data description method (SVDD)\cite{SVDD11-AD}, and 2S-GLRT\cite{2S-GLRT-AD}.
In addition, some methodscombined sparse representation with the dual window and achieved  higher detection precision, such as collaborative representation  (CR) \cite{CR-AD}, constrained sparse representation  (CSR) \cite{CSR-AD},  and dual collaborative representation (DCR)\cite{DCR-AD}.
However, these dual-window-based methods have an inherent limitation.
The size of the dual window needs to be set manually by users according to the spatial information of targets.
Generally, the inner window size should be larger than the target size.
In addition, the best size of the dual window is also related to the shape and spacing of targets.
This problem limits the generalization capability of these AD methods.
Multiple-window AD (MWAD) \cite{MW-AD} and superpixel-based dual window RX (SPDWRX) \cite{SPDWRX-AD} improved the dual window, but this problem has not been solved well.

Many researchers were not limited to the dual-window framework.
Chang  \cite{TD-AD1,TD-AD2,TD-AD3} proposed dummy variable trick  to convert hyperspectral target detection methods to AD methods.
Tu et al. \cite{DP-AD} introduced density peak clustering \cite{DP} to hyperspectral AD task.
Tao et al. \cite{FrFE-AD} employed fractional Fourier transform (FrFT) as pre-processing to enhance discrimination between anomalies and background.
But these methods neglecte spatial information and can not deal with non-stationary background or large targets well.
Some methods try to use spatial features to improve detection performance, such as attribute and edge-preserving filtering-based detection method (AED) \cite{AED-AD}, spectral-spatial feature extraction (SSFE) \cite{SSFE-AD}, structure tensor and guided filter (STGF) \cite{STGF-AD}, segmentation-based weighting strategy \cite{SW-AD}, and kernel isolation forest-based detection method (KIFD) \cite{KIFD-AD}.
Besides,  tensor-based methods pay attention on the 3D structure in HSI and extracted anomaly tensors from HSI cube data by tensor decomposition, e.g.,  tensor decomposition-based method (TDAD) \cite{TD-AD}, tensor principal component analysis  (TPCA) \cite{TPCA-AD}, and prior-based tensor approximation (PTA) \cite{PTA-AD}.
In addition, low-rank and sparsity-matrix decomposition (LRaSMD) \cite{zhou2011godec}  has also received widespread attention and derived many AD methods such as OSP-GoDec \cite{OSP-GoDec-AD},  OSP-AD \cite{OSP-AD},  SDP \cite{SDP-AD}, component decomposition analysis (CDA) \cite{CDA-AD},  and effective anomaly space (EAS) \cite{EAS-AD}.
However, the parameters of the above methods still need to be fine-tuned  on different test scenes.

Lately, deep learning is booming in many fields and brings more novel ideas for hyperspectral AD.
Generative models such as  Autoencoder (AE) and generative adversarial network (GAN) \cite{GAN} can capture nonlinear and potential characteristics and have been successfully applied to the AD task.
A common practice of AE-based methods is to take the vectors from the output space or the latent space of AE as the input of other anomaly detectors \cite{HADSDA-AD,E2E-LIADE-AD,sparseHAD-AD,LREN-AD,NF-PDRD-AD}.
In particular, in order to adapt the learned latent representation to a specific density estimation, low-rank embedded network (LREN) \cite{LREN-AD} introduces a trainable density estimation module into the latent space.
Another common practice is to reconstruct the spectra under certain constraints and regard the spectral reconstruction error as the detection anomaly score \cite{DAE-AD, OTCMA-AD, RGAE-AD, GRVAE-AD}.
This kind of methods usually use full connection layers to achieve spectral-level self-supervised learning, and need to introduce additional spatial information to improve detection accuracy. 
For example, robust graph autoencoder method (RGAE)\cite{RGAE-AD} is embedded with a superpixel segmentation-based graph regularization term to preserve the geometric structure and the local spatial consistency of HSI.
An exception to this kind of methods is AutoAD \cite{Auto-AD}, which adopts convolution autoencoder to reconstruct the original HSI from pure noise input.
Although some methods mentioned above (such as AutoAD) do not require any manual selection of parameters, they must be retrained when applying to test scenes.

In summary, two kinds of problems restrict the generalization capability of current AD methods:
(1) Some parameters need to be manually set according to the characteristics of targets in each test scene.
(2) Most deep-learning-based methods need to be retrained  to apply to new test scenes. 
To tackle these issues, we aim to develop a general network for hyperspectral AD.
Such a network should be trained only once without using ground-truth labels and can be generalized to unseen test scenes without retraining.
It is worth noting that there is very little prior study on this topic. 
Two recent methods, WeaklyAD \cite{wealyAD-AD} and dual-frequency autoencoder (DFAE) \cite{DFAE-AD}, explore this generalization issue.
Specifically, WeaklyAD uses a spectral-constrained GAN to enhance the discrimination between anomalies and background in a weakly supervised way.
DFAE transforms the original HSI into high-frequency and low-frequency components and then detects anomalies from these two components in parallel.
Both these two methods are trained on a public test HSI and achieve competitive results on several test scenes without retraining.
However, these methods are still far from the AD framework we expect. 
WeaklyAD needs to extract anomalous spectra for adversarial training, while DFAE still requires manual parameter setting.
In addition, these two methods have not been verified on a large-scale dataset.

In this paper, we introduce an image-level training paradigm and  a simple data augmentation strategy called Random Mask to solve the generalization problem of  hyperspectral AD.
The previous deep-learning-based methods almost perform self-supervised learning on the 1D spectral level,  while our network  performs self-supervised learning on a set of 3D hyperspectral images.
This allows our network to learn the spatial context information directly without requiring additional spatial constraints.
Trained on  anomaly-free hyperspectral images with random masks, our network, called \textbf{AETNet}, can achieve \textbf{A}nomaly \textbf{E}nhancement \textbf{T}ransformation on unseen test images.
With the cooperation of  the Random Mask strategy and a structural-difference-based loss function, our AETNet can learn the spatial-spectral context relationship between anomalies and background in an unsupervised way.
Besides, a simple plug-and-play model selection module is introduced to guide the training process and search for a transform domain that is more suitable for AD task than the original data.  
Once trained,  our method can project unseen test images into the searched transform domain, and achieve parameter-free but high-accuracy detection on these transformed test images.

Moreover, since existing methods generally report their best results after fine-tuning on a limited number of test images, a more comprehensive evaluation benchmark is necessary.
In this paper, we further develop a new hyperspectral AD dataset to evaluate current methods.
Our dataset, called HAD100, contains 100 test HSIs with various anomaly targets.
In addition to the proposed AETNet, we select 16 representative AD methods as solid baselines to build a comprehensive benchmark.
The contributions of this paper can be summarized as follows: 
\begin{enumerate}
		\item  We introduce an image-level training paradigm to develop a general anomaly enhancement network for hyperspectral AD.
		Once trained, our network can perform inference directly on unseen test images.
		As far as we know,  we are the first attempt to train the network on a set of hyperspectral images in this field.
		\item  We provide a simple data augmentation strategy called Random Mask for hyperspectral AD.
		This strategy can help our network learn  the context relationship between anomalies and background from  anomaly-free hyperspectral data in an unsupervised way.
 	\item We develop a new hyperspectral AD dataset and comprehensively evaluate the performance of both classic and SOTA AD methods. To the best of our knowledge, this is the first unified large-scale benchmark in the hyperspectral AD field.
 	\item Extensive experiments show that our proposed AETNet achieves the best balance between detection accuracy and inference speed among the SOTA AD methods.
 	 And our method  can still achieve competitive performance when the training and test set are captured by different sensor devices.
\end{enumerate}

The rest of this paper is organized as follows. \secref{sec:RelateWork} reviews the research basic related to our network. \secref{sec:AETNet} describes the implementation details of the proposed method. Our dataset and experiments are presented in \secref{sec:experiments}, followed by the conclusion in \secref{sec:Cnl}.

\section{Related Work}
\label{sec:RelateWork}
In this section, we briefly introduce two studies involved in our network: UNet and Vision Transformer.
\subsection{U-shaped Network}
\label{sec: UNet}
UNet\cite{unet}, named after its symmetrical U-shaped structure, is a simple but widely used segmentation network. 
UNet can be split into a contracting path and an expansive path.
The contracting path consists of a series of down-sampling modules, regarded as encoder layers.
The expansive path consists of a series of up-sampling modules, regarded as decoder layers and symmetric with the encoder layers.
Adding skip connections between the encoder and decoder layers, UNet mitigates the loss of spatial information due to down-sampling.
The fusion of multi-scale contextual features enhances image reconstruction process and significantly improves the segmentation performance.
Nowadays, UNet and its abundant variations are widely applied to many tasks such as object segmentation\cite{DNANet}, image super-resolution\cite{sr-unet}, image reconstruction \cite{mst}, and so on.

\subsection{Vision Transformer}
\label{sec:Transformer}
Transformer \cite{transformer} is first proposed in the natural language processing and outperforms the previous complicated recurrent models and CNN models.
The success of Transformer has attracted  the great  attention of researchers in the computer vision community.
Dosovitskiy et al. \cite{ViT} are the first to propose Vision Transformer (ViT) for the image classification task. 
Compared with CNN, ViT has a better capability to model the long-range dependencies and achieves a global receptive field in the shallow layers \cite{raghu2021vision}.
Therefore, CNN frameworks have been approached and even surpassed by ViT and its variations\cite{ToTViT, PVT, Swin} on the image classification task.

Nevertheless, expensive training cost makes ViT unsuitable for dense prediction tasks such as image segmentation.
To this end, PVT \cite{PVT} adds the pyramid structure to ViT to acquire multiscale feature maps. 
Furthermore, Swin Transformer \cite{Swin} limits self-attention computation to non-overlapping local windows and achieves linear computational complexity with respect to image size.
With the help of a shifted window strategy, Swin Transformer achieves cross-window information interaction. It surpasses the SOTA methods in many tasks, such as object detection, instance segmentation, and semantic segmentation.
Subsequently, Cao et al. \cite{swin-unet} combined Swin Transformer with UNet and propose the first pure Transformer-based U-shaped architecture, named Swin UNet. 
Swin UNet outperforms CNN or the combination of Transformer and CNN in medical image segmentation and is successfully applied to other dense prediction tasks \cite{swin-unet1}. 

\section{Methodology}
In this section, we first introduce the overview of our method.
Then, we describe the implementation details of each component of our method. 
\label{sec:AETNet}
\subsection{Overview}
\begin{figure*}[htb]
	\centering
	\includegraphics[width=7in]{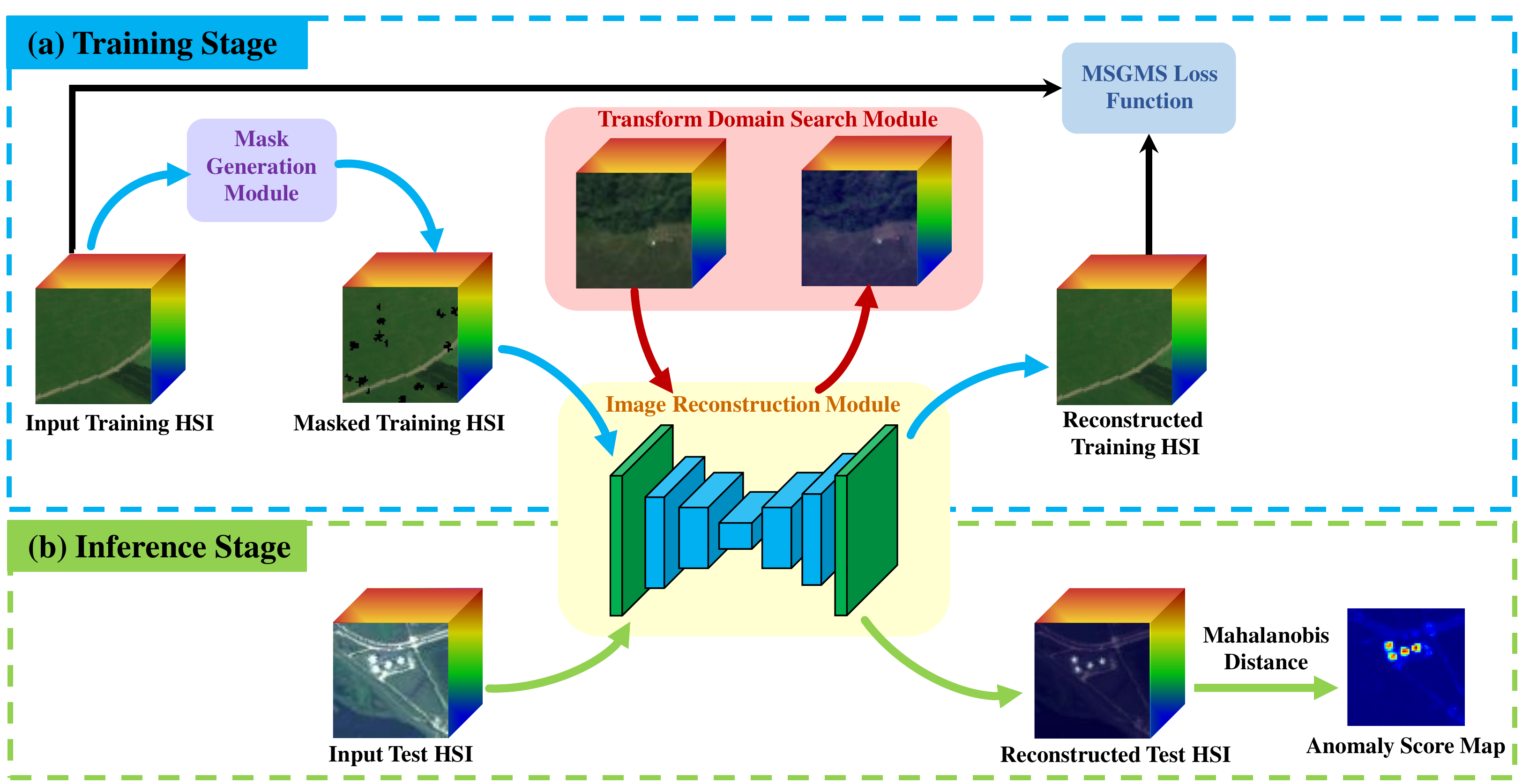}
	\caption{Overview of our proposed method. }
	\label{fig:overview}
\end{figure*} 
 
\begin{figure*}[htb]
	\centering
	\includegraphics[width=5.5in]{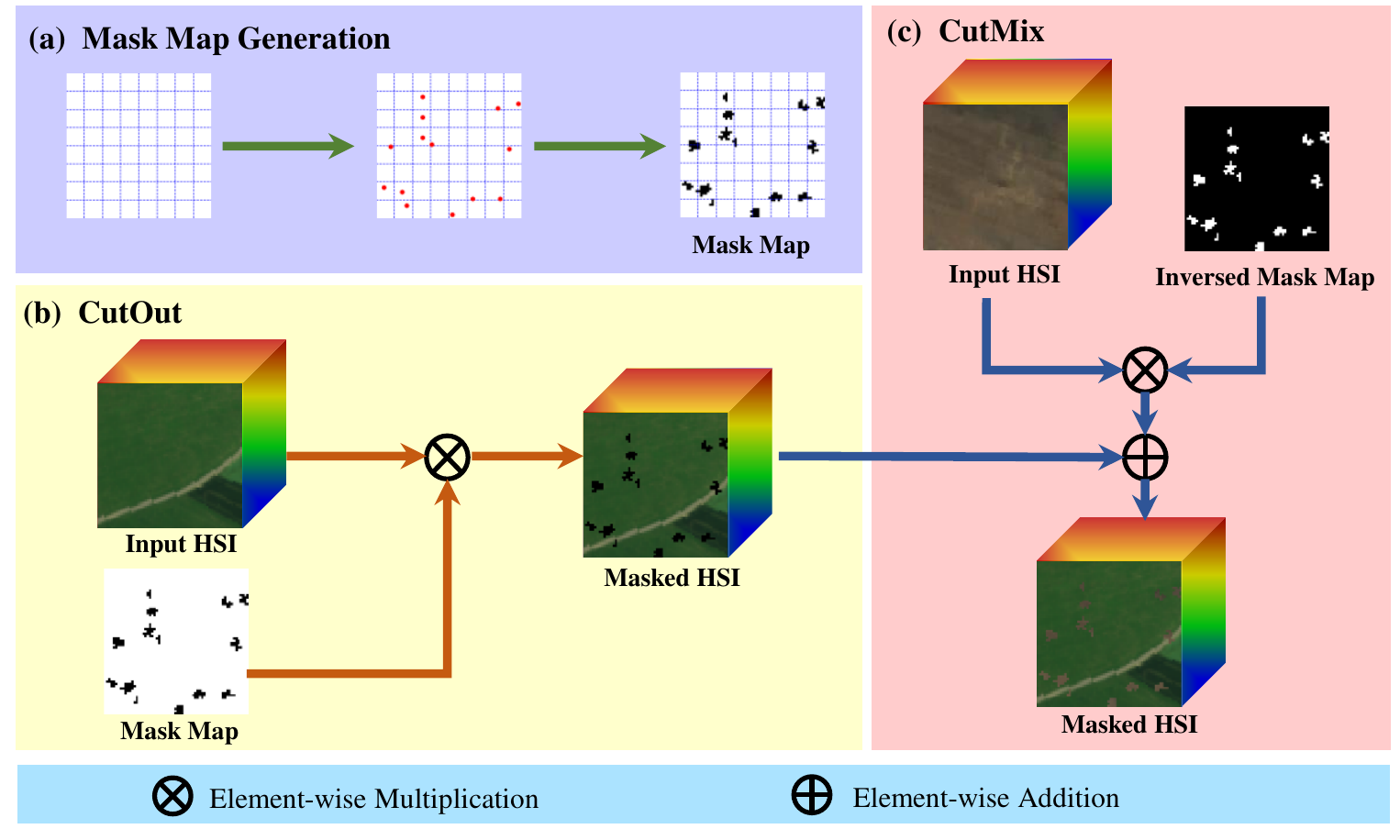}
	\caption{Diagram of the proposed Random Mask strategy. }
	\label{fig:RandomMask}
\end{figure*} 
\begin{figure*}[htb]
\centering
\includegraphics[width=6in]{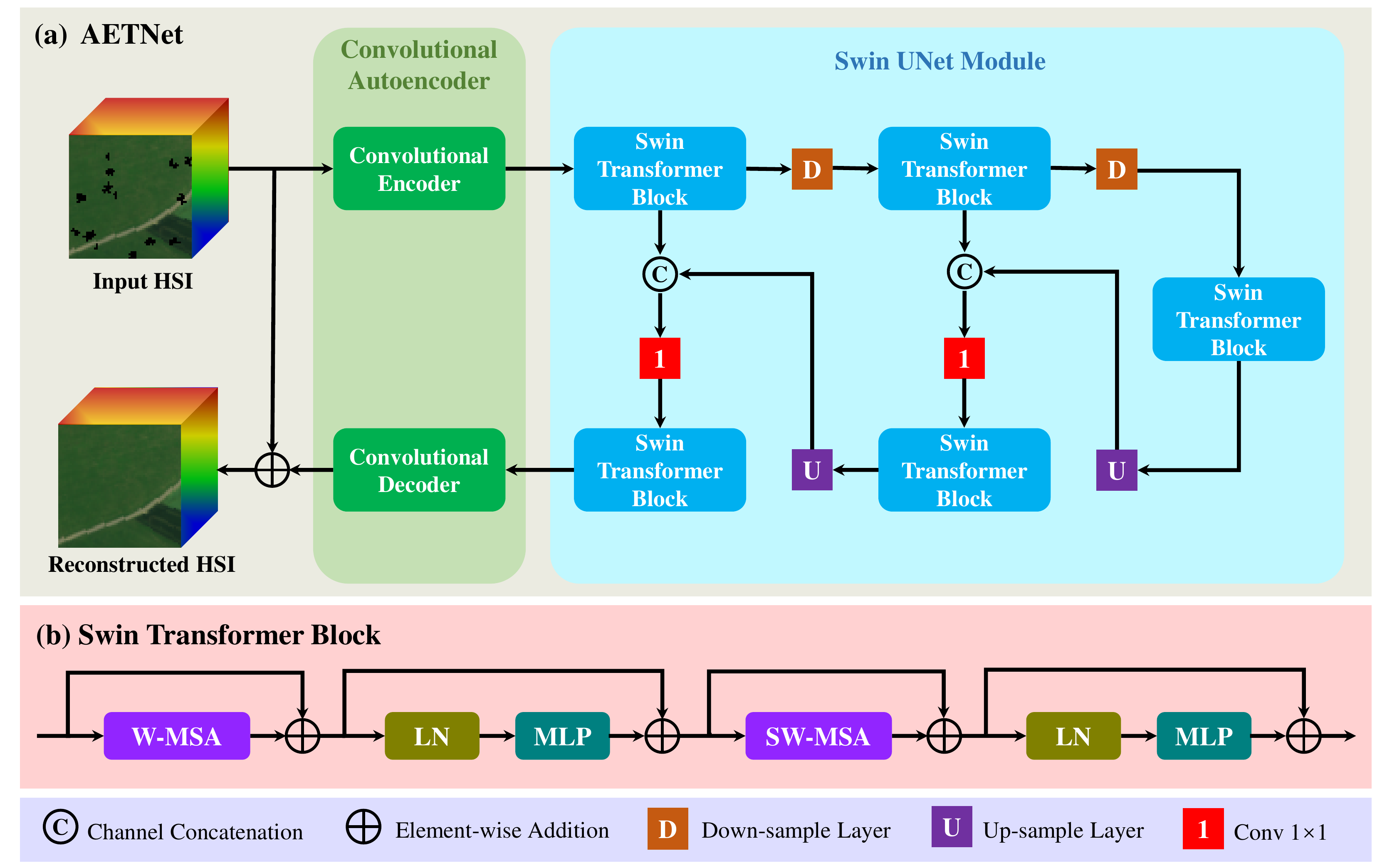}
\caption{Network structure of the proposed AETNet. }
\label{fig:Network}
\end{figure*} 
Our proposed method consists of three major components: mask generation module, image reconstruction module, and  transform domain search module.
The overview architecture of AETNet is shown in \figref{fig:overview}.
The backbone of the image reconstruction module is based on Swin UNet and convolutional autoencoder (CAE),  and the input of our network is a 3D HSI.
The training and test data are both hyperspectral cubes with the same spatial size and band number.
The training set is defined as $\boldsymbol{X}=\left\{ \boldsymbol{X}_1,\boldsymbol{X}_2,\cdots ,\boldsymbol{X}_n\right\} \subset \mathbb{R} ^{H\times W\times B}$,
where $n$ is the number of HSIs in the training set. $H$, $W$, and $B$ denote a single hyperspectral cube's height, width, and band number.
The test set is defined as $\boldsymbol{Y}=\left\{ \boldsymbol{Y}_1,\boldsymbol{Y}_2,\cdots ,\boldsymbol{Y}_m\right\} \subset \mathbb{R} ^{H\times W\times B}$, where $m$ is the number of HSIs in the test set.
 
In the training stage, the mask generation module generates random masks for each training HSI $\boldsymbol{X}_i\in \boldsymbol{X}$.
Then HSIs with  random masks are fed to the image reconstruction module.
After every training epoch, the transform domain search module is used to evaluate the enhancement performance of the reconstruction module and decide whether to terminate the training.
 
In the inference stage,  each test HSI $\boldsymbol{Y}_i\in \boldsymbol{Y}$ is sent to the image reconstruction module, and then we obtain a reconstructed HSI $\boldsymbol{Y}_r$.
Next, anomaly detectors such as GRX  take $\boldsymbol{Y}_r$ as input to generate the anomaly score map of $\boldsymbol{Y}_i$.

\subsection{Random Mask Strategy}

Considering that there is no target label in the training phase, we introduce a single data augmentation strategy called Random Mask. 
Random masking is a popular tool in the CV community \cite{MAE} and has been successfully applied to the industrial AD task \cite{RIAD}.
Different from masks in \cite{RIAD}, random masks in our paper have irregular shapes and random sizes and can simulate the spatial morphology of anomaly targets.

It is usually believed that anomaly targets in hyperspectral remote sensing imagery have the following characteristics:
(1) Anomaly targets are embedded to background in the form of irregular blocks.
(2) The spectra of anomaly targets are different from those of background.
(3) The area ratio of anomaly targets is low in the total image.
As shown in \figref{fig:RandomMask}, we divide the Random Mask strategy into mask map generation and masked region filling, simulating the spatial and spectral characteristics of anomaly targets, respectively.
\subsubsection{Mask Map Generation}
The mask map $\boldsymbol{M}$ has the same height $H$ and weight $W$ as the input training HSI $\boldsymbol{X}_i$. 
The height and weight are equally divided by $K$, and then $\boldsymbol{M}$ is divided into $K^2$ non-overlapping patches.
The number of random masks $N$ is randomly generated from the integer range of $\left[ N_{\min}, N_{\max} \right]$, where $N_{\max}$ should be less than $K^2$.
We randomly select $N$ patches from all the patches and then randomly select a pixel position in each selected patch. 
Next, an iterative expansion method is used to generate the random mask at each selected pixel position.
We regard the selected pixel position as the start point of a random mask to be generated.
The four-connected pixels of the start point are randomly merged into the random mask, and the probability of every connected pixel being selected is $50\%$.
The  mask is continuously expanded by randomly merging the updated four-connected pixels until the area reaches $A$, where $A$ is randomly sampled from the integer range of $\left[ A_{\min}, A_{\max} \right]$.
After traversing all the selected pixel positions, we get $N$ masks with different areas and random shapes.
The final mask map $\boldsymbol{M}$ is a binary map where the masked  regions are set to 0,  and the remaining pixels are set to 1.
\subsubsection{Mask Region Filling}
Considering the spectral characteristic of anomaly targets, we introduce an additional cube $\boldsymbol{I}\in \mathbb{R} ^{H\times W\times B}$ to fill the generated masks.
 The masked input HSI cube $\boldsymbol{X}_M$ can be attained by 
\begin{equation}
\label{eq:mask}
\boldsymbol{X}_M=\boldsymbol{X}_i\otimes \boldsymbol{M} +\boldsymbol{I}\otimes \overline{\boldsymbol{M}},
\end{equation}
where $\otimes$ denotes element-wise multiplication and $\overline{\boldsymbol{M}}$ represents the inversed mask map of $ \boldsymbol{M}$.
Inspired by recent data augmentation approaches, we introduce two mask-filling methods, namely CutOut\cite{Cutout} and CutMix\cite{CutMix}.
 
\textbf{CutOut:} CutOut replaces masked regions in RGB images with gray or black pixels.
In this paper, we set $\boldsymbol{I}$ to a three-dimensional zero matrix.

\textbf{CutMix:}
 CutMix replaces masked regions in RGB images with the corresponding regions from other images.
 In this paper, we randomly choose another training HSI from a different capture flight line as $\boldsymbol{I}$.

\subsection{Network Architecture}
As shown in \figref{fig:Network}(a), our network consists of a convolutional encoder, a Swin UNet module,  and a convolutional decoder.
The detailed network structure is further described below.

\subsubsection{Convolutional Autoencoder}
The previous deep-learning -based methods mainly use fully connected layers to perform self-supervised learning on the spectral level, resulting in the loss of spatial information.
The most simple improvement is to use CAE to perform self-supervised learning on the hyperspectral cubes.
CAE combines AE with convolution and pooling operations and can extract spatial features of images.

We use a  simple CAE.
The convolutional encoder is a $3\times 3$ convolutional layer, which changes the spectral channel number of the input cube $\boldsymbol{X}_m$ from $B$ to $C$. 
The convolutional decoder is also a $3\times 3$ convolutional layer, which restores the channel number from $C$ to $B$. 

Common CAEs consist of several convolutional layers, down-sample layers, deconvolutional layers, and up-sample layers.
 Since ViT can model long-range dependencies better than CNN, we abandon CAE's down-sample and up-sample operations and only apply the channel dimension transformation.
Meanwhile, we insert a Swin UNet module between the convolutional encoder and decoder as the middle layer of CAE to extract spatial and spectral features better.
 
\subsubsection{Swin UNet Module}
As shown in \figref{fig:Network}(a), the Swin UNet module can also be divided into an encoder, a bottleneck, and a decoder.
Unlike the structure in \cite{swin-unet}, we adopt a non-classical structure that directly processes latent feature maps without patch partition, merging, and expanding operations.

\textbf{Encoder:}
The encoder consists of two stages, and each stage contains a Swin Transformer block and a down-sample layer.
The down-sample layer is a $4\times 4$ strided convolutional layer which halves the spatial resolution of feature maps and doubles the channels.
In the first stage of the encoder, the feature maps from the convolutional encoder with the size of $H\times W\times C$ are sent into the first Swin Transformer block to perform representation learning. Next, the feature maps are fed into a down-sample layer, and the size changes into $\frac{H}{2}\times \frac{W}{2}\times 2C$.
Then, the feature maps are sent into the Swin Transformer block and the down-sample layer in the second stage. The size of the feature maps becomes $\frac{H}{4}\times \frac{W}{4}\times 4C$.

\textbf{Bottleneck:}
The bottleneck of the Swin UNet module is a single Swin Transformer block and learns the deep feature representation.
 In the bottleneck, the feature maps' channel number and spatial resolution remain unchanged.
 
\textbf{Decoder:}
Symmetric to the encoder, the decoder also has two stages.
In each stage, the feature maps are first fed into an up-sample layer, a $2\times 2$ strided deconvolutional layer, to halve the channels and double the spatial resolution of feature maps.
Then, we use the skip connection to fuse the up-sampled feature maps with the feature map of the same size from the encoder.
Channel concatenation on the shallow and deep feature maps can reduce the information loss caused by down-sampling.
After a $1\times 1$ convolutional layer, the concatenated feature maps have the same channel number as the up-sampled feature maps.
 A Swin Transformer block at the end of each stage is used for the representation learning of the deep feature maps.
 After the decoder, the size of the feature maps returns to $H\times W\times C$.
 
\textbf{Swin Transformer Block:}
Swin Transformer is a shifted-window-based ViT and contains two different attention modules, called Window-based Multi-head Self Attention (W-MSA) module and Shifted Window-based Multi-head Self Attention (SW-MSA) module, respectively.
More details of W-MSA and SW-MSA can be found in \cite{Swin}.
As shown in \figref{fig:Network}(b), the difference between our Swin Transformer block and the classical structure in \cite{Swin} is that LayerNorm (LN) layers in front of the two MSA modules are removed, which is validated in our network.

\subsubsection{Residual Connection}
The only difference between the masked input HSI $\boldsymbol{X}_M$ and the expected output HSI $\boldsymbol{X}_i$ is the masked region whose area accounts for a small portion of the whole HSI.
In other words, our network only needs to restore the masks to the original spectra rather than restore the whole HSI.
Therefore, we introduce a residual connection to connect the input $\boldsymbol{X}_M$ and output $\boldsymbol{X}_r$ of our network.
In this way, the network can pay more attention to the random masks and converge faster during training.
\subsection{Loss Function}
In the previous AD networks, spectrum-wise L2 loss is widely used.
However, we are more interested in the spatial texture relationship between anomaly targets and background rather than the spectral reconstruction accuracy.
Spectrum-wise loss functions usually ignore the spatial dependence of neighboring pixels.
Therefore, a multi-scale gradient magnitude similarity (MSGMS) loss \cite{RIAD} is introduced to penalize structural differences between the reconstructed HSI $\boldsymbol{X}_r$ and the original HSI $\boldsymbol{X}_i$.

The MSGMS is the multi-scale expansion of the gradient magnitude similarity (GMS) \cite{gms}.
The gradient magnitude is defined as the root mean square error of image directional gradients along two orthogonal directions in \cite{gms, RIAD}.
To acquire more accurate gradient magnitude maps, we adopt Sobel filters along horizontal, vertical, and diagonal dimensions.
The gradient magnitude map of $\boldsymbol{X}_i$ can be calculated according to:
 \begin{equation}
 	\label{eq:G_i}
 \begin{split}
&\boldsymbol{G}_i=\\
&\sqrt{\left( \boldsymbol{X}_i\circledast \boldsymbol{s}_x \right) ^2+\left( \boldsymbol{X}_i\circledast \boldsymbol{s}_y \right) ^2+\left( \boldsymbol{X}_i\circledast \boldsymbol{s}_{d1} \right) ^2+\left( \boldsymbol{X}_i\circledast \boldsymbol{s}_{d2} \right) ^2},
\end{split}
 \end{equation}
where $\circledast$ is the convolution operation, and $\boldsymbol{s}_x$, $\boldsymbol{s}_y$, $\boldsymbol{s}_{d1}$, and $\boldsymbol{s}_{d2}$ are $3\times 3$ Sobel filters along the $x$ dimension, $y$ dimension, $45^{\circ}$ diagonal, and $135^{\circ}$ diagonal.
Similarly, the gradient magnitude map of $\boldsymbol{X}_r$ can be calculated according to:
 \begin{equation}
 		\label{eq:G_r}
	\begin{split}
		&\boldsymbol{G}_r=\\
		&\sqrt{\left( \boldsymbol{X}_r\circledast \boldsymbol{s}_x \right) ^2+\left( \boldsymbol{X}_r\circledast \boldsymbol{s}_y \right) ^2+\left( \boldsymbol{X}_r\circledast \boldsymbol{s}_{d1} \right) ^2+\left( \boldsymbol{X}_r\circledast \boldsymbol{s}_{d2} \right) ^2}.
	\end{split}
\end{equation}

The gradient magnitude similarity map between $\boldsymbol{X}_r$ and $\boldsymbol{X}_i$ is then generated by:
 \begin{equation}
 		\label{eq:GMS}
\mathrm{GMS}\left( \boldsymbol{X}_i,\boldsymbol{X}_r \right) =\frac{2\boldsymbol{G}_i\boldsymbol{G}_r+c}{{\boldsymbol{G}_i}^2+{\boldsymbol{G}_r}^2+c},
\end{equation}
where $c$ is a constant that ensures numerical stability. 
Subsequently, down-sampled images at different scales of $\boldsymbol{X}_r$ and $\boldsymbol{X}_i$ are generated by $2\times 2$ strided average pooling operation.
The GMS maps between down-sampled $\boldsymbol{X}_r$ and $\boldsymbol{X}_i$ are calculated again according to Eqs. (2), (3), and (4). 
Finally, the MSGMS loss between down-sampled $\boldsymbol{X}_r$ and $\boldsymbol{X}_i$ is defined as: 
 \begin{equation}
	\label{eq:MSGMS}
		\begin{split}
	L&\left( \boldsymbol{X}_i,\boldsymbol{X}_r \right) =\\
	&\frac{1}{S}\sum_{l=1}^S{\frac{1}{H_lW_l}\sum_{a=1}^{H_l}{\sum_{b=1}^{W_l}{\left( 1-\mathrm{GMS}_{a,b}\left( {\boldsymbol{X}_i}^l,{\boldsymbol{X}_r}^l \right) \right)}}},
		\end{split}
\end{equation}
where $S$ is the number of reconstructed HSIs at different scales, and ${\boldsymbol{X}_i}^l$ and ${\boldsymbol{X}_j}^l$ are the original HSI and the reconstructed HSI at the $l$-th scale.
$H_l$ and $W_l$ denote the height and width of the HSI at the $l$-th scale, and $\mathrm{GMS}_{a,b}\left( \cdot \right) $ represents the value of the GMS map at pixel $\left(a, b\right)$.

The reason why we choose MSGMS instead of GMS is that the size of anomaly targets is various.
The receptive field of GMS is only $3\times3$, and thus large masked regions can only show edges on the gradient magnitude maps.
MSGMS enlarges the receptive field by constructing an image pyramid and can penalize masked regions of different sizes.

\subsection{Transform Domain Search}
In the image reconstruction module, masked input HSIs are regarded as damaged images and need to be reconstructed to original mask-free images.
General image inpainting networks aim to generate undamaged images accurately, but we are not interested in that.
The phenomenon that the optimized network can restore the masked training HSIs to the original HSIs indicates that the network can utilize the context information of  masks and change their spectra in the reconstructed images.
Hence, we believe that the image reconstruction process is a domain transformation, and regard the output space of our network as a spatial-spectral transform domain.
In the transform domain, the anomaly targets should be enhanced and have more significant differences from background.

After each training epoch, the network with updated parameters generates a new transform domain.
We need an automatic approach to select an appropriate transform domain.
For most image inpainting networks, the reconstruction loss is an important basis for the termination of training. 
However, a good image reconstruction performance does not mean a good detection performance. 
In the target detection task, researchers select trained models according to the detection accuracy on the verification set.
But no target label is available for training in hyperspectral AD. 
As a result, we introduce a distance measurement that implicitly reflects the detection performance after each training epoch.
Given a certain test HSI $\boldsymbol{Y}_v\in \boldsymbol{Y}$, we send it to the image reconstruction module that undergoes $j$ training epochs:
 \begin{equation}
	\tilde{\boldsymbol{Y}}_{v}^{j}=f\left( \boldsymbol{Y}_v;\boldsymbol{\theta }_j \right),
\end{equation}
where $f\left( \cdot \right) $ denotes the network structure, $\boldsymbol{\theta }_j$ is the updated parameters after the $j$-th training epoch, and $\tilde{\boldsymbol{Y}}_{v}^{j}\in \mathbb{R} ^{B\times 1}$ is the reconstructed HSI of $\boldsymbol{Y}_v$.
For each spectrum $\tilde{\boldsymbol{y}}_j$ on $\tilde{\boldsymbol{Y}}_{v}^{j}$, the global Mahalanobis distance is calculated according to:
 \begin{equation}
 		\label{eq:GRX}
 	\boldsymbol{M}\left( \tilde{\boldsymbol{y}}_j\right) =\left( \tilde{\boldsymbol{y}}_j-\boldsymbol{\mu }_j \right) ^T\boldsymbol{\varLambda }_{j}^{-1}\left( \tilde{\boldsymbol{y}}_j-\boldsymbol{\mu }_j \right),
 \end{equation}
 where $\boldsymbol{\mu }_j\in \mathbb{R} ^{B\times 1}$ and $\boldsymbol{\varLambda }_{j}^{-1} \in \mathbb{R} ^{B\times B}$ are the mean vector and the inverse covariance matrix of all the spectra on $\boldsymbol{Y}_{v}^{j}$, respectively. 
The maximum of global Mahalanobis distances $\mathcal{M}$ can reflect the difference between the most anomalous spectrum and  background on $\tilde{\boldsymbol{Y}}_{v}^{j}$.
A larger $\mathcal{M}$ represents a better domain transformation of our network.
The model corresponding to the peak value of $\mathcal{M}$ is used for inference on the test set. 
 
\subsection{Inference}
The searched network by the transform domain search module can be used for inference on the other test HSIs in $\boldsymbol{Y}$.
For each test HSI $\boldsymbol{Y}_i\in \boldsymbol{Y}$, its reconstructed HSI $\boldsymbol{Y}_r$ can be obtained by:
 \begin{equation}
 	\label{eq:AET}
 	\boldsymbol{Y}_r=f\left( \boldsymbol{Y}_i;\tilde{\boldsymbol{\theta}} \right), 
 \end{equation}
 where $\tilde{\boldsymbol{\theta}}$ represents the optimized parameters of the image reconstruction module. 
Then, GRX detector takes detection on $\boldsymbol{Y}_r$ to obtain the anomaly score map of $\boldsymbol{Y}_i$.
 As same as \eqrefnew{eq:GRX}, the anomaly score of each spectrum ${\boldsymbol{y}}$ on $\boldsymbol{Y}_r$ is calculated according to:
 \begin{equation}
 	 		\label{eq:GRX1}
\boldsymbol{M}\left( \boldsymbol{y} \right) =\left( \boldsymbol{y}-\boldsymbol{\mu } \right) ^T\boldsymbol{\varLambda }^{-1}\left( \boldsymbol{y}-\boldsymbol{\mu } \right), 
\end{equation}
where $\boldsymbol{\mu }\in \mathbb{R} ^{B\times 1}$ and $\boldsymbol{\varLambda }^{-1} \in \mathbb{R} ^{B\times B}$ are the mean vector and the inverse covariance matrix of all the spectra on $\boldsymbol{Y}_r$, respectively. 
We can also replace GRX detector with other anomaly detectors. Since GRX detector is efficient and parameter-free, we specify that \eqrefnew{eq:AET} and \eqrefnew{eq:GRX1} are standard procedures  of our method.

\section{Experiments}
\label{sec:experiments}

In this section, we first introduce our developed hyperspectral AD dataset and two metrics used to evaluate methods on our dataset.
Then, we introduce the experimental settings  and  analyze the experimental results in detail.

\begin{figure}[t]
	\centering
	\subfloat[\label{sfig:train_aviris_ng1}]{\includegraphics[width=0.6in]{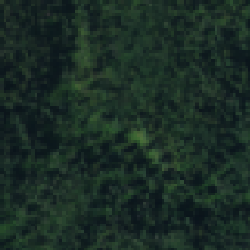}}
	\hfil
	\subfloat[\label{sfig:train_aviris_ng2}]{\includegraphics[width=0.6in]{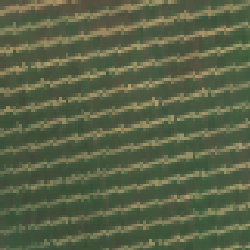}}
	\hfil
	\subfloat[\label{sfig:train_aviris_ng3}]{\includegraphics[width=0.6in]{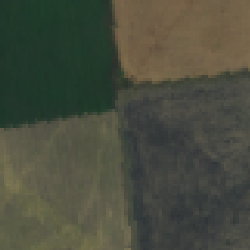}}
	\hfil
	\subfloat[\label{sfig:train_aviris_ng4}]{\includegraphics[width=0.6in]{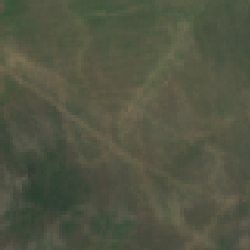}}
	\hfil
	\subfloat[\label{sfig:train_aviris_ng5}]{\includegraphics[width=0.6in]{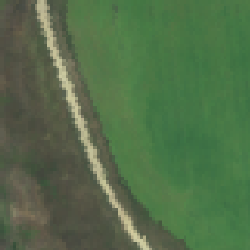}}
	\hfil
	\\
	\subfloat[\label{sfig:train_aviris_ng6}]{\includegraphics[width=0.6in]{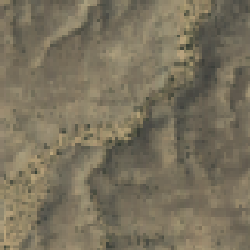}}
	\hfil
	\subfloat[\label{sfig:train_aviris_ng7}]{\includegraphics[width=0.6in]{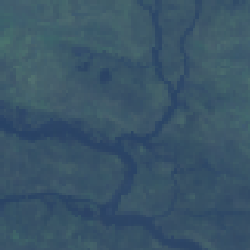}}
	\hfil
	\subfloat[\label{sfig:train_aviris_ng8}]{\includegraphics[width=0.6in]{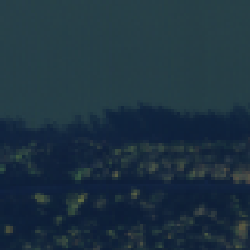}}
	\hfil
	\subfloat[\label{sfig:train_aviris_ng9}]{\includegraphics[width=0.6in]{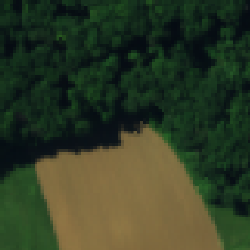}}
	\hfil
	\subfloat[\label{sfig:train_aviris_ng10}]{\includegraphics[width=0.6in]{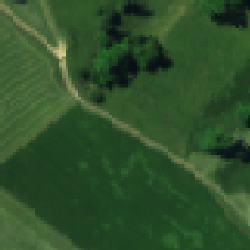}}
	\hfil
	\caption{Example scenes on the AVIRIS-NG training set. }
	\label{fig:train_aviris_ng}
\end{figure}	

\begin{figure}[t]
	\centering
	\subfloat[\label{sfig:train_aviris1}]{\includegraphics[width=0.6in]{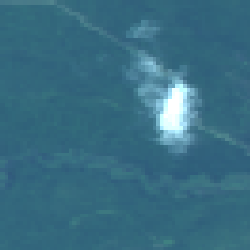}}
	\hfil
	\subfloat[\label{sfig:train_aviris2}]{\includegraphics[width=0.6in]{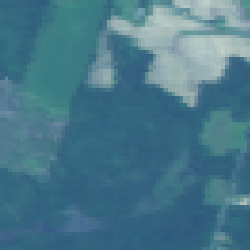}}
	\hfil
	\subfloat[\label{sfig:train_aviris3}]{\includegraphics[width=0.6in]{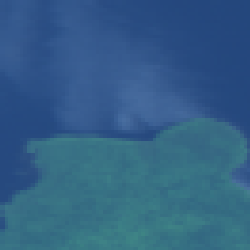}}
	\hfil
	\subfloat[\label{sfig:train_aviris4}]{\includegraphics[width=0.6in]{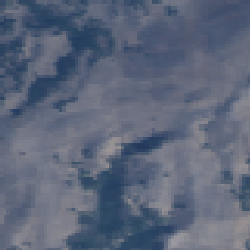}}
	\hfil
	\subfloat[\label{sfig:train_aviris5}]{\includegraphics[width=0.6in]{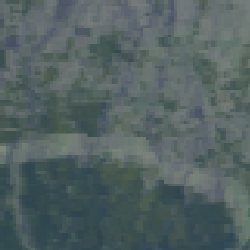}}
	\hfil
	\\
	\subfloat[\label{sfig:train_aviris6}]{\includegraphics[width=0.6in]{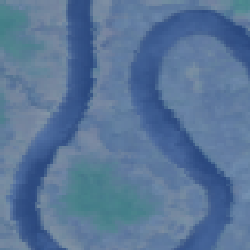}}
	\hfil
	\subfloat[\label{sfig:train_aviris7}]{\includegraphics[width=0.6in]{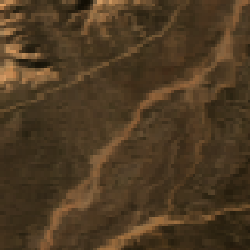}}
	\hfil
	\subfloat[\label{sfig:train_aviris8}]{\includegraphics[width=0.6in]{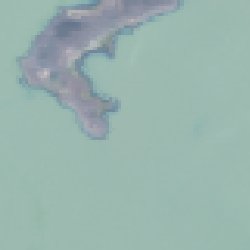}}
	\hfil
	\subfloat[\label{sfig:train_aviris9}]{\includegraphics[width=0.6in]{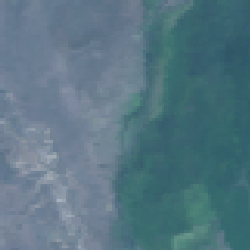}}
	\hfil
	\subfloat[\label{sfig:train_aviris10}]{\includegraphics[width=0.6in]{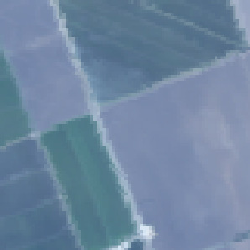}}
	\hfil
	\caption{Example scenes on the AVIRIS-Classic training set. }
	\label{fig:train_aviris}
\end{figure}	

\begin{figure*}[t]
	\centering
	\renewcommand{\thesubfigure}{\arabic{subfigure}}
	\subfloat[\label{sfig:test1}]{\includegraphics[width=0.5in]{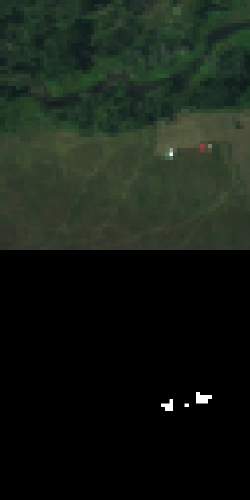}}
	\hfil
	\subfloat[\label{sfig:test2}]{\includegraphics[width=0.5in]{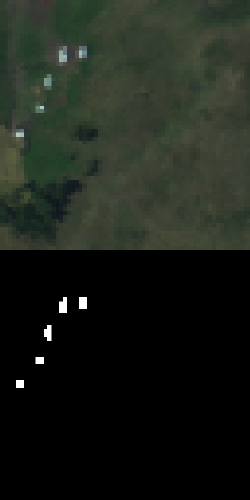}}
	\hfil
	\subfloat[\label{sfig:test3}]{\includegraphics[width=0.5in]{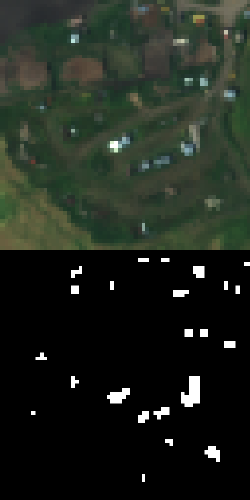}}
	\hfil
	\subfloat[\label{sfig:test4}]{\includegraphics[width=0.5in]{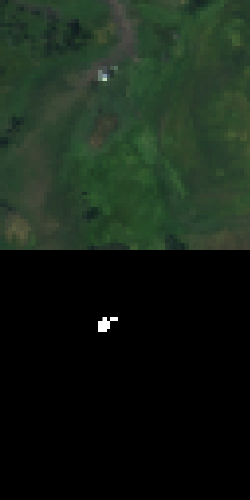}}
	\hfil
	\subfloat[\label{sfig:test5}]{\includegraphics[width=0.5in]{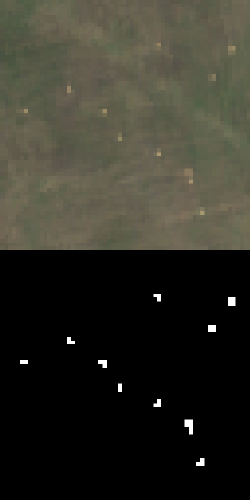}}
	\hfil
	\subfloat[\label{sfig:test6}]{\includegraphics[width=0.5in]{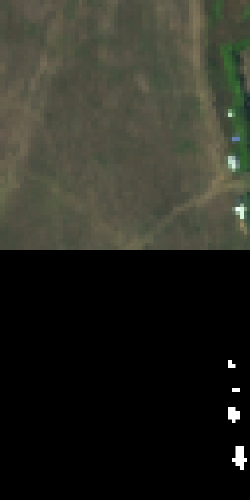}}
	\hfil
	\subfloat[\label{sfig:test7}]{\includegraphics[width=0.5in]{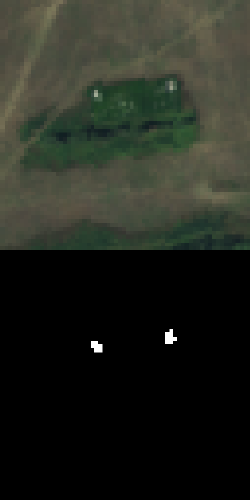}}
	\hfil
	\subfloat[\label{sfig:test8}]{\includegraphics[width=0.5in]{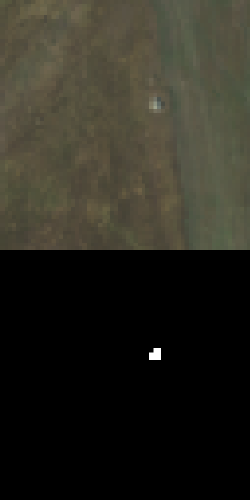}}
	\hfil
	\subfloat[\label{sfig:test9}]{\includegraphics[width=0.5in]{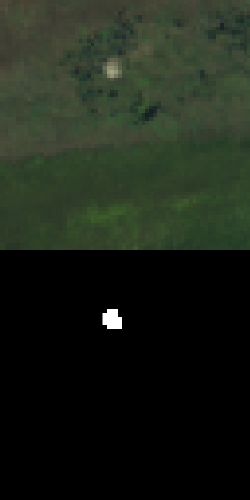}}
	\hfil
	\subfloat[\label{sfig:test10}]{\includegraphics[width=0.5in]{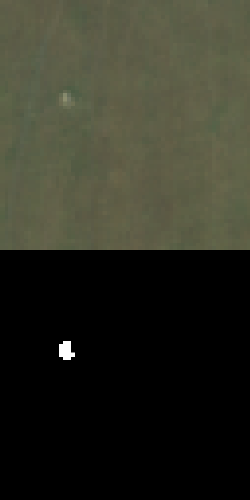}}
	\hfil
	\\
	\vspace{-3mm}
	\subfloat[\label{sfig:test11}]{\includegraphics[width=0.5in]{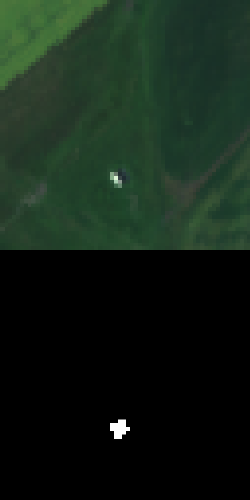}}
	\hfil
	\subfloat[\label{sfig:test12}]{\includegraphics[width=0.5in]{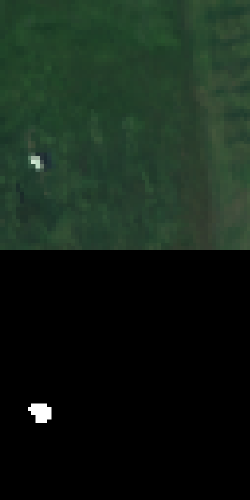}}
	\hfil
	\subfloat[\label{sfig:test13}]{\includegraphics[width=0.5in]{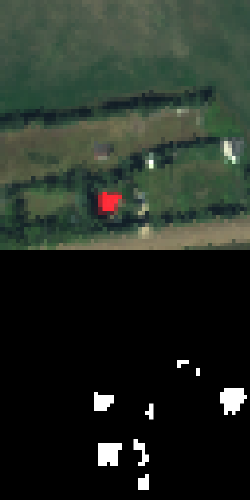}}
	\hfil
	\subfloat[\label{sfig:test14}]{\includegraphics[width=0.5in]{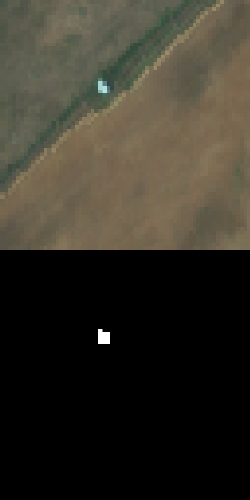}}
	\hfil
	\subfloat[\label{sfig:test15}]{\includegraphics[width=0.5in]{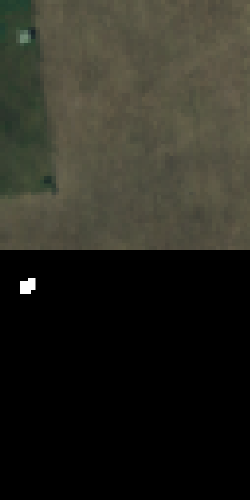}}
	\hfil
	\subfloat[\label{sfig:test16}]{\includegraphics[width=0.5in]{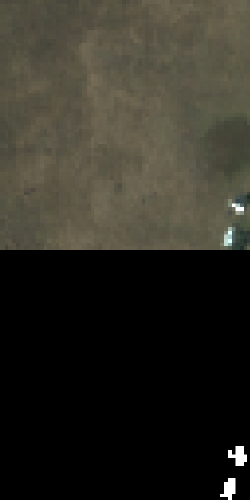}}
	\hfil
	\subfloat[\label{sfig:test17}]{\includegraphics[width=0.5in]{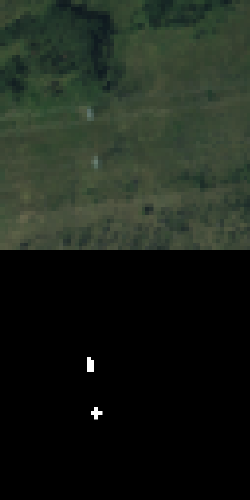}}
	\hfil
	\subfloat[\label{sfig:test18}]{\includegraphics[width=0.5in]{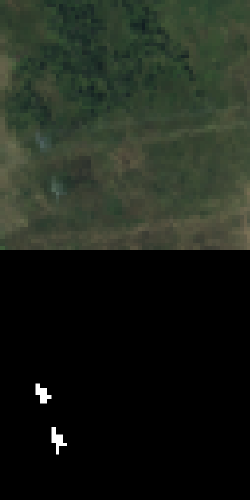}}
	\hfil
	\subfloat[\label{sfig:test19}]{\includegraphics[width=0.5in]{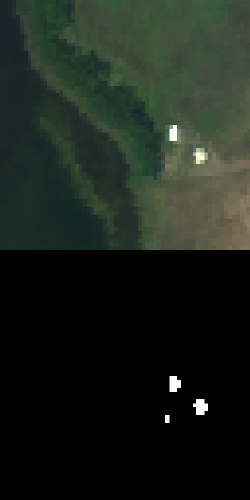}}
	\hfil
	\subfloat[\label{sfig:test20}]{\includegraphics[width=0.5in]{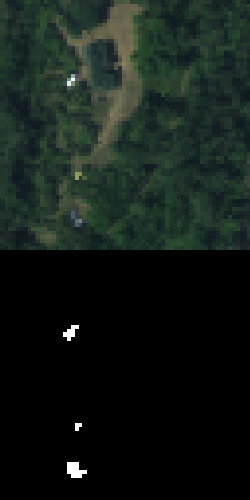}}
	\hfil
	\\
	\vspace{-3mm}
	\subfloat[\label{sfig:test21}]{\includegraphics[width=0.5in]{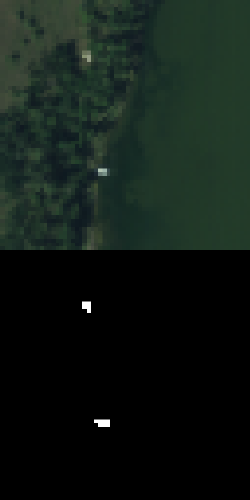}}
	\hfil
	\subfloat[\label{sfig:test22}]{\includegraphics[width=0.5in]{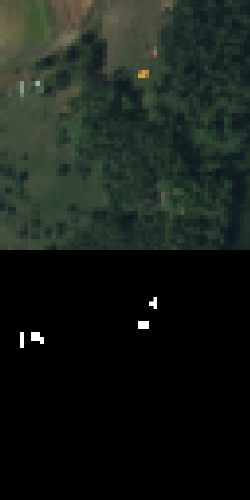}}
	\hfil
	\subfloat[\label{sfig:test23}]{\includegraphics[width=0.5in]{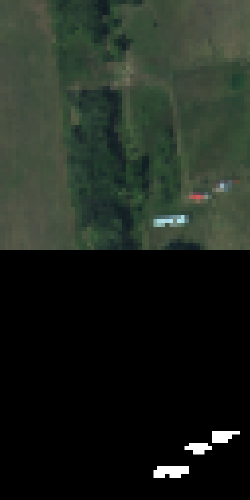}}
	\hfil
	\subfloat[\label{sfig:test24}]{\includegraphics[width=0.5in]{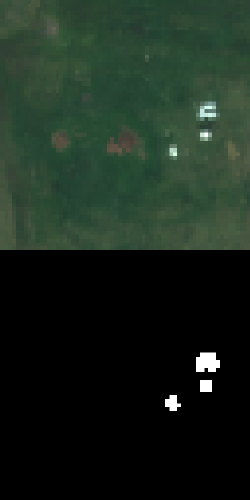}}
	\hfil
	\subfloat[\label{sfig:test25}]{\includegraphics[width=0.5in]{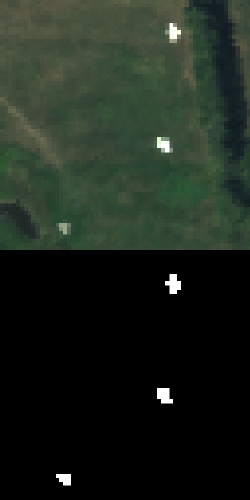}}
	\hfil
	\subfloat[\label{sfig:test26}]{\includegraphics[width=0.5in]{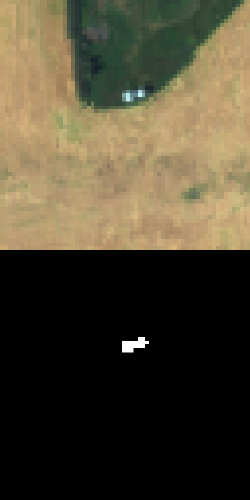}}
	\hfil
	\subfloat[\label{sfig:test27}]{\includegraphics[width=0.5in]{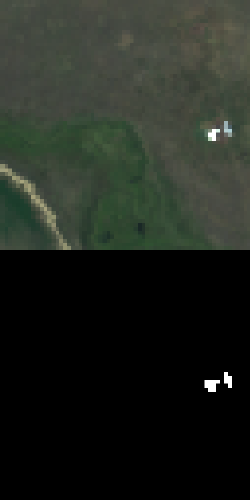}}
	\hfil
	\subfloat[\label{sfig:test28}]{\includegraphics[width=0.5in]{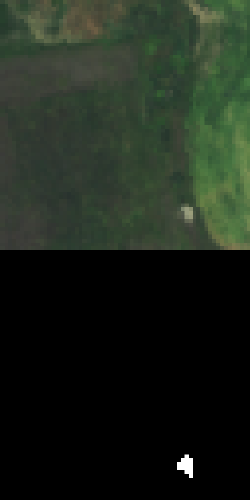}}
	\hfil
	\subfloat[\label{sfig:test29}]{\includegraphics[width=0.5in]{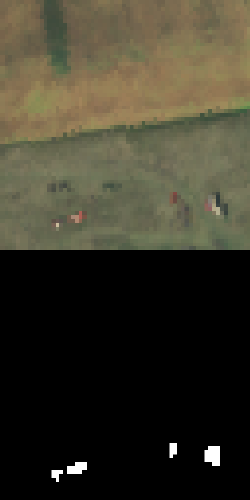}}
	\hfil	
	\subfloat[\label{sfig:test30}]{\includegraphics[width=0.5in]{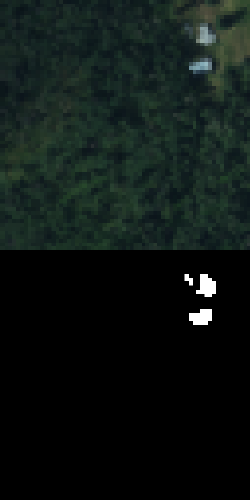}}
	\hfil
	\caption{False color maps and ground truth of example scenes (the test scenes 1-30) on the HAD100 dataset. }
	\label{fig:testset1}
\end{figure*}

\begin{table}[t]
		\renewcommand{\arraystretch}{1.1}
	\setlength\tabcolsep{4pt}
	\centering
	\caption{Data Capture Information on the AVIRIS-Classic Training Set}
	\begin{tabular}{ccccc}
		\hline
		\multirow{2}[0]{*}{Flight Line ID}& \multirow{2}[0]{*}{Date} & Sensor & \multirow{2}[0]{*}{Resolution} & Training\\
	 				&  & Device &  & Images \\
	 		\hline			
		F080709t01p00r15 & 2008/7/9 & AVIRIS-Classic & 17.0m & 3 \\
		F080723t01p00r06 & 2008/7/23 & AVIRIS-Classic & 17.2m & 1 \\
		F080723t01p00r07 & 2008/7/23 & AVIRIS-Classic & 17.0m & 2 \\
		F090706t01p00r06 & 2009/7/6 & AVIRIS-Classic & 16.6m & 12 \\
		F090710t01p00r10 & 2009/7/10 & AVIRIS-Classic & 16.8m & 42 \\
		F090715t01p00r11 & 2009/7/15 & AVIRIS-Classic & 16.2m & 13 \\
		F100825t01p00r05 & 2010/8/25 & AVIRIS-Classic & 16.0m & 6 \\
		F131030t01p00r14 & 2013/10/30 & AVIRIS-Classic & 16.3m & 18 \\
		F170119t01p00r09 & 2017/1/19 & AVIRIS-Classic & 13.4m & 20 \\
		F170507t01p00r09 & 2017/5/7 & AVIRIS-Classic & 16.8m & 26 \\
		F170507t01p00r10 & 2017/5/7 & AVIRIS-Classic & 16.9m & 16 \\
		F180111t01p00r06 & 2018/1/11 & AVIRIS-Classic & 15.6m & 12 \\
		F180118t01p00r19 & 2018/1/18 & AVIRIS-Classic & 17.0m & 16 \\
		F180601t01p00r09 & 2018/6/1 & AVIRIS-Classic & 15.0m & 75 \\
				\hline
		Total & -  & -  & -  & 262 \\
				\hline
	\end{tabular}%
	\label{tab:aviris}%
\end{table}%

\begin{table}[t]
 \renewcommand{\arraystretch}{1.1}
	\setlength\tabcolsep{2pt}
	\centering
	\caption{Data Capture Information on the AVIRIS-NG Training and Test Set}
	\begin{tabular}{cccccc}
			\hline
		\multirow{2}[0]{*}{Flight Line ID} & \multirow{2}[0]{*}{Date} & Sensor & \multirow{2}[0]{*}{Resolution} & Training& Test\\
	&  & Device &  & Images & Images \\
		\hline
		Ang20170821t183707 & 2017/8/21 & AVIRIS-NG & 2.3m & 90 & 38 \\
		Ang20170821t195229 & 2017/8/21 & AVIRIS-NG & 2.5m & 20 & 8 \\
		Ang20170825t173426 & 2017/8/25 & AVIRIS-NG & 3.0m & 13 & 5 \\
		Ang20170908t225309 & 2017/9/8 & AVIRIS-NG & 2.0m & 30 & 13 \\
		Ang20171012t194435 & 2017/10/12 & AVIRIS-NG & 2.0m & 22 & 12 \\
		Ang20180227t082814 & 2018/2/27 & AVIRIS-NG & 4.0m & 14 & 6 \\
		Ang20191004t185054 & 2019/10/4 & AVIRIS-NG & 8.4m & 21 & 6 \\
		Ang20191004t215336 & 2019/10/4 & AVIRIS-NG & 8.4m & 10 & 3 \\
		Ang20191027t204454 & 2019/10/27 & AVIRIS-NG & 5.1m & 9  & 1 \\
		Ang20210614t141018 & 2021/6/14 & AVIRIS-NG & 1.9m & 31 & 8 \\
			\hline
		Total & -  & -  & -  & 260 & 100 \\
			\hline
	\end{tabular}%
	\label{tab:aviris-ng}%
\end{table}%

\subsection{HAD100 Dataset}
\label{sec:dataset}
\begin{table}[t]
	 \renewcommand{\arraystretch}{1.1}
	\centering
	\caption{Differences between AVIRIS-Classic and AVIRIS-NG}
	\begin{tabular}{ccccc}
		\hline
		Sensor & \multirow{2}[0]{*}{Spectral Range} & Spectral & Original & Reserved\\
		Device &  & Resolution & Bands & Bands \\
		\hline
		AVIRIS-Classic & 366-2497nm & 10nm & 224 & 162 \\
		AVIRIS-NG & 377-2500nm & 5nm & 425 & 276 \\
		\hline
	\end{tabular}%
	\label{tab:sensor}%
\end{table}%
 Although many novel AD methods have been proposed in recent years, there is no large-scale dataset that can comprehensively evaluate the performance of these methods.
Existing methods generally report their best results after fine-tuning on different test scenes.
However, fine-tuned parameters cannot be guaranteed to be applicable to other scenes.
 Some recent AD methods claimed that they  could perform inference on unseen test scenes  but were only evaluated on a limited number of test scenes.
 
Consequently, building a unified and comprehensive evaluation benchmark of hyperspectral AD is necessary.
For this purpose, we develop a large-scale \textbf{H}yperspectral \textbf{A}nomaly \textbf{D}etection dataset that contains \textbf{100} real remote sensing test scenes, called \textbf{HAD100} dataset. 
Our dataset has more test scenes than the current popular ABU dataset \cite{AED-AD}, which only contains 13 test scenes. 
All the test HSIs are downloaded from AVIRIS-NG (Airborne Visible InfraRed Imaging Spectrometer-Next Generation) website\footnote{https://avirisng.jpl.nasa.gov/} and uniformly cropped to patches of size $64\times64$.
\figref{fig:testset1}  shows the false color maps and ground truth of example scenes in the test set.
The marked targets are mainly compact manufactured objects such as vehicles, boats, and buildings.
The size of targets ranges from 1 pixel to 69 pixels.
Furthermore, our dataset contains various backgrounds, including grassland, forest, farmland, desert, lake, river, and coast.
 The HAD100  dataset can be downloaded from our website\footnote{https://ZhaoxuLi123.github.io/HAD100}.

In order to inspire new ideas for hyperspectral AD, we also provide two training sets which are captured by AVIRIS-NG and AVIRIS-Classic\footnote{https://aviris.jpl.nasa.gov/}, respectively.
 As shown in Figs. 5 and 6, the HSIs in the training set are only cropped on the background areas, with the space sizes ranging from $66\times66$ to $130\times130$.
As shown in \tabref{tab:aviris-ng}, the AVIRIS-NG training set contains 260 normal HSIs, which are captured by the same sensor device and capture flight lines as the test set. 
It means that most background spectra are common to both the AVIRIS-NG training set and the test set.
As shown in \tabref{tab:aviris}, the AVIRIS-Classic training set contains 262 normal HSIs, which are captured by a different sensor device from the test set.
Within a similar data scale, the AVIRIS-NG training set and the AVIRIS-Classic training set can be used for the generalization evaluation of AD methods to different degrees.
 
\tabref{tab:sensor} shows the differences between AVIRIS-Classic and AVIRIS-NG.
 AVIRIS-NG is the upgraded version of AVIRIS-Classic. 
 Under the condition that the spectral range is almost unchanged, AVIRIS-NG has higher spectral resolution and more bands than AVIRIS-Classic.
We remove water absorption bands and noise bands in the training and test sets.
AVIRIS-NG data retains 276 bands, of which ids are 16-109, 119-145, 159-187, 228-274, and 329-407.
AVIRIS-Classic data retains 162 bands, of which ids are 8-57, 66-79, 86-104, 123-149, and 173-224.

\subsection{Evaluation Metrics}
In this paper, we use two evaluation metrics to compare the detection performance of different AD methods on our HAD100 dataset.
The receiver operating characteristic (ROC) area under the curve (AUC) is the most widely used evaluation tool in the hyperspectral AD task and can directly give quantitative results.
Besides, researchers usually use Box-whisker Plot to  compare the anomaly-background separability of AD methods qualitatively.
We replace Box-whisker Plot with our improved version of signal-to-noise probability ratio (SNPR) \cite{3droc} to  evaluate the anomaly-background separability quantitatively.
\subsubsection{AUC}
ROC curve shows the corresponding relationship between detection probability $P_d$ and false alarm probability $P_f$.
Given a segmentation threshold $\tau$ and an anomaly score map $\boldsymbol{M}$, we specify that pixels whose anomaly scores greater than or equal to $\tau$ are classified as positive samples, and $P_d$ and $P_f$ can be calculated according to:
 \begin{equation}
	P_d=\frac{\mathrm{TP}}{\mathrm{TP}+\mathrm{FN}}, 
\end{equation}
 \begin{equation}
	P_f=\frac{\mathrm{FP}}{\mathrm{TN}+\mathrm{FP}}, 
\end{equation}
where TP is the number of target pixels that are predicted as positive samples, FN is the number of target pixels that are predicted as negative samples,
FP is the number of background pixels that are predicted as positive samples, and TN is the number of background pixels that are predicted as negative samples.
Each $\tau$ can produce a pair of $P_d$ and $P_f$ as a point ($P_d$, $P_f$) on the ROC curve.
We traverse every anomaly score value on $\boldsymbol{M}$ as $\tau$ and generate a discrete ROC curve.
Then, the AUC value of $\boldsymbol{M}$ is calculated by:
 \begin{equation}
 	 	\label{eq:auc}
	\mathrm{AUC}=\frac{1}{2} \sum_{l=1}^{p-1}\left(P_f^{l+1}-P_f^l\right)\left(P_d^{l+1}+P_d^l\right),
\end{equation}
where $\left( P_f^l, P_d^l\right)$ denotes the $l$-th point on the discrete ROC curve and $p$ is the number of points.
 A larger AUC value means a better detection performance.
 In particular, the AUC value is specified as 0.5 when the anomaly score of each pixel is equal.
 We use the mean AUC value (mAUC) on all the test HSIs in HAD100 dataset to evaluate the AD methods throughout the rest of this paper.

\subsubsection{ASNPR}
A recent study \cite{3droc} deeply analyzed the evaluation tools for hyperspectral target detection, and proposed signal-to-noise probability ratio (SNPR) to evaluate background suppressibility. 
SNPR is calculated based on the 3D ROC curve.
In addition to $P_d$ and $P_f$, the 3D ROC curve introduces the segmentation threshold $\tau$ as the third variable, which is sampled from the anomaly score map with Maxmin Normalization.
A 3D ROC curve can generate three 2D ROC curves, which take $\left( P_d,P_f \right)$, $\left( P_d,\tau \right)$, and $\left( P_f,\tau \right)$ as variables, respectively.
Similar to \eqrefnew{eq:auc}, the AUC value urder the 2D ROC curve $\left( P_d,\tau \right)$ is calculated according to:
\begin{equation}
\mathrm{AUC}_{d,\tau}=\frac{1}{2}\sum_{l=1}^{p-1}{\left( \tau ^{l+1}-\tau ^l \right) \left( P_{d}^{l+1}+P_{d}^{l} \right)},
\end{equation}
and the AUC value urder the 2D ROC curve $\left( P_f,\tau \right)$ is calculated according to:
\begin{equation}
	\mathrm{AUC}_{f,\tau}=\frac{1}{2}\sum_{l=1}^{p-1}{\left( \tau ^{l+1}-\tau ^l \right) \left( P_{f}^{l+1}+P_{f}^{l} \right)}.
\end{equation}
Then, the SNPR value is calculated according to:
\begin{equation}
	\mathrm{SNPR}=\frac{\mathrm{AUC}_{d,\tau}}{\mathrm{AUC}_{f,\tau}}.
\end{equation}
A larger SNPR value means a better background suppressibility of AD methods.

In practice, we find that the above 3D ROC-based evaluation metrics are unsuitable for scenes with multiple types of targets.
When a certain target has an extremely high anomaly score, other targets can have low values in the normalized anomaly score map.
Whether these targets are separated from background or not, the $\mathrm{AUC}_{f,\tau}$ value remains low.
To solve this problem, we develop adaptive versions of $\mathrm{AUC}_{d,\tau}$, $\mathrm{AUC}_{f, \tau}$, and $\mathrm{SNPR}$.
Specifically, we set the median value of anomaly scores on the real targets as the upper bound of the anomaly score map.
Calculated on the truncated anomaly score map, $\mathrm{AUC}_{d,\tau}$, $\mathrm{AUC}_{f, \tau}$ and $\mathrm{SNPR}$ are robust to the maximum anomaly response value.
The mean value of the adaptive $\mathrm{SNPR}$ on all the test HSIs is calculated according to:
\begin{equation}
	\mathrm{mASNPR}=\sum_{k=1}^n{10\log _{10}\frac{\mathrm{AAUC}_{d,\tau}^{k}}{\mathrm{AAUC}_{f,\tau}^{k}}},
\end{equation}
where $n$ is the total number of the test HSIs, and $\mathrm{AAUC}_{d,\tau}^{k}$ and $\mathrm{AAUC}_{f,\tau}^{k}$ are the adaptive $\mathrm{AUC}_{d,\tau}$ and $\mathrm{AUC}_{f, \tau}$ on the $k$-th test HSI, respectively.
Since the value range of ASNPR is $\left[ 0,+\infty \right) $, we introduce logarithmic operation to avoid over-large ASNPR values of individual test HSIs.
In particular, the logarithmic value of ASNPR is specified as 0 dB when the anomaly score of each pixel is equal.

\begin{table}[t]
	\renewcommand{\arraystretch}{1.3}
	\setlength\tabcolsep{4pt}
	\centering
	\caption{The Best Dual Window Sizes of Different AD Methods}
	\begin{tabular}{cccccccc}
			\hline
		Parameter & LRX & KRX & QLRX & 2S-GLRT & KSVDD & CR & KCSR \\
			\hline
		$\omega_\text{in}$ & 13 & 3  & 11 & 3  & 3  & 5  & 3 \\
			\hline
		$\omega_\text{out}$ & 29 & 7  & 29 & 29 & 29 & 29 & 29 \\
			\hline
	\end{tabular}%
	\label{tab:window}%
\end{table}%

\begin{table*}[htbp]
	\renewcommand{\arraystretch}{1.3}
	\setlength\tabcolsep{8pt}
	\centering
	\caption{Detection Performance and Inference Time (in Seconds) of Different AD Methods on the Frist 50, 100, and 200 Bands of HAD100 Dataset.
		The bold black, red, and blue  font in each column indicate the top three values.}
	\begin{tabular}{cccrccrccr}
		\hline
		\multirow{2}[0]{*}{Method} & \multicolumn{3}{c}{\cellcolor[rgb]{ .749, .749, .749}The First 50 Bands} & \multicolumn{3}{c}{\cellcolor[rgb]{ .553, .706, .886}The First 100 Bands} & \multicolumn{3}{c}{\cellcolor[rgb]{ .847, .894, .737} The First 200 Bands} \\
	 & \cellcolor[rgb]{ .749, .749, .749}mAUC & \cellcolor[rgb]{ .749, .749, .749}mASNPR & \cellcolor[rgb]{ .749, .749, .749}time~~& \cellcolor[rgb]{ .553, .706, .886}mAUC & \cellcolor[rgb]{ .553, .706, .886}mASNPR & \cellcolor[rgb]{ .553, .706, .886}time~ & \cellcolor[rgb]{ .847, .894, .737}mAUC & \cellcolor[rgb]{ .847, .894, .737}mASNPR & \cellcolor[rgb]{ .847, .894, .737}time~ \\
		\hline
		GRX & \cellcolor[rgb]{ .749, .749, .749}0.9799 & \cellcolor[rgb]{ .749, .749, .749}7.93 & \cellcolor[rgb]{ .749, .749, .749}\textbf{0.022} & \cellcolor[rgb]{ .553, .706, .886}0.9714 & \cellcolor[rgb]{ .553, .706, .886}7.03 & \cellcolor[rgb]{ .553, .706, .886}\textbf{0.031} & \cellcolor[rgb]{ .847, .894, .737}0.9649 & \cellcolor[rgb]{ .847, .894, .737}6.59 & \cellcolor[rgb]{ .847, .894, .737}\textbf{0.058} \\
		\hline
		LRX & \cellcolor[rgb]{ .749, .749, .749}0.9775 & \cellcolor[rgb]{ .749, .749, .749}11.21~ & \cellcolor[rgb]{ .749, .749, .749}2.492 & \cellcolor[rgb]{ .553, .706, .886}0.9692 & \cellcolor[rgb]{ .553, .706, .886}\textcolor[rgb]{ 1, 0, 0}{\textbf{10.92~}} & \cellcolor[rgb]{ .553, .706, .886}6.272 & \cellcolor[rgb]{ .847, .894, .737}0.9720 & \cellcolor[rgb]{ .847, .894, .737}\textbf{11.79~} & \cellcolor[rgb]{ .847, .894, .737}19.560 \\
		KRX & \cellcolor[rgb]{ .749, .749, .749}0.9616 & \cellcolor[rgb]{ .749, .749, .749}10.59~ & \cellcolor[rgb]{ .749, .749, .749}1.918 & \cellcolor[rgb]{ .553, .706, .886}0.9329 & \cellcolor[rgb]{ .553, .706, .886}7.54 & \cellcolor[rgb]{ .553, .706, .886}2.435 & \cellcolor[rgb]{ .847, .894, .737}0.9282 & \cellcolor[rgb]{ .847, .894, .737}7.30 & \cellcolor[rgb]{ .847, .894, .737}2.506 \\
		QLRX & \cellcolor[rgb]{ .749, .749, .749}0.9801 & \cellcolor[rgb]{ .749, .749, .749}7.85 & \cellcolor[rgb]{ .749, .749, .749}1.091 & \cellcolor[rgb]{ .553, .706, .886}0.9715 & \cellcolor[rgb]{ .553, .706, .886}6.96 & \cellcolor[rgb]{ .553, .706, .886}1.725 & \cellcolor[rgb]{ .847, .894, .737}0.9682 & \cellcolor[rgb]{ .847, .894, .737}6.86 & \cellcolor[rgb]{ .847, .894, .737}2.947 \\
		2S-GLRT & \cellcolor[rgb]{ .749, .749, .749}0.9853 & \cellcolor[rgb]{ .749, .749, .749}9.38 & \cellcolor[rgb]{ .749, .749, .749}5.207 & \cellcolor[rgb]{ .553, .706, .886}0.9683 & \cellcolor[rgb]{ .553, .706, .886}8.39 & \cellcolor[rgb]{ .553, .706, .886}11.765 & \cellcolor[rgb]{ .847, .894, .737}0.9310 & \cellcolor[rgb]{ .847, .894, .737}7.43 & \cellcolor[rgb]{ .847, .894, .737}30.830 \\
		KSVDD & \cellcolor[rgb]{ .749, .749, .749}0.9845 & \cellcolor[rgb]{ .749, .749, .749}6.47 & \cellcolor[rgb]{ .749, .749, .749}43.563 & \cellcolor[rgb]{ .553, .706, .886}0.9737 & \cellcolor[rgb]{ .553, .706, .886}4.87 & \cellcolor[rgb]{ .553, .706, .886}75.653 & \cellcolor[rgb]{ .847, .894, .737}0.9706 & \cellcolor[rgb]{ .847, .894, .737}4.42 & \cellcolor[rgb]{ .847, .894, .737}140.750\\
		CR & \cellcolor[rgb]{ .749, .749, .749}\textcolor[rgb]{ 0, 0, 1}{\textbf{0.9929}} & \cellcolor[rgb]{ .749, .749, .749}11.24~ & \cellcolor[rgb]{ .749, .749, .749}27.175 & \cellcolor[rgb]{ .553, .706, .886}0.9859 & \cellcolor[rgb]{ .553, .706, .886}10.24~ & \cellcolor[rgb]{ .553, .706, .886}24.361 & \cellcolor[rgb]{ .847, .894, .737}0.9849 & \cellcolor[rgb]{ .847, .894, .737}10.12~ & \cellcolor[rgb]{ .847, .894, .737}25.882 \\
		KCSR & \cellcolor[rgb]{ .749, .749, .749}\textcolor[rgb]{ 1, 0, 0}{\textbf{0.9939}} & \cellcolor[rgb]{ .749, .749, .749}10.48~ & \cellcolor[rgb]{ .749, .749, .749}6.006 & \cellcolor[rgb]{ .553, .706, .886}\textcolor[rgb]{ 0, 0, 1}{\textbf{0.9865}} & \cellcolor[rgb]{ .553, .706, .886}7.73 & \cellcolor[rgb]{ .553, .706, .886}9.475 & \cellcolor[rgb]{ .847, .894, .737}\textcolor[rgb]{ 1, 0, 0}{\textbf{0.9860}} & \cellcolor[rgb]{ .847, .894, .737}7.13 & \cellcolor[rgb]{ .847, .894, .737}17.286 \\
		\hline
		KIFD & \cellcolor[rgb]{ .749, .749, .749}0.9829 & \cellcolor[rgb]{ .749, .749, .749}8.14 & \cellcolor[rgb]{ .749, .749, .749}35.725 & \cellcolor[rgb]{ .553, .706, .886}0.9822 & \cellcolor[rgb]{ .553, .706, .886}7.89 & \cellcolor[rgb]{ .553, .706, .886}34.681 & \cellcolor[rgb]{ .847, .894, .737}0.9774 & \cellcolor[rgb]{ .847, .894, .737}7.61 & \cellcolor[rgb]{ .847, .894, .737}34.267 \\
		PTA & \cellcolor[rgb]{ .749, .749, .749}0.9031 & \cellcolor[rgb]{ .749, .749, .749}6.32 & \cellcolor[rgb]{ .749, .749, .749}4.170 & \cellcolor[rgb]{ .553, .706, .886}0.8976 & \cellcolor[rgb]{ .553, .706, .886}4.72 & \cellcolor[rgb]{ .553, .706, .886}14.201 & \cellcolor[rgb]{ .847, .894, .737}0.8686 & \cellcolor[rgb]{ .847, .894, .737}4.34 & \cellcolor[rgb]{ .847, .894, .737}34.728 \\
		FrFE & \cellcolor[rgb]{ .749, .749, .749}0.9476 & \cellcolor[rgb]{ .749, .749, .749}7.80 & \cellcolor[rgb]{ .749, .749, .749}1.292 & \cellcolor[rgb]{ .553, .706, .886}0.9371 & \cellcolor[rgb]{ .553, .706, .886}7.05 & \cellcolor[rgb]{ .553, .706, .886}3.091 & \cellcolor[rgb]{ .847, .894, .737}0.9164 & \cellcolor[rgb]{ .847, .894, .737}6.56 & \cellcolor[rgb]{ .847, .894, .737}11.955 \\
		AED & \cellcolor[rgb]{ .749, .749, .749}0.9859 & \cellcolor[rgb]{ .749, .749, .749}8.79 & \cellcolor[rgb]{ .749, .749, .749}0.050 & \cellcolor[rgb]{ .553, .706, .886}0.9645 & \cellcolor[rgb]{ .553, .706, .886}8.47 & \cellcolor[rgb]{ .553, .706, .886}\textcolor[rgb]{ 0, 0, 1}{\textbf{0.054}} & \cellcolor[rgb]{ .847, .894, .737}0.9625 & \cellcolor[rgb]{ .847, .894, .737}8.38 & \cellcolor[rgb]{ .847, .894, .737}\textcolor[rgb]{ 1, 0, 0}{\textbf{0.067}} \\
		\hline
		RGAE & \cellcolor[rgb]{ .749, .749, .749}0.9003 & \cellcolor[rgb]{ .749, .749, .749}9.98 & \cellcolor[rgb]{ .749, .749, .749}43.599 & \cellcolor[rgb]{ .553, .706, .886}0.8876 & \cellcolor[rgb]{ .553, .706, .886}7.37 & \cellcolor[rgb]{ .553, .706, .886}47.748 & \cellcolor[rgb]{ .847, .894, .737}0.8823 & \cellcolor[rgb]{ .847, .894, .737}7.29 & \cellcolor[rgb]{ .847, .894, .737}54.039 \\
		AutoAD & \cellcolor[rgb]{ .749, .749, .749}0.9688 & \cellcolor[rgb]{ .749, .749, .749}10.17~ & \cellcolor[rgb]{ .749, .749, .749}6.305 & \cellcolor[rgb]{ .553, .706, .886}0.9661 & \cellcolor[rgb]{ .553, .706, .886}10.27~ & \cellcolor[rgb]{ .553, .706, .886}6.360 & \cellcolor[rgb]{ .847, .894, .737}0.9642 & \cellcolor[rgb]{ .847, .894, .737}\textcolor[rgb]{ 0, 0, 1}{\textbf{10.86~}} & \cellcolor[rgb]{ .847, .894, .737}6.304 \\
		LREN & \cellcolor[rgb]{ .749, .749, .749}0.9583 & \cellcolor[rgb]{ .749, .749, .749}5.85 & \cellcolor[rgb]{ .749, .749, .749}35.427 & \cellcolor[rgb]{ .553, .706, .886}0.9286 & \cellcolor[rgb]{ .553, .706, .886}4.24 & \cellcolor[rgb]{ .553, .706, .886}35.703 & \cellcolor[rgb]{ .847, .894, .737}0.9309 & \cellcolor[rgb]{ .847, .894, .737}4.84 & \cellcolor[rgb]{ .847, .894, .737}35.546 \\
		$\text{WeaklyAD}_{\mathcal{A}}$&\cellcolor[rgb]{ .749, .749, .749}0.9862 & \cellcolor[rgb]{ .749, .749, .749}11.33~ & \cellcolor[rgb]{ .749, .749, .749}1.887 & \cellcolor[rgb]{ .553, .706, .886}0.9788 & \cellcolor[rgb]{ .553, .706, .886}10.04~ & \cellcolor[rgb]{ .553, .706, .886}1.768 & \cellcolor[rgb]{ .847, .894, .737}0.9773 & \cellcolor[rgb]{ .847, .894, .737}9.39 & \cellcolor[rgb]{ .847, .894, .737}1.782 \\
		\hline
		$\text{WeaklyAD}_{\mathcal{B}}$ & \cellcolor[rgb]{ .749, .749, .749}0.9887 & \cellcolor[rgb]{ .749, .749, .749}\textbf{13.75~} & \cellcolor[rgb]{ .749, .749, .749}\textcolor[rgb]{ 0, 0, 1}{\textbf{0.046}} & \cellcolor[rgb]{ .553, .706, .886}0.9838 & \cellcolor[rgb]{ .553, .706, .886}\textbf{12.76~} & \cellcolor[rgb]{ .553, .706, .886}0.085 & \cellcolor[rgb]{ .847, .894, .737}\textcolor[rgb]{ 0, 0, 1}{\textbf{0.9850}} & \cellcolor[rgb]{ .847, .894, .737}\textcolor[rgb]{ 1, 0, 0}{\textbf{11.46~}} & \cellcolor[rgb]{ .847, .894, .737}0.118 \\
		AETNet & \cellcolor[rgb]{ .749, .749, .749}0.9925 & \cellcolor[rgb]{ .749, .749, .749}\textcolor[rgb]{ 0, 0, 1}{\textbf{11.72~}} & \cellcolor[rgb]{ .749, .749, .749}\textcolor[rgb]{ 1, 0, 0}{\textbf{0.035}} & \cellcolor[rgb]{ .553, .706, .886}\textcolor[rgb]{ 1, 0, 0}{\textbf{0.9875}} & \cellcolor[rgb]{ .553, .706, .886}9.96 & \cellcolor[rgb]{ .553, .706, .886}\textcolor[rgb]{ 1, 0, 0}{\textbf{0.043}} & \cellcolor[rgb]{ .847, .894, .737}0.9818 & \cellcolor[rgb]{ .847, .894, .737}8.06 & \cellcolor[rgb]{ .847, .894, .737}\textcolor[rgb]{ 0, 0, 1}{\textbf{0.070}} \\
		AETNet-KCSR & \cellcolor[rgb]{ .749, .749, .749}\textbf{0.9948} & \cellcolor[rgb]{ .749, .749, .749}\textcolor[rgb]{ 1, 0, 0}{\textbf{13.15~}} & \cellcolor[rgb]{ .749, .749, .749}6.815 & \cellcolor[rgb]{ .553, .706, .886}\textbf{0.9897} & \cellcolor[rgb]{ .553, .706, .886}\textcolor[rgb]{ 0, 0, 1}{\textbf{10.64~}} & \cellcolor[rgb]{ .553, .706, .886}11.169 & \cellcolor[rgb]{ .847, .894, .737}\textbf{0.9869} & \cellcolor[rgb]{ .847, .894, .737}10.63~~& \cellcolor[rgb]{ .847, .894, .737}20.348 \\
		\hline
	\end{tabular}%
	\label{tab:result}%
\end{table*}%

\begin{table*}[htbp]\tiny
	\vspace{-5mm}
	\centering
	\tabcolsep=0.1cm
	\renewcommand\arraystretch{0.95}
	\caption{The AUC performance of different AD methods achieved on the first 50 bands of the HAD100 dataset. The bold black, red, and blue  font in each row indicate the top three values. The gray cells represent that the AUC values are less than 0.95.}
	\vspace{-1mm}
	\scalebox{1.15}{
		\begin{tabular}{cccccccccccccccccccc}
			\hline
			\multirow{2}[0]{*}{Scene} & \multirow{2}[0]{*}{GRX} & \multirow{2}[0]{*}{LRX} & \multirow{2}[0]{*}{KRX} & \multirow{2}[0]{*}{QLRX} & 2S- & \multirow{2}[0]{*}{KSVDD} & \multirow{2}[0]{*}{CR} & \multirow{2}[0]{*}{KCSR} & \multirow{2}[0]{*}{KIFD} & \multirow{2}[0]{*}{PTA} & \multirow{2}[0]{*}{FrFE} & \multirow{2}[0]{*}{AED} & \multirow{2}[0]{*}{RGAE} & \multirow{2}[0]{*}{AutoAD} & \multirow{2}[0]{*}{LREN} & Weakly- & Weakly- & \multirow{2}[0]{*}{AETNet} & AETNet- \\
			&  &  &  &  & GLRT &  &  &  &  &  &  &  &  &  &  & $\text{AD}_{\mathcal{A}}$ & $\text{AD}_{\mathcal{B}}$ & & {KCSR} \\
			\hline
	 1  & \textbf{0.99962} & 0.99144 & \cellcolor[rgb]{ .753, .753, .753}0.94310 & \textcolor[rgb]{ 1, 0, 0}{\textbf{0.99953}} & 0.99897 & 0.98063 & 0.99531 & 0.99469 & 0.98213 & 0.96223 & \textcolor[rgb]{ 0, 0, 1}{\textbf{0.99902}} & \cellcolor[rgb]{ .753, .753, .753}0.90824 & 0.95829 & 0.96457 & 0.97865 & 0.99394 & 0.99804 & 0.99876 & 0.99828 \\
	2  & \textbf{0.99998} & 0.99867 & 0.99910 & \textcolor[rgb]{ 0, 0, 1}{\textbf{0.99997}} & 0.99655 & 0.99500 & 0.99869 & 0.99988 & 0.99728 & 0.99995 & 0.99996 & 0.99995 & 0.99965 & 0.99377 & \textbf{0.99998} & 0.99975 & 0.99864 & 0.99866 & 0.99916 \\
	3  & 0.98819 & 0.96344 & 0.97025 & \textcolor[rgb]{ 0, 0, 1}{\textbf{0.99026}} & 0.97194 & \textcolor[rgb]{ 1, 0, 0}{\textbf{0.99046}} & 0.98986 & \textbf{0.99348} & 0.98666 & \cellcolor[rgb]{ .753, .753, .753}0.93605 & 0.98756 & 0.96613 & \cellcolor[rgb]{ .753, .753, .753}0.94636 & \cellcolor[rgb]{ .753, .753, .753}0.88084 & \cellcolor[rgb]{ .753, .753, .753}0.94661 & 0.97081 & 0.98281 & 0.98082 & 0.98651 \\
	4  & \textbf{0.99978} & 0.99724 & 0.98952 & \textcolor[rgb]{ 1, 0, 0}{\textbf{0.99978}} & 0.99615 & 0.98778 & 0.99835 & 0.99731 & 0.99786 & 0.98932 & 0.99826 & 0.99326 & 0.98553 & 0.98976 & 0.99875 & 0.99784 & \textcolor[rgb]{ 0, 0, 1}{\textbf{0.99902}} & 0.99895 & 0.99878 \\
	5  & 0.97158 & 0.96190 & \textbf{0.99865} & 0.97484 & 0.99273 & 0.98637 & 0.98629 & 0.99371 & 0.98044 & 0.98283 & 0.97653 & 0.99495 & 0.96140 & 0.97070 & \cellcolor[rgb]{ .753, .753, .753}0.87241 & 0.96894 & 0.95974 & \textcolor[rgb]{ 0, 0, 1}{\textbf{0.99601}} & \textcolor[rgb]{ 1, 0, 0}{\textbf{0.99692}} \\
	6  & 0.98724 & 0.97592 & 0.98630 & 0.98177 & 0.99273 & 0.97856 & \textbf{0.99875} & \textcolor[rgb]{ 1, 0, 0}{\textbf{0.99831}} & 0.98991 & 0.97179 & 0.98197 & 0.99387 & 0.95500 & 0.97218 & 0.98563 & 0.99486 & 0.99559 & 0.99581 & \textcolor[rgb]{ 0, 0, 1}{\textbf{0.99719}} \\
	7  & 0.99595 & 0.99195 & 0.95488 & 0.99639 & \textbf{0.99949} & 0.98995 & 0.99722 & 0.99603 & 0.99564 & \cellcolor[rgb]{ .753, .753, .753}0.86162 & 0.99650 & \textcolor[rgb]{ 0, 0, 1}{\textbf{0.99897}} & \cellcolor[rgb]{ .753, .753, .753}0.72763 & 0.95963 & \cellcolor[rgb]{ .753, .753, .753}0.94748 & 0.98644 & 0.98914 & \textcolor[rgb]{ 1, 0, 0}{\textbf{0.99912}} & 0.99895 \\
	8  & \textcolor[rgb]{ 1, 0, 0}{\textbf{0.99994}} & 0.99991 & 0.99737 & \textcolor[rgb]{ 1, 0, 0}{\textbf{0.99994}} & \textcolor[rgb]{ 1, 0, 0}{\textbf{0.99994}} & 0.99761 & 0.99988 & 0.99982 & 0.99957 & 0.99875 & \textbf{1.00000} & 0.99957 & 0.98667 & 0.99991 & 0.98489 & 0.99985 & \textcolor[rgb]{ 1, 0, 0}{\textbf{0.99994}} & \textcolor[rgb]{ 1, 0, 0}{\textbf{0.99994}} & 0.99979 \\
	9  & 0.99975 & \textbf{0.99988} & 0.99878 & 0.99977 & 0.99949 & 0.98921 & 0.99882 & 0.99902 & 0.99932 & 0.99762 & 0.99803 & 0.99732 & \textcolor[rgb]{ 0, 0, 1}{\textbf{0.99978}} & 0.95918 & 0.99825 & 0.99901 & 0.99716 & \textcolor[rgb]{ 1, 0, 0}{\textbf{0.99981}} & 0.99898 \\
	10 & 0.98105 & 0.99640 & 0.99228 & 0.97837 & 0.99673 & 0.99454 & 0.99687 & \textbf{0.99886} & 0.99776 & 0.99741 & 0.98115 & 0.99808 & 0.98854 & 0.97806 & 0.99312 & \textcolor[rgb]{ 1, 0, 0}{\textbf{0.99857}} & 0.99792 & 0.99787 & \textcolor[rgb]{ 0, 0, 1}{\textbf{0.99850}} \\
	11 & 0.98941 & 0.99733 & 0.98850 & 0.99825 & 0.99881 & 0.99643 & 0.99891 & 0.99905 & 0.99844 & \cellcolor[rgb]{ .753, .753, .753}0.94335 & 0.99265 & \textbf{0.99968} & \cellcolor[rgb]{ .753, .753, .753}0.89225 & 0.99493 & 0.99815 & 0.99677 & 0.99819 & \textcolor[rgb]{ 1, 0, 0}{\textbf{0.99919}} & \textcolor[rgb]{ 0, 0, 1}{\textbf{0.99913}} \\
	12 & 0.99664 & \textcolor[rgb]{ 1, 0, 0}{\textbf{0.99903}} & 0.99208 & 0.99541 & 0.99894 & 0.99434 & 0.99869 & 0.99740 & 0.96041 & 0.96143 & 0.99769 & \textcolor[rgb]{ 0, 0, 1}{\textbf{0.99897}} & \cellcolor[rgb]{ .753, .753, .753}0.92498 & 0.99268 & 0.99853 & 0.99837 & 0.99697 & \textbf{0.99907} & 0.99846 \\
	13 & 0.97049 & \cellcolor[rgb]{ .753, .753, .753}0.94425 & \cellcolor[rgb]{ .753, .753, .753}0.93150 & 0.97065 & 0.95857 & 0.96385 & \textcolor[rgb]{ 1, 0, 0}{\textbf{0.99263}} & \textbf{0.99351} & 0.98869 & \cellcolor[rgb]{ .753, .753, .753}0.85529 & 0.97340 & 0.97768 & \cellcolor[rgb]{ .753, .753, .753}0.86210 & \cellcolor[rgb]{ .753, .753, .753}0.91637 & \cellcolor[rgb]{ .753, .753, .753}0.91588 & 0.96675 & 0.98574 & 0.97647 & \textcolor[rgb]{ 0, 0, 1}{\textbf{0.99012}} \\
	14 & 0.99988 & 0.99998 & 0.99444 & 0.99958 & 0.99963 & 0.99567 & \textbf{1.00000} & 0.99995 & \cellcolor[rgb]{ .753, .753, .753}0.92974 & 0.99344 & \textbf{1.00000} & 0.99998 & 0.98458 & 0.99941 & 0.99868 & 0.99993 & \textbf{1.00000} & 0.99985 & \textbf{1.00000} \\
	15 & 0.99540 & 0.99896 & 0.99813 & 0.99297 & 0.97528 & 0.99037 & \textcolor[rgb]{ 1, 0, 0}{\textbf{0.99942}} & 0.99891 & 0.99885 & 0.99322 & 0.99791 & \textbf{0.99957} & 0.95164 & 0.98704 & 0.98636 & 0.99889 & \textcolor[rgb]{ 0, 0, 1}{\textbf{0.99936}} & 0.99923 & 0.99902 \\
	16 & \textcolor[rgb]{ 1, 0, 0}{\textbf{0.99980}} & 0.97738 & 0.99752 & 0.99729 & 0.98697 & 0.99160 & 0.99956 & 0.99939 & 0.99638 & \cellcolor[rgb]{ .753, .753, .753}0.50000 & \textbf{0.99988} & 0.99436 & 0.98428 & 0.99468 & \textcolor[rgb]{ 0, 0, 1}{\textbf{0.99974}} & 0.99914 & 0.99918 & 0.99907 & 0.99919 \\
	17 & \textcolor[rgb]{ 0, 0, 1}{\textbf{0.99980}} & 0.99845 & 0.96054 & 0.99969 & 0.99955 & 0.99586 & 0.99941 & 0.99804 & 0.99955 & \cellcolor[rgb]{ .753, .753, .753}0.92207 & 0.99855 & 0.99861 & \cellcolor[rgb]{ .753, .753, .753}0.92163 & 0.99925 & \textbf{1.00000} & 0.99892 & 0.99888 & \textcolor[rgb]{ 1, 0, 0}{\textbf{0.99984}} & 0.99869 \\
	18 & 0.98536 & 0.98373 & \cellcolor[rgb]{ .753, .753, .753}0.91600 & 0.97662 & 0.96292 & \cellcolor[rgb]{ .753, .753, .753}0.94792 & 0.99218 & 0.98846 & 0.99230 & \cellcolor[rgb]{ .753, .753, .753}0.50000 & 0.98712 & \textcolor[rgb]{ 0, 0, 1}{\textbf{0.99708}} & \cellcolor[rgb]{ .753, .753, .753}0.79820 & \textcolor[rgb]{ 1, 0, 0}{\textbf{0.99769}} & \textbf{0.99907} & 0.98800 & 0.97605 & 0.99294 & 0.97998 \\
	19 & \textcolor[rgb]{ 1, 0, 0}{\textbf{0.99993}} & 0.99510 & \textbf{0.99994} & \textcolor[rgb]{ 0, 0, 1}{\textbf{0.99987}} & 0.99912 & 0.99666 & 0.99957 & 0.99969 & 0.99823 & 0.99789 & 0.99985 & 0.99888 & 0.99839 & 0.97921 & 0.99693 & 0.99975 & 0.99901 & 0.99799 & 0.99935 \\
	20 & \cellcolor[rgb]{ .753, .753, .753}0.94822 & \textbf{0.99907} & \cellcolor[rgb]{ .753, .753, .753}0.90231 & \cellcolor[rgb]{ .753, .753, .753}0.94849 & 0.99521 & 0.99445 & \textcolor[rgb]{ 0, 0, 1}{\textbf{0.99854}} & 0.99734 & 0.99048 & \cellcolor[rgb]{ .753, .753, .753}0.89033 & 0.95114 & \textcolor[rgb]{ 1, 0, 0}{\textbf{0.99889}} & \cellcolor[rgb]{ .753, .753, .753}0.87891 & 0.98504 & 0.99391 & 0.97613 & 0.99287 & 0.99568 & 0.99756 \\
	21 & \textbf{1.00000} & \textbf{1.00000} & 0.99443 & \textbf{1.00000} & 0.99943 & 0.99947 & 0.99978 & 0.99973 & 0.99700 & 0.99082 & 0.99998 & 0.99680 & 0.98925 & 0.99029 & \textbf{1.00000} & 0.99976 & 0.99986 & 0.99908 & 0.99971 \\
	22 & \textcolor[rgb]{ 0, 0, 1}{\textbf{0.99885}} & 0.99668 & 0.97693 & \textcolor[rgb]{ 1, 0, 0}{\textbf{0.99920}} & 0.99350 & 0.99125 & 0.99848 & 0.99685 & 0.99882 & 0.97656 & 0.99791 & \textbf{0.99923} & 0.96742 & 0.99351 & 0.99794 & 0.99539 & 0.99672 & 0.99815 & 0.99881 \\
	23 & 0.99772 & 0.99365 & 0.96288 & 0.99710 & 0.99349 & 0.99216 & 0.99751 & 0.99764 & \textbf{0.99863} & 0.97279 & 0.99806 & 0.99715 & \cellcolor[rgb]{ .753, .753, .753}0.93201 & 0.98372 & 0.98918 & 0.99221 & 0.98934 & \textcolor[rgb]{ 1, 0, 0}{\textbf{0.99828}} & \textcolor[rgb]{ 0, 0, 1}{\textbf{0.99822}} \\
	24 & 0.99683 & \cellcolor[rgb]{ .753, .753, .753}0.91625 & 0.99427 & 0.99596 & 0.99641 & 0.98810 & 0.99427 & 0.99646 & 0.99821 & 0.99357 & 0.99549 & 0.99195 & 0.99395 & \cellcolor[rgb]{ .753, .753, .753}0.94782 & \textcolor[rgb]{ 0, 0, 1}{\textbf{0.99889}} & 0.99756 & 0.99844 & \textbf{0.99967} & \textcolor[rgb]{ 1, 0, 0}{\textbf{0.99952}} \\
	25 & 0.99977 & 0.99972 & 0.99886 & 0.99986 & 0.99664 & \textcolor[rgb]{ 1, 0, 0}{\textbf{0.99998}} & \textbf{0.99999} & \textcolor[rgb]{ 0, 0, 1}{\textbf{0.99996}} & 0.99839 & 0.98939 & 0.99986 & 0.99680 & 0.99298 & 0.95643 & 0.99653 & 0.99990 & 0.99995 & 0.99748 & 0.99923 \\
	26 & 0.98420 & 0.99946 & 0.97166 & 0.99292 & 0.99880 & 0.97699 & 0.99685 & 0.99514 & 0.98316 & 0.95011 & \cellcolor[rgb]{ .753, .753, .753}0.40540 & \textcolor[rgb]{ 0, 0, 1}{\textbf{0.99954}} & \cellcolor[rgb]{ .753, .753, .753}0.80332 & 0.98965 & \textcolor[rgb]{ 1, 0, 0}{\textbf{0.99960}} & 0.99921 & 0.99702 & \textbf{0.99962} & 0.99891 \\
	27 & 0.99889 & \textbf{1.00000} & 0.99611 & 0.99922 & 0.99673 & 0.99982 & \textcolor[rgb]{ 0, 0, 1}{\textbf{0.99995}} & \textcolor[rgb]{ 1, 0, 0}{\textbf{0.99997}} & \cellcolor[rgb]{ .753, .753, .753}0.93484 & 0.99683 & 0.99886 & 0.99982 & 0.99606 & 0.98412 & 0.99234 & 0.99938 & 0.99944 & 0.99874 & 0.99948 \\
	28 & 0.96821 & 0.99531 & 0.98412 & 0.98528 & 0.98033 & 0.98765 & \textcolor[rgb]{ 1, 0, 0}{\textbf{0.99773}} & 0.99706 & 0.99628 & 0.95083 & 0.96640 & \textbf{0.99799} & 0.95107 & 0.97956 & \textcolor[rgb]{ 0, 0, 1}{\textbf{0.99766}} & 0.98691 & 0.98625 & 0.99418 & 0.99716 \\
	29 & 0.98342 & 0.99553 & \cellcolor[rgb]{ .753, .753, .753}0.93009 & 0.98861 & \textcolor[rgb]{ 1, 0, 0}{\textbf{0.99833}} & \textbf{0.99901} & 0.99637 & 0.99481 & 0.99187 & \cellcolor[rgb]{ .753, .753, .753}0.84638 & \cellcolor[rgb]{ .753, .753, .753}0.38223 & 0.99739 & \cellcolor[rgb]{ .753, .753, .753}0.74893 & 0.99461 & 0.95807 & 0.96628 & 0.97618 & 0.99755 & \textcolor[rgb]{ 0, 0, 1}{\textbf{0.99763}} \\
	30 & 0.99909 & 0.98587 & 0.99416 & 0.99835 & 0.99503 & 0.98703 & \textcolor[rgb]{ 0, 0, 1}{\textbf{0.99923}} & \textcolor[rgb]{ 1, 0, 0}{\textbf{0.99967}} & 0.99896 & 0.99060 & 0.99851 & 0.98595 & 0.99278 & 0.99095 & \textbf{0.99998} & 0.99916 & 0.99884 & 0.99765 & 0.99854 \\
	31 & 0.97919 & 0.99170 & 0.98455 & 0.98719 & 0.98818 & 0.98679 & 0.99327 & \textcolor[rgb]{ 0, 0, 1}{\textbf{0.99555}} & 0.98562 & \cellcolor[rgb]{ .753, .753, .753}0.91163 & 0.97443 & 0.99288 & \cellcolor[rgb]{ .753, .753, .753}0.83842 & 0.96620 & \cellcolor[rgb]{ .753, .753, .753}0.94122 & 0.99331 & 0.98942 & \textbf{0.99658} & \textcolor[rgb]{ 1, 0, 0}{\textbf{0.99581}} \\
	32 & \textbf{0.99993} & 0.99985 & 0.99900 & \textcolor[rgb]{ 0, 0, 1}{\textbf{0.99989}} & 0.99654 & 0.99729 & 0.99065 & 0.99862 & 0.96146 & 0.99974 & \textcolor[rgb]{ 1, 0, 0}{\textbf{0.99989}} & 0.97384 & 0.99977 & 0.98622 & 0.99978 & 0.99907 & 0.99723 & 0.99682 & 0.99668 \\
	33 & 0.97990 & \cellcolor[rgb]{ .753, .753, .753}0.94724 & 0.95419 & 0.97960 & 0.98056 & \textcolor[rgb]{ 0, 0, 1}{\textbf{0.99597}} & 0.99467 & \textbf{0.99740} & 0.98383 & \cellcolor[rgb]{ .753, .753, .753}0.88990 & 0.98260 & 0.99170 & \cellcolor[rgb]{ .753, .753, .753}0.91677 & 0.96149 & 0.95070 & 0.97607 & 0.97899 & 0.98949 & \textcolor[rgb]{ 1, 0, 0}{\textbf{0.99612}} \\
	34 & \textbf{0.99897} & 0.96678 & 0.99798 & \textcolor[rgb]{ 1, 0, 0}{\textbf{0.99858}} & 0.98209 & 0.97668 & 0.97969 & 0.99647 & 0.99126 & 0.99206 & \textcolor[rgb]{ 0, 0, 1}{\textbf{0.99821}} & 0.99389 & 0.96830 & 0.96299 & 0.98821 & 0.99706 & 0.99218 & 0.99481 & 0.99419 \\
	35 & 0.96907 & 0.95702 & \textbf{0.99157} & 0.97493 & \cellcolor[rgb]{ .753, .753, .753}0.93485 & 0.97933 & 0.97358 & \textcolor[rgb]{ 0, 0, 1}{\textbf{0.98091}} & \cellcolor[rgb]{ .753, .753, .753}0.94390 & 0.96827 & 0.96657 & \cellcolor[rgb]{ .753, .753, .753}0.94784 & \cellcolor[rgb]{ .753, .753, .753}0.94997 & \cellcolor[rgb]{ .753, .753, .753}0.91764 & \cellcolor[rgb]{ .753, .753, .753}0.80658 & 0.95588 & \cellcolor[rgb]{ .753, .753, .753}0.92418 & 0.97264 & \textcolor[rgb]{ 1, 0, 0}{\textbf{0.98684}} \\
	36 & \textcolor[rgb]{ 0, 0, 1}{\textbf{0.99949}} & 0.99469 & 0.99792 & 0.99939 & \textcolor[rgb]{ 1, 0, 0}{\textbf{0.99959}} & 0.99063 & 0.99892 & 0.99888 & \textbf{0.99976} & 0.99584 & 0.99925 & 0.98749 & 0.99482 & 0.98755 & 0.98545 & 0.99820 & 0.99902 & 0.99918 & 0.99929 \\
	37 & \textcolor[rgb]{ 0, 0, 1}{\textbf{0.99992}} & \textcolor[rgb]{ 1, 0, 0}{\textbf{0.99996}} & 0.96736 & \textcolor[rgb]{ 0, 0, 1}{\textbf{0.99992}} & 0.99963 & 0.99055 & 0.99984 & 0.99967 & 0.99756 & \cellcolor[rgb]{ .753, .753, .753}0.84943 & 0.99976 & 0.99698 & \cellcolor[rgb]{ .753, .753, .753}0.73362 & 0.99919 & \textbf{1.00000} & 0.99723 & 0.99756 & 0.99980 & 0.99971 \\
	38 & 0.99538 & 0.97900 & 0.99148 & 0.99167 & 0.98413 & 0.99846 & 0.99630 & \textcolor[rgb]{ 0, 0, 1}{\textbf{0.99930}} & 0.99906 & 0.96334 & 0.99737 & 0.99855 & 0.98074 & 0.98428 & \textbf{0.99971} & 0.99793 & \textcolor[rgb]{ 1, 0, 0}{\textbf{0.99953}} & 0.99590 & 0.99767 \\
	39 & 0.95263 & \cellcolor[rgb]{ .753, .753, .753}0.91497 & \cellcolor[rgb]{ .753, .753, .753}0.90757 & 0.95166 & 0.95629 & \cellcolor[rgb]{ .753, .753, .753}0.92645 & \textcolor[rgb]{ 0, 0, 1}{\textbf{0.97465}} & \textcolor[rgb]{ 1, 0, 0}{\textbf{0.97695}} & 0.96248 & \cellcolor[rgb]{ .753, .753, .753}0.87893 & 0.95023 & \textbf{0.98202} & \cellcolor[rgb]{ .753, .753, .753}0.86175 & \cellcolor[rgb]{ .753, .753, .753}0.89516 & \cellcolor[rgb]{ .753, .753, .753}0.93031 & \cellcolor[rgb]{ .753, .753, .753}0.90263 & 0.95383 & 0.96726 & 0.97376 \\
	40 & 0.99304 & 0.97355 & 0.98878 & 0.99484 & 0.99366 & 0.99207 & 0.98971 & \textbf{0.99695} & 0.99176 & 0.95623 & 0.99588 & \textcolor[rgb]{ 1, 0, 0}{\textbf{0.99675}} & \cellcolor[rgb]{ .753, .753, .753}0.92802 & 0.96890 & 0.99229 & \textcolor[rgb]{ 0, 0, 1}{\textbf{0.99649}} & 0.99173 & 0.99565 & 0.99505 \\
	41 & 0.98623 & 0.98887 & \cellcolor[rgb]{ .753, .753, .753}0.90939 & 0.98158 & \textcolor[rgb]{ 1, 0, 0}{\textbf{0.99928}} & 0.99175 & 0.99862 & 0.99719 & 0.99769 & \cellcolor[rgb]{ .753, .753, .753}0.82164 & 0.98738 & 0.99856 & \cellcolor[rgb]{ .753, .753, .753}0.64898 & 0.99730 & 0.96261 & 0.98767 & 0.99725 & \textbf{0.99952} & \textcolor[rgb]{ 0, 0, 1}{\textbf{0.99870}} \\
	42 & \cellcolor[rgb]{ .753, .753, .753}0.86529 & 0.95220 & \cellcolor[rgb]{ .753, .753, .753}0.92777 & \cellcolor[rgb]{ .753, .753, .753}0.87971 & \textbf{0.98779} & 0.96413 & 0.97880 & 0.97480 & \cellcolor[rgb]{ .753, .753, .753}0.93840 & \cellcolor[rgb]{ .753, .753, .753}0.94111 & \cellcolor[rgb]{ .753, .753, .753}0.90340 & 0.96942 & \cellcolor[rgb]{ .753, .753, .753}0.86217 & \cellcolor[rgb]{ .753, .753, .753}0.93191 & \cellcolor[rgb]{ .753, .753, .753}0.90532 & \cellcolor[rgb]{ .753, .753, .753}0.93758 & \cellcolor[rgb]{ .753, .753, .753}0.93248 & \textcolor[rgb]{ 1, 0, 0}{\textbf{0.98751}} & \textcolor[rgb]{ 0, 0, 1}{\textbf{0.98689}} \\
	43 & \textbf{0.99995} & 0.99840 & 0.97129 & \textbf{0.99995} & 0.99886 & 0.99111 & 0.99870 & 0.99769 & 0.98565 & 0.97398 & 0.99935 & 0.99834 & 0.96648 & 0.99902 & 0.98407 & 0.99649 & 0.99456 & \textcolor[rgb]{ 0, 0, 1}{\textbf{0.99935}} & 0.99842 \\
	44 & \cellcolor[rgb]{ .753, .753, .753}0.94703 & \cellcolor[rgb]{ .753, .753, .753}0.93335 & \cellcolor[rgb]{ .753, .753, .753}0.92137 & 0.95961 & 0.95522 & 0.96731 & 0.97171 & \textcolor[rgb]{ 0, 0, 1}{\textbf{0.98254}} & \cellcolor[rgb]{ .753, .753, .753}0.86910 & \cellcolor[rgb]{ .753, .753, .753}0.74919 & \cellcolor[rgb]{ .753, .753, .753}0.91652 & 0.97214 & \cellcolor[rgb]{ .753, .753, .753}0.48556 & \cellcolor[rgb]{ .753, .753, .753}0.94864 & \cellcolor[rgb]{ .753, .753, .753}0.80487 & 0.95306 & \cellcolor[rgb]{ .753, .753, .753}0.93383 & \textcolor[rgb]{ 1, 0, 0}{\textbf{0.99018}} & \textbf{0.99123} \\
	45 & 0.98849 & 0.98272 & 0.95195 & 0.98973 & 0.98802 & 0.99361 & \textcolor[rgb]{ 1, 0, 0}{\textbf{0.99710}} & 0.99481 & 0.99668 & \cellcolor[rgb]{ .753, .753, .753}0.50000 & \cellcolor[rgb]{ .753, .753, .753}0.32088 & 0.98789 & \cellcolor[rgb]{ .753, .753, .753}0.81380 & 0.97481 & 0.95597 & 0.98275 & 0.97223 & \textcolor[rgb]{ 0, 0, 1}{\textbf{0.99686}} & \textbf{0.99752} \\
	46 & 0.99970 & \textcolor[rgb]{ 0, 0, 1}{\textbf{0.99989}} & 0.98660 & 0.99974 & 0.99725 & 0.98568 & \textcolor[rgb]{ 1, 0, 0}{\textbf{0.99991}} & \textcolor[rgb]{ 0, 0, 1}{\textbf{0.99989}} & 0.99770 & 0.97249 & \textbf{0.99994} & 0.99778 & 0.96667 & 0.99000 & 0.95835 & 0.99949 & 0.99977 & 0.99889 & 0.99968 \\
	47 & \textcolor[rgb]{ 0, 0, 1}{\textbf{0.97441}} & \cellcolor[rgb]{ .753, .753, .753}0.91580 & 0.97358 & 0.97179 & \cellcolor[rgb]{ .753, .753, .753}0.89034 & \cellcolor[rgb]{ .753, .753, .753}0.94786 & 0.95673 & 0.96727 & 0.95289 & \textbf{0.98311} & \cellcolor[rgb]{ .753, .753, .753}0.94116 & \cellcolor[rgb]{ .753, .753, .753}0.65446 & \cellcolor[rgb]{ .753, .753, .753}0.72823 & \cellcolor[rgb]{ .753, .753, .753}0.85456 & \cellcolor[rgb]{ .753, .753, .753}0.51047 & \cellcolor[rgb]{ .753, .753, .753}0.94881 & \cellcolor[rgb]{ .753, .753, .753}0.91471 & 0.97235 & \textcolor[rgb]{ 1, 0, 0}{\textbf{0.97923}} \\
	48 & 0.96767 & 0.97082 & 0.95510 & 0.96501 & 0.98879 & 0.96952 & 0.97796 & 0.98548 & 0.99341 & 0.95827 & 0.97505 & 0.99429 & \cellcolor[rgb]{ .753, .753, .753}0.89580 & 0.98710 & \textcolor[rgb]{ 1, 0, 0}{\textbf{0.99814}} & 0.99059 & 0.99121 & \textbf{0.99859} & \textcolor[rgb]{ 0, 0, 1}{\textbf{0.99511}} \\
	49 & \textcolor[rgb]{ 0, 0, 1}{\textbf{0.99864}} & \cellcolor[rgb]{ .753, .753, .753}0.94034 & 0.98880 & 0.99849 & 0.98880 & 0.98401 & 0.99428 & 0.99834 & 0.99830 & 0.99228 & 0.99778 & 0.99400 & 0.99388 & 0.99209 & \textbf{0.99973} & \textcolor[rgb]{ 1, 0, 0}{\textbf{0.99933}} & 0.99689 & 0.99773 & 0.99748 \\
	50 & \cellcolor[rgb]{ .753, .753, .753}0.93762 & 0.97116 & \cellcolor[rgb]{ .753, .753, .753}0.65081 & \cellcolor[rgb]{ .753, .753, .753}0.93443 & \cellcolor[rgb]{ .753, .753, .753}0.93779 & 0.97269 & \textcolor[rgb]{ 1, 0, 0}{\textbf{0.98531}} & \textcolor[rgb]{ 0, 0, 1}{\textbf{0.98422}} & 0.96274 & \cellcolor[rgb]{ .753, .753, .753}0.76715 & \cellcolor[rgb]{ .753, .753, .753}0.93198 & 0.95106 & \cellcolor[rgb]{ .753, .753, .753}0.80734 & 0.95122 & \textbf{0.98827} & 0.95800 & 0.95959 & 0.96719 & 0.97868 \\
	51 & 0.99301 & 0.97387 & 0.98953 & 0.99407 & 0.99203 & 0.99741 & 0.99377 & \textcolor[rgb]{ 1, 0, 0}{\textbf{0.99827}} & 0.99645 & 0.98845 & 0.99268 & 0.99754 & 0.98377 & 0.96006 & \textbf{0.99870} & 0.99387 & 0.99802 & \textcolor[rgb]{ 0, 0, 1}{\textbf{0.99805}} & 0.99750 \\
	52 & 0.99984 & 0.99857 & 0.98462 & 0.99930 & 0.99685 & 0.99799 & 0.99998 & \textbf{1.00000} & 0.99405 & \cellcolor[rgb]{ .753, .753, .753}0.50000 & 0.99998 & 0.99998 & \cellcolor[rgb]{ .753, .753, .753}0.92306 & 0.99997 & 0.99993 & 0.99969 & 0.99991 & \textbf{1.00000} & \textbf{1.00000} \\
	53 & 0.99743 & 0.99479 & 0.99522 & 0.99566 & 0.96779 & 0.99779 & 0.99793 & \textbf{0.99946} & 0.99862 & \textcolor[rgb]{ 1, 0, 0}{\textbf{0.99938}} & 0.99727 & 0.98837 & 0.99888 & 0.99017 & \textcolor[rgb]{ 0, 0, 1}{\textbf{0.99894}} & 0.99889 & 0.99828 & 0.98684 & 0.99265 \\
	54 & 0.99646 & 0.99754 & 0.99442 & 0.99493 & 0.97512 & 0.99686 & 0.99731 & \textcolor[rgb]{ 0, 0, 1}{\textbf{0.99835}} & 0.99688 & \textcolor[rgb]{ 1, 0, 0}{\textbf{0.99877}} & 0.99624 & 0.99088 & 0.99794 & 0.99641 & \textbf{0.99927} & 0.99750 & 0.99683 & 0.98967 & 0.99460 \\
	55 & \textbf{0.99886} & 0.99290 & \cellcolor[rgb]{ .753, .753, .753}0.94212 & 0.99767 & 0.98970 & 0.96906 & 0.99856 & \textcolor[rgb]{ 0, 0, 1}{\textbf{0.99869}} & 0.97926 & 0.97445 & \textcolor[rgb]{ 1, 0, 0}{\textbf{0.99882}} & 0.98299 & 0.97298 & 0.99330 & 0.99185 & 0.99762 & 0.99817 & 0.99508 & 0.99731 \\
	56 & \textcolor[rgb]{ 1, 0, 0}{\textbf{0.99825}} & 0.99017 & 0.98916 & 0.96390 & 0.98674 & 0.99293 & 0.99763 & 0.99743 & 0.99494 & 0.98618 & \textbf{0.99825} & 0.99417 & 0.97826 & 0.97891 & 0.98585 & 0.98828 & 0.99555 & 0.99766 & \textcolor[rgb]{ 0, 0, 1}{\textbf{0.99769}} \\
	57 & 0.98011 & 0.97916 & \cellcolor[rgb]{ .753, .753, .753}0.90563 & 0.95117 & 0.97372 & 0.97605 & \textcolor[rgb]{ 1, 0, 0}{\textbf{0.99771}} & \textcolor[rgb]{ 0, 0, 1}{\textbf{0.99401}} & 0.96457 & \cellcolor[rgb]{ .753, .753, .753}0.92044 & 0.97899 & \textbf{0.99847} & \cellcolor[rgb]{ .753, .753, .753}0.77419 & 0.98671 & \cellcolor[rgb]{ .753, .753, .753}0.88986 & 0.98373 & 0.97969 & 0.98253 & 0.98993 \\
	58 & \textbf{0.99524} & 0.96993 & 0.98088 & 0.99484 & 0.98552 & 0.96139 & 0.99376 & \textcolor[rgb]{ 1, 0, 0}{\textbf{0.99497}} & \cellcolor[rgb]{ .753, .753, .753}0.93202 & 0.98186 & \textcolor[rgb]{ 0, 0, 1}{\textbf{0.99495}} & 0.98500 & 0.98158 & 0.97451 & 0.97388 & 0.99221 & 0.98363 & 0.98959 & 0.98947 \\
	59 & 0.99264 & 0.99318 & 0.97185 & 0.99177 & 0.97286 & 0.97574 & \textbf{0.99872} & \textcolor[rgb]{ 1, 0, 0}{\textbf{0.99828}} & 0.96182 & 0.96869 & 0.99218 & \textcolor[rgb]{ 0, 0, 1}{\textbf{0.99726}} & 0.96694 & 0.98186 & 0.99489 & 0.99195 & 0.99425 & 0.99333 & 0.99576 \\
	60 & 0.99188 & 0.98165 & 0.97821 & 0.99250 & 0.97897 & 0.97146 & \textbf{0.99716} & \textcolor[rgb]{ 0, 0, 1}{\textbf{0.99553}} & \textcolor[rgb]{ 1, 0, 0}{\textbf{0.99573}} & 0.97861 & 0.99182 & 0.98433 & 0.97898 & 0.97978 & 0.97552 & 0.99234 & 0.99220 & 0.98874 & 0.99143 \\
	61 & 0.98268 & 0.98734 & 0.96314 & 0.97967 & 0.98870 & 0.96310 & \textcolor[rgb]{ 1, 0, 0}{\textbf{0.99527}} & 0.99234 & 0.96395 & 0.96727 & 0.98264 & 0.99102 & 0.96615 & 0.97864 & 0.95580 & 0.99104 & \textbf{0.99748} & 0.98889 & \textcolor[rgb]{ 0, 0, 1}{\textbf{0.99420}} \\
	62 & \textbf{1.00000} & 0.99982 & 0.97600 & \textbf{1.00000} & 0.99951 & 0.99427 & 0.99996 & 0.99994 & 0.99973 & 0.95446 & \textbf{1.00000} & 0.99971 & 0.96731 & 0.99629 & 0.97286 & \textbf{1.00000} & 0.99994 & 0.99969 & 0.99994 \\
	63 & \cellcolor[rgb]{ .753, .753, .753}0.94162 & \cellcolor[rgb]{ .753, .753, .753}0.91371 & 0.96591 & \cellcolor[rgb]{ .753, .753, .753}0.93317 & \textbf{0.99021} & \cellcolor[rgb]{ .753, .753, .753}0.92673 & 0.98532 & 0.97570 & 0.96547 & \cellcolor[rgb]{ .753, .753, .753}0.88716 & \cellcolor[rgb]{ .753, .753, .753}0.94223 & 0.98111 & \cellcolor[rgb]{ .753, .753, .753}0.88190 & \cellcolor[rgb]{ .753, .753, .753}0.91472 & \cellcolor[rgb]{ .753, .753, .753}0.83796 & 0.96189 & 0.97825 & \textcolor[rgb]{ 0, 0, 1}{\textbf{0.98611}} & \textcolor[rgb]{ 1, 0, 0}{\textbf{0.98740}} \\
	64 & 0.98897 & 0.98550 & 0.98863 & \cellcolor[rgb]{ .753, .753, .753}0.94305 & \cellcolor[rgb]{ .753, .753, .753}0.88882 & 0.98765 & 0.99567 & 0.99583 & 0.99478 & 0.99131 & 0.98938 & \textcolor[rgb]{ 1, 0, 0}{\textbf{0.99697}} & 0.98568 & \cellcolor[rgb]{ .753, .753, .753}0.88503 & \cellcolor[rgb]{ .753, .753, .753}0.94506 & \textcolor[rgb]{ 0, 0, 1}{\textbf{0.99681}} & \textbf{0.99794} & 0.99557 & 0.99554 \\
	65 & 0.96486 & \cellcolor[rgb]{ .753, .753, .753}0.93660 & \cellcolor[rgb]{ .753, .753, .753}0.91710 & 0.95979 & 0.96258 & 0.95879 & \textcolor[rgb]{ 0, 0, 1}{\textbf{0.98204}} & \textbf{0.98509} & 0.97553 & \cellcolor[rgb]{ .753, .753, .753}0.73960 & 0.96832 & \cellcolor[rgb]{ .753, .753, .753}0.94928 & \cellcolor[rgb]{ .753, .753, .753}0.73011 & 0.95428 & \cellcolor[rgb]{ .753, .753, .753}0.92115 & \cellcolor[rgb]{ .753, .753, .753}0.93771 & 0.97769 & 0.97669 & \textcolor[rgb]{ 1, 0, 0}{\textbf{0.98419}} \\
	66 & 0.96811 & \textbf{0.99991} & \cellcolor[rgb]{ .753, .753, .753}0.93426 & 0.98204 & 0.99749 & \textcolor[rgb]{ 1, 0, 0}{\textbf{0.99973}} & 0.99826 & 0.99895 & 0.99909 & \cellcolor[rgb]{ .753, .753, .753}0.82755 & 0.98400 & \textcolor[rgb]{ 0, 0, 1}{\textbf{0.99958}} & \cellcolor[rgb]{ .753, .753, .753}0.74505 & 0.99562 & \cellcolor[rgb]{ .753, .753, .753}0.75255 & 0.97772 & 0.99840 & 0.99913 & 0.99951 \\
	67 & \textcolor[rgb]{ 1, 0, 0}{\textbf{0.99982}} & \textcolor[rgb]{ 0, 0, 1}{\textbf{0.99974}} & 0.98686 & 0.99965 & 0.99639 & 0.99305 & 0.99948 & 0.99886 & 0.98992 & \cellcolor[rgb]{ .753, .753, .753}0.93295 & 0.99931 & 0.99902 & \cellcolor[rgb]{ .753, .753, .753}0.82781 & 0.98950 & \textbf{0.99983} & 0.99625 & 0.99542 & 0.99889 & 0.99798 \\
	68 & \textbf{1.00000} & \textcolor[rgb]{ 0, 0, 1}{\textbf{0.99993}} & 0.98252 & \textbf{1.00000} & 0.99967 & 0.99794 & 0.99949 & 0.99876 & 0.99913 & \cellcolor[rgb]{ .753, .753, .753}0.86125 & \cellcolor[rgb]{ .753, .753, .753}0.49701 & 0.99832 & \cellcolor[rgb]{ .753, .753, .753}0.78582 & 0.99634 & 0.99993 & 0.99771 & 0.99918 & 0.99986 & 0.99969 \\
	69 & 0.99860 & 0.98509 & \cellcolor[rgb]{ .753, .753, .753}0.94514 & 0.99869 & 0.99691 & 0.99717 & \textbf{0.99953} & \textcolor[rgb]{ 1, 0, 0}{\textbf{0.99914}} & 0.99897 & \cellcolor[rgb]{ .753, .753, .753}0.77945 & \cellcolor[rgb]{ .753, .753, .753}0.92934 & 0.97941 & \cellcolor[rgb]{ .753, .753, .753}0.69234 & 0.98703 & \textcolor[rgb]{ 0, 0, 1}{\textbf{0.99911}} & 0.99226 & 0.99595 & 0.99875 & 0.99860 \\
	70 & 0.97177 & 0.99714 & \cellcolor[rgb]{ .753, .753, .753}0.93108 & 0.97968 & 0.99616 & 0.98700 & 0.99744 & 0.99603 & \textcolor[rgb]{ 0, 0, 1}{\textbf{0.99854}} & \cellcolor[rgb]{ .753, .753, .753}0.76559 & \textcolor[rgb]{ 1, 0, 0}{\textbf{0.99882}} & 0.99792 & \cellcolor[rgb]{ .753, .753, .753}0.70946 & 0.96650 & 0.97091 & 0.99176 & \textbf{0.99885} & 0.99529 & 0.99719 \\
	71 & \cellcolor[rgb]{ .753, .753, .753}0.88996 & \cellcolor[rgb]{ .753, .753, .753}0.94212 & \cellcolor[rgb]{ .753, .753, .753}0.92381 & \cellcolor[rgb]{ .753, .753, .753}0.89023 & 0.99170 & 0.97839 & 0.98953 & 0.98197 & 0.96186 & \cellcolor[rgb]{ .753, .753, .753}0.84525 & \cellcolor[rgb]{ .753, .753, .753}0.50734 & \textcolor[rgb]{ 0, 0, 1}{\textbf{0.99187}} & \cellcolor[rgb]{ .753, .753, .753}0.81777 & 0.96189 & 0.99097 & 0.95845 & 0.98835 & \textcolor[rgb]{ 1, 0, 0}{\textbf{0.99689}} & \textbf{0.99695} \\
	72 & \cellcolor[rgb]{ .753, .753, .753}0.92788 & 0.95995 & \cellcolor[rgb]{ .753, .753, .753}0.86920 & \cellcolor[rgb]{ .753, .753, .753}0.93852 & 0.97297 & 0.98538 & 0.99526 & 0.99282 & 0.98815 & \cellcolor[rgb]{ .753, .753, .753}0.76534 & 0.98047 & 0.99119 & \cellcolor[rgb]{ .753, .753, .753}0.79534 & \cellcolor[rgb]{ .753, .753, .753}0.94838 & \textbf{0.99914} & 0.98384 & 0.98910 & \textcolor[rgb]{ 1, 0, 0}{\textbf{0.99891}} & \textcolor[rgb]{ 0, 0, 1}{\textbf{0.99770}} \\
	73 & \cellcolor[rgb]{ .753, .753, .753}0.88235 & \cellcolor[rgb]{ .753, .753, .753}0.93601 & \cellcolor[rgb]{ .753, .753, .753}0.90893 & \cellcolor[rgb]{ .753, .753, .753}0.86449 & 0.99403 & 0.97101 & 0.96647 & 0.96856 & 0.96363 & \cellcolor[rgb]{ .753, .753, .753}0.83900 & 0.96233 & \cellcolor[rgb]{ .753, .753, .753}0.92578 & \cellcolor[rgb]{ .753, .753, .753}0.85134 & 0.97070 & \textcolor[rgb]{ 0, 0, 1}{\textbf{0.99572}} & 0.98772 & 0.98507 & \textbf{0.99757} & \textcolor[rgb]{ 1, 0, 0}{\textbf{0.99682}} \\
	74 & 0.97037 & 0.98035 & \cellcolor[rgb]{ .753, .753, .753}0.90796 & 0.97134 & 0.98873 & 0.99512 & 0.99439 & 0.99690 & 0.99423 & \cellcolor[rgb]{ .753, .753, .753}0.50000 & 0.99223 & \textcolor[rgb]{ 0, 0, 1}{\textbf{0.99696}} & \cellcolor[rgb]{ .753, .753, .753}0.80992 & \cellcolor[rgb]{ .753, .753, .753}0.92194 & 0.99412 & 0.99561 & 0.99689 & \textcolor[rgb]{ 1, 0, 0}{\textbf{0.99769}} & \textbf{0.99866} \\
	75 & \cellcolor[rgb]{ .753, .753, .753}0.92498 & \cellcolor[rgb]{ .753, .753, .753}0.89053 & \cellcolor[rgb]{ .753, .753, .753}0.94546 & \cellcolor[rgb]{ .753, .753, .753}0.91954 & 0.97106 & 0.98050 & 0.99014 & \textcolor[rgb]{ 0, 0, 1}{\textbf{0.99258}} & 0.95321 & \cellcolor[rgb]{ .753, .753, .753}0.79161 & \cellcolor[rgb]{ .753, .753, .753}0.21224 & 0.98910 & \cellcolor[rgb]{ .753, .753, .753}0.74136 & 0.96868 & \cellcolor[rgb]{ .753, .753, .753}0.90336 & 0.96421 & 0.99171 & \textcolor[rgb]{ 1, 0, 0}{\textbf{0.99773}} & \textbf{0.99775} \\
	76 & 0.95124 & \textcolor[rgb]{ 0, 0, 1}{\textbf{0.99800}} & \cellcolor[rgb]{ .753, .753, .753}0.75151 & 0.97025 & 0.97250 & 0.99701 & \textbf{0.99849} & \textcolor[rgb]{ 1, 0, 0}{\textbf{0.99811}} & 0.96835 & \cellcolor[rgb]{ .753, .753, .753}0.68676 & 0.97485 & 0.98480 & \cellcolor[rgb]{ .753, .753, .753}0.59581 & 0.98464 & \cellcolor[rgb]{ .753, .753, .753}0.93003 & 0.97200 & 0.99687 & 0.97495 & 0.99014 \\
	77 & 0.99983 & 0.99954 & 0.99931 & 0.99983 & 0.99810 & 0.99915 & 0.99968 & 0.99991 & 0.99559 & \textbf{0.99999} & 0.99983 & 0.99873 & \textcolor[rgb]{ 1, 0, 0}{\textbf{0.99997}} & 0.99808 & 0.99975 & 0.99994 & 0.99952 & 0.99815 & \textcolor[rgb]{ 1, 0, 0}{\textbf{0.99997}} \\
	78 & 0.99997 & 0.99998 & \textbf{1.00000} & 0.99997 & 0.99962 & 0.99902 & 0.99997 & 0.99989 & \cellcolor[rgb]{ .753, .753, .753}0.81375 & \textbf{1.00000} & 0.99997 & \textbf{1.00000} & \textbf{1.00000} & 0.99884 & 0.99989 & 0.99971 & 0.99977 & 0.99997 & \textbf{1.00000} \\
	79 & 0.99893 & 0.99862 & 0.99985 & 0.99954 & 0.99972 & 0.99761 & 0.99979 & \textbf{0.99991} & 0.99966 & 0.97208 & 0.99893 & 0.99905 & 0.96129 & \textbf{0.99991} & 0.96710 & 0.95536 & 0.99826 & 0.99985 & \textbf{0.99991} \\
	80 & \textbf{1.00000} & 0.99972 & 0.99988 & \textbf{1.00000} & 0.99939 & 0.99982 & 0.99972 & 0.99985 & 0.99942 & 0.99994 & \textbf{1.00000} & 0.99141 & 0.99985 & 0.99985 & \cellcolor[rgb]{ .753, .753, .753}0.92557 & 0.99648 & 0.99991 & 0.99988 & 0.99982 \\
	81 & \textbf{1.00000} & \textbf{1.00000} & \textbf{1.00000} & \textbf{1.00000} & 0.99817 & 0.99991 & \textbf{1.00000} & 0.99991 & 0.99896 & 0.99979 & \textbf{1.00000} & 0.98939 & 0.99982 & \textbf{1.00000} & 0.98098 & 0.99988 & \textbf{1.00000} & \textbf{1.00000} & \textbf{1.00000} \\
	82 & 0.95723 & 0.99053 & 0.99778 & 0.96609 & 0.99918 & 0.98337 & 0.97994 & \textcolor[rgb]{ 0, 0, 1}{\textbf{0.99957}} & 0.99547 & 0.99125 & 0.96292 & \textbf{0.99988} & 0.98398 & 0.99206 & \cellcolor[rgb]{ .753, .753, .753}0.92516 & 0.99641 & 0.99339 & 0.99676 & \textcolor[rgb]{ 1, 0, 0}{\textbf{0.99965}} \\
	83 & \cellcolor[rgb]{ .753, .753, .753}0.84093 & \cellcolor[rgb]{ .753, .753, .753}0.89191 & \cellcolor[rgb]{ .753, .753, .753}0.87625 & \cellcolor[rgb]{ .753, .753, .753}0.84095 & \cellcolor[rgb]{ .753, .753, .753}0.91313 & \textcolor[rgb]{ 0, 0, 1}{\textbf{0.96640}} & \cellcolor[rgb]{ .753, .753, .753}0.92502 & \cellcolor[rgb]{ .753, .753, .753}0.94422 & 0.96415 & \cellcolor[rgb]{ .753, .753, .753}0.76452 & \cellcolor[rgb]{ .753, .753, .753}0.88209 & 0.96409 & \cellcolor[rgb]{ .753, .753, .753}0.46200 & \cellcolor[rgb]{ .753, .753, .753}0.94748 & \textbf{0.99235} & 0.95967 & \textcolor[rgb]{ 1, 0, 0}{\textbf{0.97450}} & \cellcolor[rgb]{ .753, .753, .753}0.91208 & \cellcolor[rgb]{ .753, .753, .753}0.94677 \\
	84 & 0.99274 & \cellcolor[rgb]{ .753, .753, .753}0.90761 & 0.98272 & 0.98264 & 0.98719 & 0.98370 & 0.99019 & \textcolor[rgb]{ 1, 0, 0}{\textbf{0.99704}} & \textbf{0.99818} & 0.99084 & 0.99314 & 0.95448 & 0.98963 & \cellcolor[rgb]{ .753, .753, .753}0.92667 & 0.99318 & 0.99290 & 0.99365 & 0.99614 & \textcolor[rgb]{ 0, 0, 1}{\textbf{0.99698}} \\
	85 & 0.99726 & 0.97773 & 0.98828 & 0.99321 & 0.99646 & 0.99035 & 0.99411 & 0.99698 & 0.99409 & 0.99274 & 0.99703 & 0.98765 & 0.99161 & 0.96819 & 0.96453 & \textcolor[rgb]{ 0, 0, 1}{\textbf{0.99811}} & 0.99799 & \textbf{0.99857} & \textcolor[rgb]{ 1, 0, 0}{\textbf{0.99856}} \\
	86 & 0.99447 & 0.99225 & 0.98129 & 0.99602 & 0.98868 & 0.99382 & \textbf{0.99918} & \textcolor[rgb]{ 0, 0, 1}{\textbf{0.99768}} & 0.99191 & \cellcolor[rgb]{ .753, .753, .753}0.94169 & 0.99595 & 0.99424 & \cellcolor[rgb]{ .753, .753, .753}0.92858 & 0.96300 & \cellcolor[rgb]{ .753, .753, .753}0.92408 & 0.98654 & 0.96886 & 0.99612 & \textcolor[rgb]{ 1, 0, 0}{\textbf{0.99768}} \\
	87 & \cellcolor[rgb]{ .753, .753, .753}0.88410 & \textbf{0.99614} & \cellcolor[rgb]{ .753, .753, .753}0.81167 & \textcolor[rgb]{ 0, 0, 1}{\textbf{0.99065}} & \textcolor[rgb]{ 1, 0, 0}{\textbf{0.99400}} & 0.95207 & 0.96672 & 0.97376 & 0.96918 & \cellcolor[rgb]{ .753, .753, .753}0.56200 & \cellcolor[rgb]{ .753, .753, .753}0.92318 & 0.96191 & \cellcolor[rgb]{ .753, .753, .753}0.60734 & \cellcolor[rgb]{ .753, .753, .753}0.91820 & \cellcolor[rgb]{ .753, .753, .753}0.47409 & 0.95891 & \cellcolor[rgb]{ .753, .753, .753}0.94874 & \cellcolor[rgb]{ .753, .753, .753}0.87797 & \cellcolor[rgb]{ .753, .753, .753}0.93903 \\
	88 & 0.98583 & 0.98510 & 0.98177 & \textbf{0.99212} & 0.96837 & 0.97750 & 0.97618 & 0.98710 & 0.98863 & \textcolor[rgb]{ 1, 0, 0}{\textbf{0.99141}} & 0.98837 & 0.98684 & 0.99035 & \cellcolor[rgb]{ .753, .753, .753}0.79181 & 0.98075 & \textcolor[rgb]{ 0, 0, 1}{\textbf{0.99082}} & 0.99028 & 0.99025 & 0.99034 \\
	89 & 0.99944 & \textcolor[rgb]{ 1, 0, 0}{\textbf{0.99963}} & 0.99850 & \textcolor[rgb]{ 0, 0, 1}{\textbf{0.99958}} & 0.99504 & 0.99871 & 0.99943 & \textbf{0.99970} & 0.99757 & 0.99588 & 0.99948 & 0.99842 & 0.99117 & 0.98001 & 0.97916 & 0.99924 & 0.99898 & 0.99823 & 0.99871 \\
	90 & \cellcolor[rgb]{ .753, .753, .753}0.83511 & \cellcolor[rgb]{ .753, .753, .753}0.82214 & \cellcolor[rgb]{ .753, .753, .753}0.84415 & \cellcolor[rgb]{ .753, .753, .753}0.79736 & \cellcolor[rgb]{ .753, .753, .753}0.93217 & \cellcolor[rgb]{ .753, .753, .753}0.88720 & 0.96791 & \cellcolor[rgb]{ .753, .753, .753}0.93930 & 0.96588 & \cellcolor[rgb]{ .753, .753, .753}0.86656 & \cellcolor[rgb]{ .753, .753, .753}0.84141 & \textcolor[rgb]{ 0, 0, 1}{\textbf{0.97382}} & \cellcolor[rgb]{ .753, .753, .753}0.71027 & \cellcolor[rgb]{ .753, .753, .753}0.87168 & \cellcolor[rgb]{ .753, .753, .753}0.82263 & \cellcolor[rgb]{ .753, .753, .753}0.88812 & 0.96388 & \textcolor[rgb]{ 1, 0, 0}{\textbf{0.97861}} & \textbf{0.98151} \\
	91 & 0.99436 & 0.98946 & 0.99344 & 0.99557 & 0.99251 & 0.98360 & 0.98901 & \textcolor[rgb]{ 0, 0, 1}{\textbf{0.99756}} & 0.99560 & 0.99147 & 0.99334 & 0.99139 & 0.99139 & 0.97607 & 0.98290 & 0.99696 & 0.99292 & \textcolor[rgb]{ 1, 0, 0}{\textbf{0.99831}} & \textbf{0.99907} \\
	92 & 0.99985 & 0.99984 & \textcolor[rgb]{ 0, 0, 1}{\textbf{0.99990}} & 0.99985 & 0.99839 & 0.99962 & 0.99914 & 0.99974 & 0.95788 & \cellcolor[rgb]{ .753, .753, .753}0.50000 & 0.99985 & 0.99975 & \textcolor[rgb]{ 1, 0, 0}{\textbf{0.99990}} & 0.99485 & \textbf{0.99994} & 0.99958 & 0.99869 & 0.99917 & 0.99968 \\
	93 & 0.96483 & \cellcolor[rgb]{ .753, .753, .753}0.94694 & 0.98138 & 0.96672 & 0.98350 & 0.98659 & 0.99053 & \textcolor[rgb]{ 1, 0, 0}{\textbf{0.99084}} & 0.97518 & \cellcolor[rgb]{ .753, .753, .753}0.80236 & 0.96430 & \textcolor[rgb]{ 0, 0, 1}{\textbf{0.99059}} & 0.95937 & \cellcolor[rgb]{ .753, .753, .753}0.87783 & 0.96516 & 0.96703 & 0.95910 & 0.98862 & \textbf{0.99264} \\
	94 & 0.99988 & 0.99266 & 0.99958 & 0.99984 & 0.99794 & 0.99654 & 0.99933 & 0.99947 & 0.99042 & \textcolor[rgb]{ 0, 0, 1}{\textbf{0.99999}} & 0.99988 & 0.99758 & \textbf{1.00000} & 0.99511 & \textbf{1.00000} & 0.99975 & 0.99910 & 0.99970 & 0.99973 \\
	95 & \textbf{1.00000} & 0.99883 & 0.99453 & \textcolor[rgb]{ 0, 0, 1}{\textbf{0.99997}} & 0.99665 & 0.99606 & 0.99977 & 0.99993 & 0.99948 & 0.99843 & 0.99994 & 0.99984 & 0.99840 & 0.99438 & \textbf{1.00000} & 0.99981 & 0.99869 & 0.99742 & 0.99888 \\
	96 & 0.99353 & 0.96564 & 0.97143 & \textcolor[rgb]{ 0, 0, 1}{\textbf{0.99449}} & 0.98413 & 0.99198 & \textcolor[rgb]{ 1, 0, 0}{\textbf{0.99493}} & 0.99388 & \textbf{0.99734} & 0.98226 & 0.99204 & 0.99272 & 0.98393 & \cellcolor[rgb]{ .753, .753, .753}0.93232 & 0.96657 & 0.98298 & 0.97014 & 0.98889 & 0.99284 \\
	97 & 0.99293 & \cellcolor[rgb]{ .753, .753, .753}0.90117 & 0.97337 & 0.99056 & 0.98289 & 0.97327 & 0.99406 & 0.99153 & \textbf{0.99595} & 0.95605 & 0.99190 & 0.98374 & 0.97354 & 0.96027 & 0.96665 & 0.98781 & 0.99374 & \textcolor[rgb]{ 0, 0, 1}{\textbf{0.99445}} & \textcolor[rgb]{ 1, 0, 0}{\textbf{0.99472}} \\
	98 & \textbf{1.00000} & 0.99991 & 0.98462 & \textbf{1.00000} & 0.99856 & 0.99972 & 0.99988 & 0.99997 & 0.99936 & 0.99645 & \textbf{1.00000} & \textbf{1.00000} & 0.99398 & 0.99113 & 0.98719 & 0.99951 & 0.99804 & 0.99804 & 0.99994 \\
	99 & \textbf{1.00000} & 0.99250 & 0.97899 & \textbf{1.00000} & 0.99314 & 0.98020 & 0.99825 & 0.99678 & 0.99991 & 0.98220 & \textbf{1.00000} & 0.99925 & 0.98628 & 0.99893 & \cellcolor[rgb]{ .753, .753, .753}0.93975 & 0.99970 & 0.99657 & 0.99702 & 0.99802 \\
	100 & 0.98000 & 0.98740 & \cellcolor[rgb]{ .753, .753, .753}0.85947 & 0.98138 & 0.99003 & 0.98584 & \textcolor[rgb]{ 0, 0, 1}{\textbf{0.99719}} & 0.99575 & 0.99070 & \cellcolor[rgb]{ .753, .753, .753}0.62044 & 0.98083 & 0.99055 & \cellcolor[rgb]{ .753, .753, .753}0.65218 & 0.96071 & \cellcolor[rgb]{ .753, .753, .753}0.93395 & 0.99012 & 0.98517 & \textbf{0.99917} & \textcolor[rgb]{ 1, 0, 0}{\textbf{0.99777}} \\
	\hline
	mean & 0.97986 & 0.97750 & 0.96159 & 0.98006 & 0.98530 & 0.98452 & \textcolor[rgb]{ 0, 0, 1}{\textbf{0.99289}} & \textcolor[rgb]{ 1, 0, 0}{\textbf{0.99390}} & 0.98288 &\cellcolor[rgb]{ .753, .753, .753}0.90307 & \cellcolor[rgb]{ .753, .753, .753}0.94764 & 0.98593 & \cellcolor[rgb]{ .753, .753, .753}0.90029 & 0.96881 & 0.95834 & 0.98622 & 0.98866 & 0.99249 & \textbf{ 0.99481} \\
	\hline
\end{tabular}}%
\label{tab:auc_50}%
\end{table*}

\begin{figure*}[htb]
	\centering
	\captionsetup[subfloat]{labelsep=none,format=plain,labelformat=empty} %
	\subfloat[False Color\label{sfig:t83_color}]{ \includegraphics[width=0.8in]{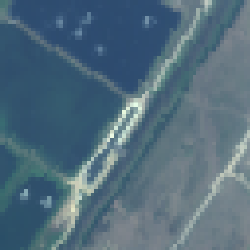}} 
	\hfil
	\subfloat[Ground Truth\label{sfig:t83_gt}]{\includegraphics[width=0.8in]{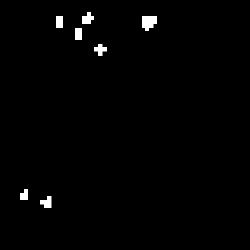}}
	\hfil
	\subfloat[GRX\label{sfig:t83_GRX}]{\includegraphics[width=0.8in]{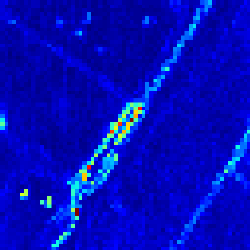}}
	\hfil
	\subfloat[LRX\label{sfig:t83_LRX}]{\includegraphics[width=0.8in]{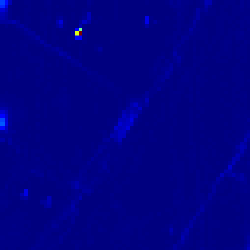}}
	\hfil
	\subfloat[KRX\label{sfig:t83_KRX}]{\includegraphics[width=0.8in]{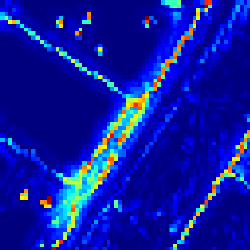}}
	\hfil
	\subfloat[QLRX\label{sfig:t83_QLRX}]{\includegraphics[width=0.8in]{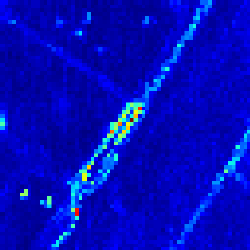}}
	\hfil
	\subfloat[2S-GLRT\label{sfig:t83_2S-GLRT}]{\includegraphics[width=0.8in]{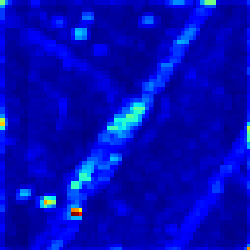}}
	\hfil
	\\
	\vspace{-3mm}
	\subfloat[KSVDD\label{sfig:t83_KSVDD}]{ \includegraphics[width=0.8in]{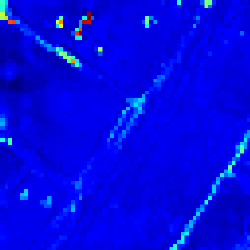}} 
	\hfil
	\subfloat[CR\label{sfig:t83_CR}]{\includegraphics[width=0.8in]{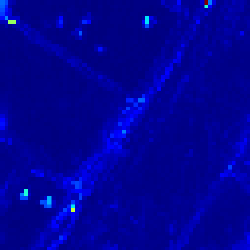}}
	\hfil
	\subfloat[KCSR\label{sfig:t83_KCSR}]{\includegraphics[width=0.8in]{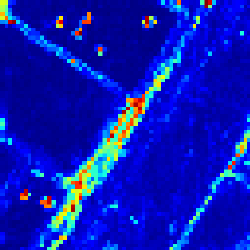}}
	\hfil
	\subfloat[KIFD\label{sfig:t83_KIFD}]{\includegraphics[width=0.8in]{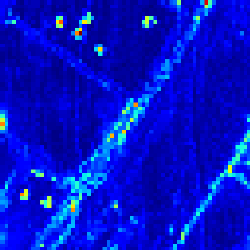}}
	\hfil
	\subfloat[PTA\label{sfig:t83_PTA}]{\includegraphics[width=0.8in]{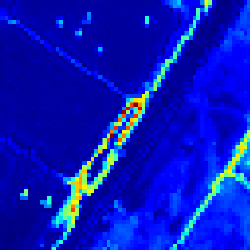}}
	\hfil
	\subfloat[FrFE\label{sfig:t83_FrFE}]{\includegraphics[width=0.8in]{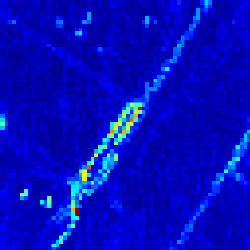}}
	\hfil
	\subfloat[AED\label{sfig:t83_AED}]{\includegraphics[width=0.8in]{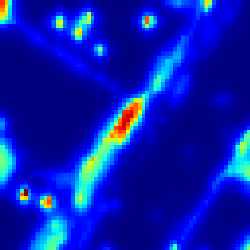}}
	\\
	\vspace{-3mm}
	\subfloat[RGAE\label{sfig:t83_RGAE}]{\includegraphics[width=0.8in]{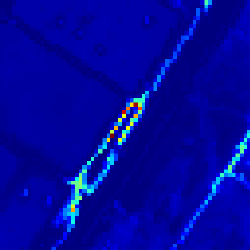}}
	\hfil
	\subfloat[AutoAD\label{sfig:t83_AutoAD}]{\includegraphics[width=0.8in]{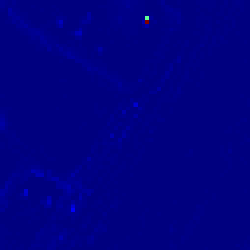}}
	\hfil
	\subfloat[LREN\label{sfig:t83_LREN}]{\includegraphics[width=0.8in]{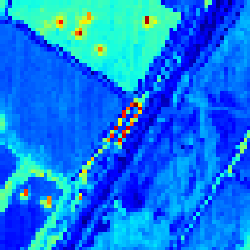}}
	\hfil
	\subfloat[$\text{WeaklyAD}_{\mathcal{A}}$\label{sfig:t83_WeaklyAD}]{\includegraphics[width=0.8in]{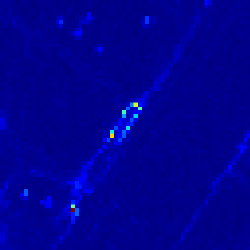}}
	\hfil
	\subfloat[$\text{WeaklyAD}_{\mathcal{B}}$\label{sfig:t83_WeaklyAD-G}]{\includegraphics[width=0.8in]{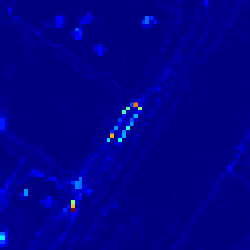}}
	\hfil
	\subfloat[AETNet\label{sfig:t83_AETNet}]{\includegraphics[width=0.8in]{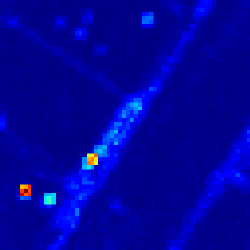}}
	\hfil
	\subfloat[AETNet-KCSR\label{sfig:t83_AETNet-KCSR}]{\includegraphics[width=0.8in]{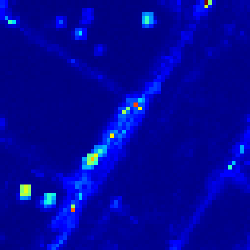}}
	\hfil
	\\
	\caption{Detection maps of different AD methods on the first 50 bands of the test scene 83.}
	\label{fig:83}
\end{figure*}

\begin{figure*}[htb]
	\centering
	\captionsetup[subfloat]{labelsep=none,format=plain,labelformat=empty} %
	\subfloat[False Color\label{sfig:t87_color}]{ \includegraphics[width=0.8in]{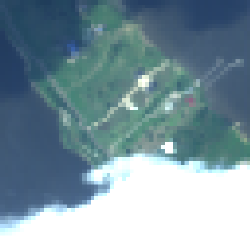}} 
	\hfil
	\subfloat[Ground Truth\label{sfig:t87_gt}]{\includegraphics[width=0.8in]{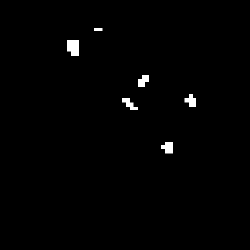}}
	\hfil
	\subfloat[GRX\label{sfig:t87_GRX}]{\includegraphics[width=0.8in]{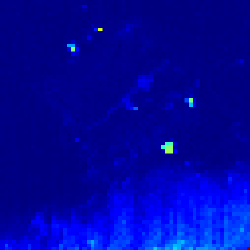}}
	\hfil
	\subfloat[LRX\label{sfig:t87_LRX}]{\includegraphics[width=0.8in]{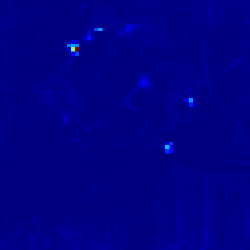}}
	\hfil
	\subfloat[KRX\label{sfig:t87_KRX}]{\includegraphics[width=0.8in]{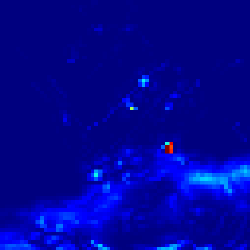}}
	\hfil
	\subfloat[QLRX\label{sfig:t87_QLRX}]{\includegraphics[width=0.8in]{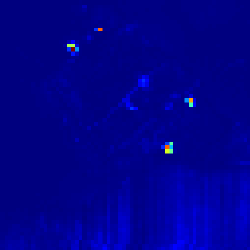}}
	\hfil
	\subfloat[2S-GLRT\label{sfig:t87_2S-GLRT}]{\includegraphics[width=0.8in]{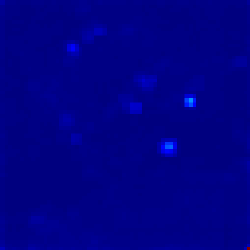}}
	\\
	\vspace{-3mm}
	\subfloat[KSVDD\label{sfig:t87_KSVDD}]{ \includegraphics[width=0.8in]{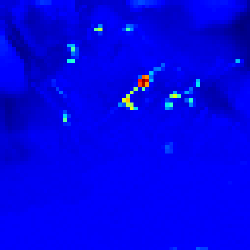}} 
	\hfil
	\subfloat[CR\label{sfig:t87_CR}]{\includegraphics[width=0.8in]{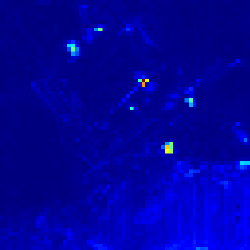}}
	\hfil
	\subfloat[KCSR\label{sfig:t87_KCSR}]{\includegraphics[width=0.8in]{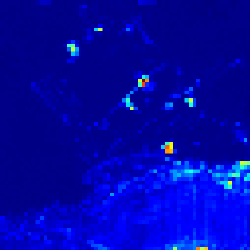}}
	\hfil
	\subfloat[KIFD\label{sfig:t87_KIFD}]{\includegraphics[width=0.8in]{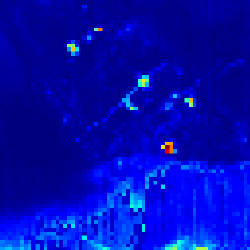}}
	\hfil
	\subfloat[PTA\label{sfig:t87_PTA}]{\includegraphics[width=0.8in]{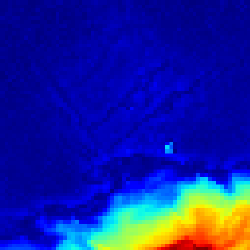}}
	\hfil
	\subfloat[FrFE\label{sfig:t87_FrFE}]{\includegraphics[width=0.8in]{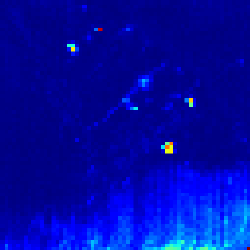}}
	\hfil
	\subfloat[AED\label{sfig:t87_AED}]{\includegraphics[width=0.8in]{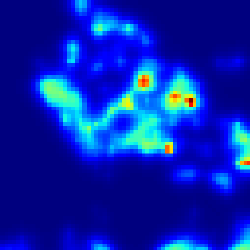}}
	\\
	\vspace{-3mm}
	\subfloat[RGAE\label{sfig:t87_RGAE}]{\includegraphics[width=0.8in]{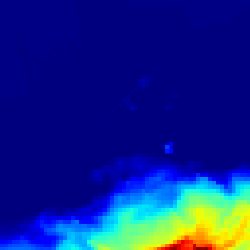}}
	\hfil
	\subfloat[AutoAD\label{sfig:t87_AutoAD}]{\includegraphics[width=0.8in]{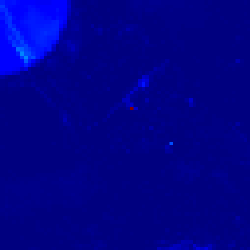}}
	\hfil
	\subfloat[LREN\label{sfig:t87_LREN}]{\includegraphics[width=0.8in]{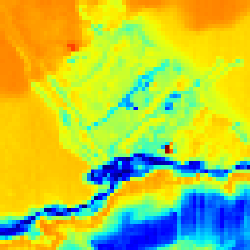}}
	\hfil
	\subfloat[$\text{WeaklyAD}_{\mathcal{A}}$\label{sfig:t87_WeaklyAD}]{\includegraphics[width=0.8in]{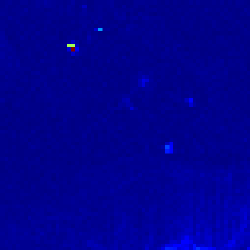}}
	\hfil
	\subfloat[$\text{WeaklyAD}_{\mathcal{B}}$\label{sfig:t87_WeaklyAD-G}]{\includegraphics[width=0.8in]{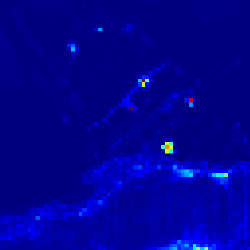}}
	\hfil
	\subfloat[AETNet\label{sfig:t87_AETNet}]{\includegraphics[width=0.8in]{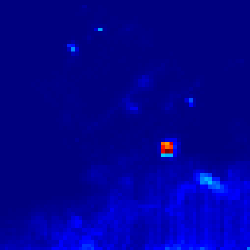}}
	\hfil
	\subfloat[AETNet-KCSR\label{sfig:t87_AETNet-KCSR}]{\includegraphics[width=0.8in]{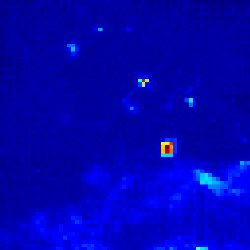}}
	\caption{Detection maps of different AD methods on the first 50 bands of the test scene 87.}
	\label{fig:87}
\end{figure*}

\begin{figure*}[htb]
	\centering
	\captionsetup[subfloat]{labelsep=none,format=plain,labelformat=empty} %
	\subfloat[False Color\label{sfig:t90_color}]{ \includegraphics[width=0.8in]{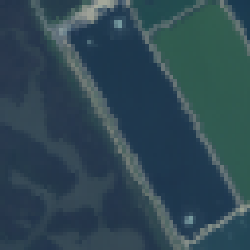}} 
	\hfil
	\subfloat[Ground Truth\label{sfig:t90_gt}]{\includegraphics[width=0.8in]{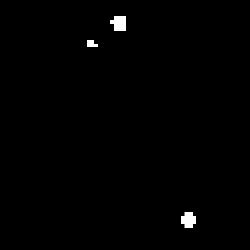}}
	\hfil
	\subfloat[GRX\label{sfig:t90_GRX}]{\includegraphics[width=0.8in]{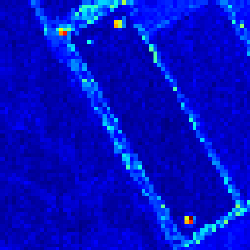}}
	\hfil
	\subfloat[LRX\label{sfig:t90_LRX}]{\includegraphics[width=0.8in]{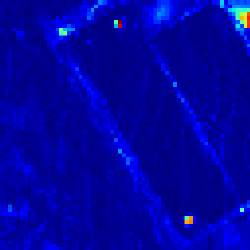}}
	\hfil
	\subfloat[KRX\label{sfig:t90_KRX}]{\includegraphics[width=0.8in]{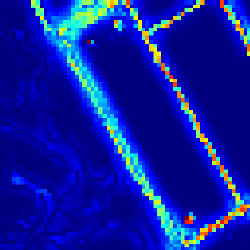}}
	\hfil
	\subfloat[QLRX\label{sfig:t90_QLRX}]{\includegraphics[width=0.8in]{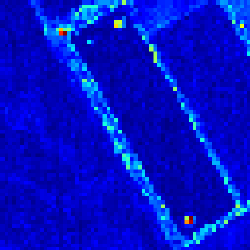}}
	\hfil
	\subfloat[2S-GLRT\label{sfig:t90_2S-GLRT}]{\includegraphics[width=0.8in]{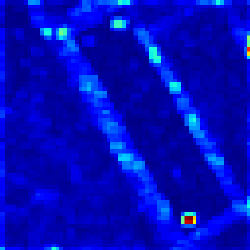}}
	\\
	\vspace{-3mm}
	\subfloat[KSVDD\label{sfig:t90_KSVDD}]{ \includegraphics[width=0.8in]{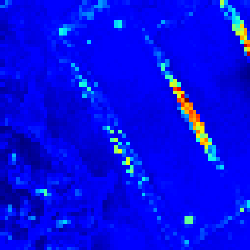}} 
	\hfil
	\subfloat[CR\label{sfig:t90_CR}]{\includegraphics[width=0.8in]{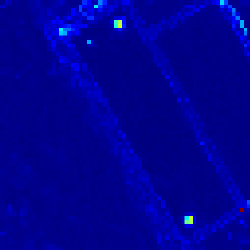}}
	\hfil
	\subfloat[KCSR\label{sfig:t90_KCSR}]{\includegraphics[width=0.8in]{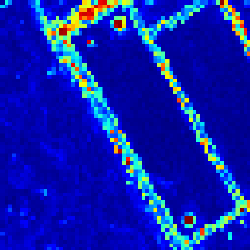}}
	\hfil
	\subfloat[KIFD\label{sfig:t90_KIFD}]{\includegraphics[width=0.8in]{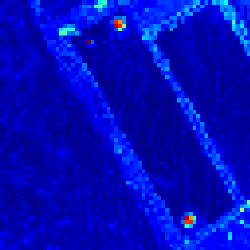}}
	\hfil
	\subfloat[PTA\label{sfig:t90_PTA}]{\includegraphics[width=0.8in]{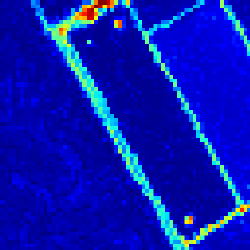}}
	\hfil
	\subfloat[FrFE\label{sfig:t90_FrFE}]{\includegraphics[width=0.8in]{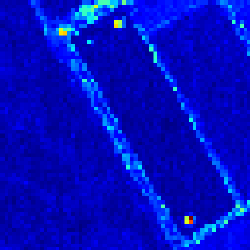}}
	\hfil
	\subfloat[AED\label{sfig:t90_AED}]{\includegraphics[width=0.8in]{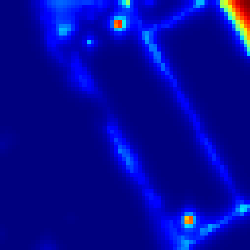}}
	\\
	\vspace{-3mm}
	\subfloat[RGAE\label{sfig:t90_RGAE}]{\includegraphics[width=0.8in]{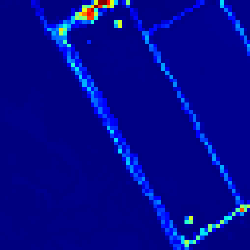}}
	\hfil
	\subfloat[AutoAD\label{sfig:t90_AutoAD}]{\includegraphics[width=0.8in]{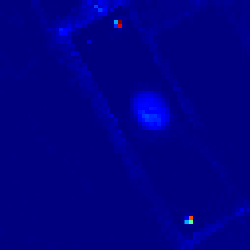}}
	\hfil
	\subfloat[LREN\label{sfig:t90_LREN}]{\includegraphics[width=0.8in]{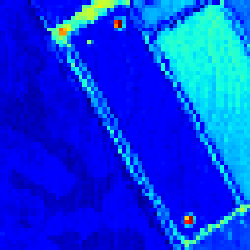}}
	\hfil
	\subfloat[$\text{WeaklyAD}_{\mathcal{A}}$\label{sfig:t90_WeaklyAD}]{\includegraphics[width=0.8in]{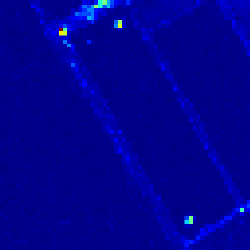}}
	\hfil
	\subfloat[$\text{WeaklyAD}_{\mathcal{B}}$\label{sfig:t90_WeaklyAD-G}]{\includegraphics[width=0.8in]{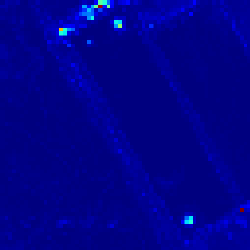}}
	\hfil
	\subfloat[AETNet\label{sfig:t90_AETNet}]{\includegraphics[width=0.8in]{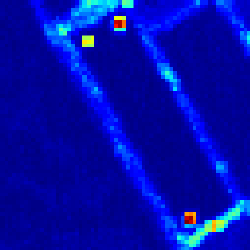}}
	\hfil
	\subfloat[AETNet-KCSR\label{sfig:t90_AETNet-KCSR}]{\includegraphics[width=0.8in]{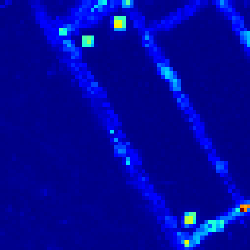}}
	\\
	\caption{Detection maps of different AD methods on the first 50 bands of the test scene 90.}
	\label{fig:90}
\end{figure*}

\subsection{Experimental Settings}

\subsubsection{Implementation Details of AETNet}

Since the training input of AETNet needs to be consistent with the test HSIs in the spatial size and band number, we generate $64\times64$ mask maps in the mask generation module and set the parameter $K$ to 8.
The number range $\left[ N_{\min}, N_{\max} \right]$ of pseudo targets in mask maps is set to $\left[1, 32 \right]$, and the area range $\left[ A_{\min}, A_{\max} \right]$ of pseudo targets is set to $\left[3, 20\right]$.
Regardless of the band number $B$ of input data, the channel number $C$ output by the convolutional encoder is fixed at 32.
The number of attention heads of five Swin Transformer blocks in AETNet is set to 2, 4, 8, 4, and 2, respectively.
The number of non-overlapping local windows in all Swin Transformer blocks is set to $8\times8$.
In the calculation of MSGMS loss, we set the constant $c$ to 1 and perform four pooling operations, i.e., the parameter $S=5$.

In the training stage, the Adam optimizer \cite{adam} with a learning rate of 1e$^{-4}$ and a weight decay of 5e$^{-6}$ is used to optimize our AETNet.
The batch size is set to 16.
Training terminates when the peak value of the measurement  $\mathcal{M}$ in the transform domain search module does not change for 30 epochs, or the epoch number reaches 200.
Then, the trained model corresponding to the peak value of $\mathcal{M}$ is used for inference.
Before being sent to AETNet, all the input data are processed with linear normalization and scaled to $\left[ -0.1,0.1 \right]$.
We select the test scene 1 in HAD100 dataset as the validation HSI in the transform domain search module.
If there is no specific statement, we use the AVIRIS-NG training set and CutOut by default. 

In addition, we use three simple data augmentation approaches to expand the training set of AETNet:

\textbf{Image Crop:} We crop a single training HSI into four $64\times64$ HSIs.
The AVIRIS-NG training set is expanded to 1048 HSIs, and the AVIRIS-Classic training set is expanded to 1040 HSIs.

\textbf{Image Rotation:} 
At the beginning of each training epoch, every input HSI is randomly rotated by $0^{\circ}$, $90^{\circ}$, $180^{\circ}$, or $270^{\circ}$.

\textbf{Image Flip:} After random rotation, each input HSI is randomly flipped along the horizontal or vertical direction.
The probabilities of horizontal flip and vertical flip are both 50\%.

\subsubsection{Comparative Methods}
Sixteen classic or SOTA hyperspectral AD methods are selected to be evaluated on our HAD100 dataset, including GRX\cite{RX-AD}, LRX\cite{GRX-LRX-AD}, KRX\cite{KRX-AD},	QLRX\cite{QLRX-AD}, 2S-GLRT\cite{2S-GLRT-AD},	KSVDD\cite{SVDD11-AD},	CR\cite{CR-AD},	KCSR\cite{CSR-AD},	KIFD\cite{KIFD-AD},	PTA\cite{PTA-AD},	FrFE\cite{FrFE-AD},	AED \cite{AED-AD},	RGAE \cite{RGAE-AD},	AutoAD\cite{Auto-AD},	LREN \cite{LREN-AD}, and WeaklyAD\cite{wealyAD-AD}.
GRX is the most classic AD method and is widely used as the AD baseline.
LRX, KRX, QLRX, 2S-GLRT, CR, and KCSR are dual-window-based AD methods. 
CR and KCSR are sparse-representation-based AD methods.
PTA is one of the most advanced tensor-approximation-based AD methods, and AET introduces morphological image processing to the hyperspectral AD field.
KRX, KSVDD, KCSR, and KIFD all use the kernel trick to process spectral samples.
RGAE, LREN, WeaklyAD, and AutoAD are deep-learning-based methods, of which the first three are spectral-level self-supervised methods.
In particular, WeaklyAD can perform inference on unseen test scenes.
 
The codes of 2S-GLRT, KCSR, KIFD, PTA, FrFE, RGAE, AutoAD, and LREN are all official versions, and the codes of GRX and AED are widely used unofficial versions.
Besides, we reproduce the codes of LRX, KRX, QLRX, KSVDD, CR, and WeaklyAD, whose detection performances are close to those in the literature.
 
Most of the above methods adopt Maxmin Normalization for data preprocessing and linearly scale each test HSI to $\left[ 0, 1 \right]$. 
But some methods adopt distinctive normalization.
2S-GLRT and PTA perform Maxmin Normalization on each band of HSI data.
AET uses principal component analysis (PCA) to reduce the dimension of HSI data and then linearly scale drop-dimensional HSI data to $\left[ 0, 1 \right]$.
 FrFE divides all values on the test HSI by the maximum value.
\subsubsection{Settings of Comparative Experiments}
To explore the influence of channel number and spectral range on the generalization ability of AD methods, we conducte three comparative experiments on the first 50 bands, the first 100 bands, and the first 200 bands of HAD100 dataset, respectively.
For WeaklyAD, we simultaneously compare its retraining version and non-retraining version, denoted as $\text{WeaklyAD}_{\mathcal{A}}$ and $\text{WeaklyAD}_{\mathcal{B}}$, respectively. 
To achieve a fair comparison with our AETNet, $\text{WeaklyAD}_{\mathcal{B}}$ extracts coarse anomaly samples on tthe test sence 1 and adopts the whole AVIRIS-NG training set as the background sample set. 
On the first 50 bands of the test set, the critical parameters of all comparative methods are fine-tuned to achieve the best mAUC values. 
Then, these carefully regulated methods are  directly evaluated on the first 100 bands and first 200 bands without parameter readjustment.
Furthermore, the dual window sizes $\left( \omega_\text{in}, \omega_\text{out} \right)$ of all the dual-window-based methods traverse from 3 to 29, and 
\tabref{tab:window} shows the best dual window sizes of different methods.
The upper bound of the outer window size $\omega_\text{out} $ is set to 29 to avoid huge calculation costs and ensure the local property of the dual-window-based methods.
Interestingly, the best outer window sizes $\omega_\text{out} $ of all the dual-window-based methods except KRX reach the upper bound of 29.
In addition to the standard version of AETNet, we also provide an advanced version that replaces \eqrefnew{eq:GRX1} with KCSR, named AETNet-KCSR.
AETNet-KCSR adopts the same parameter settings as AETNet and KCSR. 

In addition to mAUC and mASNPR, we also report the average inference time of all the comparative methods on a single test HSI.
 All the experiments are implemented on a PC with an Intel Core i9-9980XE CPU and two GeForce RTX 2080 Ti GPUs.

\subsection{Analysis of Detection Performance}
\subsubsection{Analysis of Experiments on the First 50 Bands}
The gray region (the 2nd to 5th columns) in \tabref{tab:result} shows the detection performance and inference time of difference AD methods on the first 50 bands of  HAD100 dataset.
In terms of mAUC, the top three methods, AETNet-KCSR, KCSR, and CR, are all based on sparse representation, with values of 0.9948, 0.9939, and 0.9929, respectively.
The standard AETNet ranks fourth with an mAUC value of 0.9925.
The two worst-performing methods are RGAE and PTA, of which mAUC values are 0.9003 and 0.9031, respectively. 
The mAUC value of GRX reaches 0.9799, and the local improved versions of GRX (LRX, KRX, and QLRX) do not show advantages.
\tabref{tab:auc_50} shows the AUC values of different AD methods on all the test scenes.
The colored table intuitively gives a similar conclusion as the mAUC performance in \tabref{tab:result}.
CR, KCSR, AETNet, and AETNet-KCSR have the fewest gray cells, while PTA and RGAE have the most gray cells. 
The standard and advanced versions of AETNet have obvious advantages on the test scenes 42, 44, and 90.
 Visual results of different methods on the test scene 90 are shown in \figref{fig:90}. 
Compared to other methods, our AETNet has higher response values to compact targets and lower response values to borders (the false alarms on the detection maps of most AD methods).
However, both AETNet and AETNet-KCSR performs less competitively on the test scenes 83 and 87.
As shown in \figref{fig:83}, our AETNet can effectively suppress most borders on the test scene 83 but generates a false target on a dark border.
The rest region of the dark border is mixed with the bright borders, causing AETNet not to treat this border as a whole. 
As shown in \figref{fig:87}, there is a large cloud region the test scene 87, where many methods have high anomaly response values.
Our AETNet also has high anomaly response values on a group of mixed pixels of cloud and land, which has spatial characteristics similar to those of anomaly targets.
The performance on the test scenes 83 and 87 indicates that the AETNet's ability to perceive mixed pixels needs further improvement.

In terms of mASNPR, the top three methods, $\text{WeaklyAD}_{\mathcal{B}}$, AETNet-KCSR and AETNet, reach 13.75 dB, 13.15 dB, and 11.72 dB, respectively.
These three methods are all deep-learning-based methods without retraining. 
In terms of the inference time, the standard AETNet with 0.035s ranks second after GRX with 0.022s.
Except for AED and $\text{WeaklyAD}_{\mathcal{B}}$, the remaining methods have high calculation costs. 
Compared to GRX, the standard AETNet improves the AUC performance from 0.9799 to 0.9925 with little additional time, while maintaining the parameter-free characteristic of GRX.
As shown in \figref{fig:auc_pfs}, the standard AETNet achieves the best balance between detection accuracy and efficiency with mAUC of 0.9925 and inference speed of 28.6 FPS.
\subsubsection{Analysis of Experiments on the First 100 Bands}
The blue region (the 6th to 9th columns) in \tabref{tab:result} shows the detection performance and inference time of difference AD methods on the first 100 bands.
All the detection accuracy evaluations except the  mASNPR value of AutoAD are lower than those on the first 50 bands.
GRX is a parameter-free AD method, and AET uses PAC to reduce the spectral dimension to a fixed value.
The performance degradation of these two methods indicates that the added spectral range on the first 100 bands contains distractive spectral information and increases the difficulty of AD.
 
 Our AETNet still achieves higher accuracy and efficiency than the SOTA methods. 
In terms of mAUC, AETNet-KCSR and AETNet win the top two places with 0.9897 and 0.9875, respectively.
The detection time of most methods increases dramatically, while that of the standard AETNet remains at a low level of 0.043s.

\subsubsection{Analysis of Experiments on the First 200 Bands}
The blue region (the 10th to 13th columns) in \tabref{tab:result} shows the detection performance and inference time of difference AD methods on the first 200 bands.
Compared to the first 100 bands data, the first 200 bands data has more distractive information and is more challenging for AD methods.
Except for LRX, LREN,  and $\text{WeaklyAD}_{\mathcal{B}}$, the remaining methods have lower mAUC values than those on the first 50 and 100 bands.
AETNet-KCSR is still the first place in terms of mAUC, and the standard AETNet ranks fifth with a value of 0.9818.
In terms of inference efficiency, GRX is still the fastest with inference time of 0.058s, followed by AED and AETNet with inference time of 0.067s and 0.07s, respectively.
 
\subsection{Influence of Random Mask Strategy on Generalization Capability}
\begin{table}[t]
	\renewcommand{\arraystretch}{1.3}
	\setlength\tabcolsep{13pt}
	\centering
	\caption{The mAUC Performance Achieved by Different Settings of Random Mask Strategy on the AVIRIS-NG Training Set.}
	\begin{tabular}{c|c c c}
		\hline
  Random Mask   & 50 Bands & 100 Bands & 200 Bands \\
		\hline
		CutOut & \textbf{0.9925} & \textbf{0.9875} & \textbf{0.9818} \\
		CutMix & 0.9883 & 0.9826 & 0.9770 \\
		w/o & 0.9905 & 0.9714 & 0.9793 \\
		\hline
	\end{tabular}%
	\label{tab:maskNG}%
\end{table}

\begin{table}[t]
	\renewcommand{\arraystretch}{1.3}
	\setlength\tabcolsep{10pt}
	\centering
	\caption{The mAUC Performance Achieved by Different Settings of Random Mask Strategy on the AVIRIS-Classic Training Set.}
	\begin{tabular}{c|cc}
		\hline
	 Random Mask & 50 Bands & 100 Bands \\
		\hline
		CutOut & \textbf{0.9904} & 0.9730 \\
		CutMix & 0.9864 & \textbf{0.9837} \\
		w/o & 0.9883 & 0.9715 \\
		\hline
	\end{tabular}%
	\label{tab:maskC}%
\end{table}

Both our standard AETNet and $\text{WeaklyAD}_{\mathcal{B}}$ are trained on the AVIRIS-NG training set and use GRX as the detector.
As shown in \tabref{tab:result}, AETNet has a higher gain to GRX than $\text{WeaklyAD}_{\mathcal{B}}$ on the unseen test scenes.
Compared with $\text{WeaklyAD}_{\mathcal{B}}$, AETNet achieves an image-level training paradigm and adopts the Random Mask strategy to handle  the generalization problem.
In the following, we analyze the generalization ability under different settings of the Random Mask strategy.
\subsubsection{Analysis of Cross-Scene Generalization Capability}
The AVIRIS-NG training set is collected from the same device sensor as the test set and does not include anomaly scenes.
As shown in \tabref{tab:maskNG}, we compare the mAUC values of AETNet with CutOut, AETNet with CutMix, and AETNet without Random Mask.
All the mAUC values in three  comparative  experiments with different band numbers are higher than those of GRX except AETNet without Random Mask. 
Among them, AETNet with CutOut achieves the best mAUC values on all three comparative experiments, which are 1.53\%, 1.88\%, and 2.28\% higher than those of GRX.
It can be seen that CutOut has the most obvious effect on the cross-scene (on the same device) generalization capability of AETNet.

\subsubsection{Analysis of Cross-Device Generalization Capability} 
The AVIRIS-Classic training set is collected by a different device sensor from the test set.
Since AVIRIS-Classic data only remains 162 bands, we select the first 50 and the first 100 bands of the AVIRIS-Classic training set as the training data.
Due to the different spectral intervals and resolutions, there is no common background spectrum between the training and test data.
The cross-device generalization capability is a more difficult challenge for hyperspectral AD networks.
 As shown in \tabref{tab:maskC}, the mAUC values of AETNet with CutOut are 0.9904 and 0.9730, respectively, which are lower than those in the  cross-scene generalization capability (on the same device) experiments. 
AETNet with CutMix has the lowest mAUC value of 0.9864 on the first 50 bands of the test set, but achieves the highest mAUC value of 0.9837 on the first 100 bands of the test set.
In addition, AETNet with CutMix is superior to the baseline (GRX) on these two comparative experiments, and the mAUC values are increased by 0.67\% and 1.27\%, respectively.
In summary, AETNet with CutMix is suitable for inference on the cross-device test data.
\subsection{Analysis of Training Data Scale} 

\begin{table}[t]
 \renewcommand{\arraystretch}{1.3}
\setlength\tabcolsep{6pt}
\centering
\caption{Detection Accuracy and Training Epoch Searched by the transform domain Search Module under the AVIRIS-NG Training Data of Different Scales on the First 50 Bands}
	\begin{tabular}{cccc}
						\hline
		Training Images & Data Ratio & mAUC & Training Epoch \\			
			\hline
 52 & 5\% & 0.9916 & 70 \\
	104 & 10\% & 0.9904 & 45 \\
	156 & 15\% & 0.9922 & 93 \\
	208 & 20\% & 0.9925 & 39 \\
	312 & 30\% & 0.9925 & 32 \\
	416 & 40\% & 0.9926 & 18 \\
	520 & 50\% & 0.9924 & 17 \\
	624 & 60\% & 0.9926 & 13 \\
	728 & 70\% & 0.9905 & 6 \\
	832 & 80\% & 0.9919 & 6 \\
	936 & 90\% & 0.9920 & 6 \\
	1040 & 100\% & 0.9925 & 10 \\
	\hline
	\end{tabular}%
	\label{tab:dataratio}%
\end{table}%

Compared with the previous AD methods, our AETNet can be trained on a training set that does not contain any test HSI.
The above experiments demonstrate that the image-level training paradigm is effective for hyperspectral AD.
 \tabref{tab:dataratio} shows the detection accuracy of AETNet and the training epoch that produces the peak value of the measurement $\mathcal{M}$ in the transform domain search module under the AVIRIS-NG training data of different scales.
With only 5\% of the training data,  AETNet can reach a mAUC value of 0.9916.
Using 15\% of the training set can achieve the detection accuracy obtained on the whole training set.
In addition, the epoch required for searching the transform domain decreases as the scale of the training data increases.

\begin{table*}[tb!]
	\centering
		\renewcommand{\arraystretch}{1.3}
	\setlength\tabcolsep{3pt}
	\caption{Detection Accuracy Performance of AETNet Achieved by Different Loss Functions on HAD100 Dataset.}
	\begin{tabular}{c|cc|cc|cc|cc|cc|cc|cc|cc}
			\hline
	Data	& \multicolumn{8}{c}{\cellcolor[rgb]{ .749, .749, .749}50 Bands}\vline& \multicolumn{8}{c}{\cellcolor[rgb]{ .847, .894, .737}100 Bands} \\
			\hline
		Loss & \multicolumn{2}{c}{\cellcolor[rgb]{ .749, .749, .749}L1} \vline& \multicolumn{2}{c}{\cellcolor[rgb]{ .749, .749, .749}L2} \vline& \multicolumn{2}{c}{\cellcolor[rgb]{ .749, .749, .749}SSIM} \vline& \multicolumn{2}{c}{\cellcolor[rgb]{ .749, .749, .749}MSGMS} \vline& \multicolumn{2}{c}{\cellcolor[rgb]{ .847, .894, .737}L1} \vline& \multicolumn{2}{c}{\cellcolor[rgb]{ .847, .894, .737}L2} \vline& \multicolumn{2}{c}{\cellcolor[rgb]{ .847, .894, .737}SSIM}\vline & \multicolumn{2}{c}{\cellcolor[rgb]{ .847, .894, .737}MSGMS} \\
			\hline
		Random Mask & \cellcolor[rgb]{ .749, .749, .749}$\checkmark$ & \cellcolor[rgb]{ .749, .749, .749}$\usym{2613}$ & \cellcolor[rgb]{ .749, .749, .749}$\checkmark$ & \cellcolor[rgb]{ .749, .749, .749}$\usym{2613}$ & \cellcolor[rgb]{ .749, .749, .749} $\checkmark$ & \cellcolor[rgb]{ .749, .749, .749}$\usym{2613}$& \cellcolor[rgb]{ .749, .749, .749}$\checkmark$ & \cellcolor[rgb]{ .749, .749, .749} $\usym{2613}$& \cellcolor[rgb]{ .847, .894, .737} $\checkmark$& \cellcolor[rgb]{ .847, .894, .737} $\usym{2613}$& \cellcolor[rgb]{ .847, .894, .737} $\checkmark$ & \cellcolor[rgb]{ .847, .894, .737}$\usym{2613}$& \cellcolor[rgb]{ .847, .894, .737}$\checkmark$ & \cellcolor[rgb]{ .847, .894, .737}$\usym{2613}$& \cellcolor[rgb]{ .847, .894, .737} $\checkmark$& \cellcolor[rgb]{ .847, .894, .737} $\usym{2613}$\\
			\hline
		mAUC & \cellcolor[rgb]{ .749, .749, .749}0.9799 & \cellcolor[rgb]{ .749, .749, .749}0.9799 & \cellcolor[rgb]{ .749, .749, .749}0.9875 & \cellcolor[rgb]{ .749, .749, .749}0.9799 & \cellcolor[rgb]{ .749, .749, .749}0.9880 & \cellcolor[rgb]{ .749, .749, .749}0.9799 & \cellcolor[rgb]{ .749, .749, .749}0.9925 & \cellcolor[rgb]{ .749, .749, .749}0.9905 & \cellcolor[rgb]{ .847, .894, .737}0.9811 & \cellcolor[rgb]{ .847, .894, .737}0.9714 & \cellcolor[rgb]{ .847, .894, .737}0.9856 & \cellcolor[rgb]{ .847, .894, .737}0.9775 & \cellcolor[rgb]{ .847, .894, .737}0.9858 & \cellcolor[rgb]{ .847, .894, .737}0.9716 & \cellcolor[rgb]{ .847, .894, .737}0.9875 & \cellcolor[rgb]{ .847, .894, .737}0.9714 \\ 
			\hline
	\end{tabular}%
	\label{tab:loss}%
\end{table*}%

\begin{figure*}[tb]
	\centering
	\captionsetup[subfloat]{labelsep=none,format=plain,labelformat=empty}
	(a)
	\begin{minipage}[c]{0.95\textwidth}
		\subfloat[\label{sfig:6_msgms_rm_1}]{\includegraphics[width=0.6in]{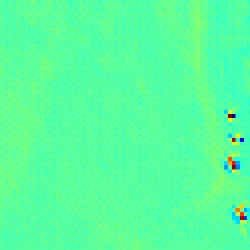}}
		\hfil
		\subfloat[\label{sfig:6_msgms_rm_2}]{\includegraphics[width=0.6in]{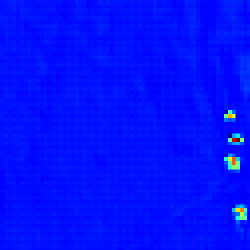}}
		\hfil
		\subfloat[\label{sfig:6_msgms_rm_3}]{\includegraphics[width=0.6in]{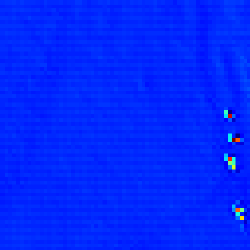}}
		\hfil
		\subfloat[\label{sfig:6_msgms_rm_4}]{\includegraphics[width=0.6in]{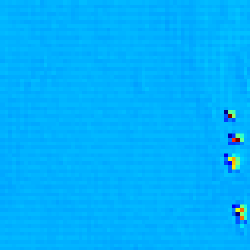}}
		\hfil
		\subfloat[\label{sfig:6_msgms_rm_5}]{\includegraphics[width=0.6in]{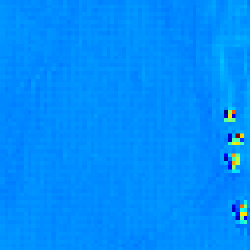}}
		\hfil
		\subfloat[\label{sfig:6_msgms_rm_6}]{\includegraphics[width=0.6in]{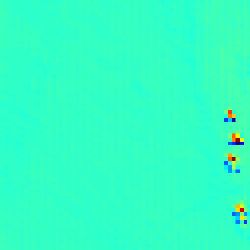}}
		\hfil
		\subfloat[\label{sfig:6_msgms_rm_7}]{\includegraphics[width=0.6in]{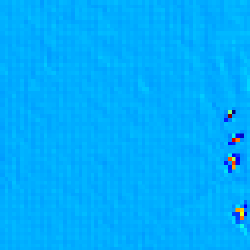}}
		\hfil
		\subfloat[\label{sfig:6_msgms_rm_8}]{\includegraphics[width=0.6in]{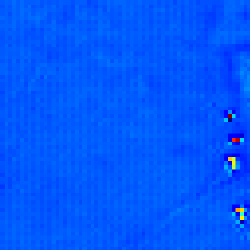}}
		\hfil
		\subfloat[\label{sfig:6_msgms_rm_9}]{\includegraphics[width=0.6in]{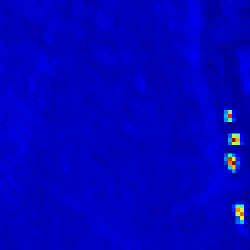}}
		\hfil
		\subfloat[\label{sfig:6_msgms_rm_10}]{\includegraphics[width=0.6in]{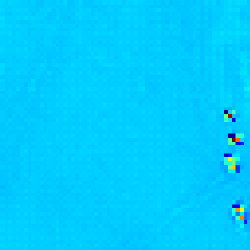}}
	\end{minipage}
	\\
	\vspace{-5mm}
	(b)
	\begin{minipage}[c]{0.95\textwidth}
		\subfloat[\label{sfig:6_msgms_worm_1}]{\includegraphics[width=0.6in]{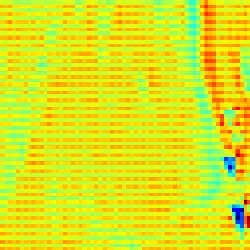}}
		\hfil
		\subfloat[\label{sfig:6_msgms_worm_2}]{\includegraphics[width=0.6in]{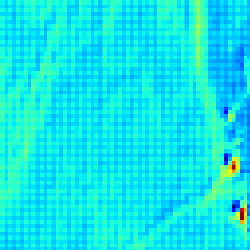}}
		\hfil
		\subfloat[\label{sfig:6_msgms_worm_3}]{\includegraphics[width=0.6in]{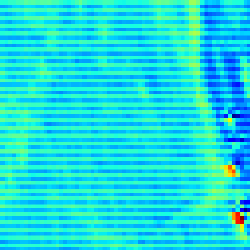}}
		\hfil
		\subfloat[\label{sfig:6_msgms_worm_4}]{\includegraphics[width=0.6in]{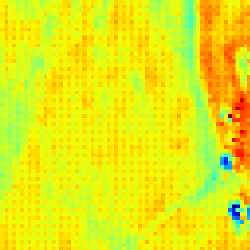}}
		\hfil
		\subfloat[\label{sfig:6_msgms_worm_5}]{\includegraphics[width=0.6in]{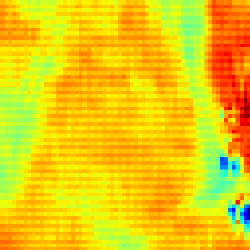}}
		\hfil
		\subfloat[\label{sfig:6_msgms_worm_6}]{\includegraphics[width=0.6in]{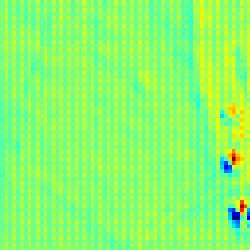}}
		\hfil
		\subfloat[\label{sfig:6_msgms_worm_7}]{\includegraphics[width=0.6in]{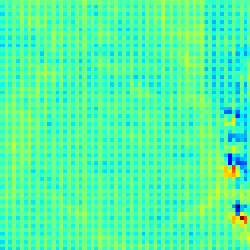}}
		\hfil
		\subfloat[\label{sfig:6_msgms_worm_8}]{\includegraphics[width=0.6in]{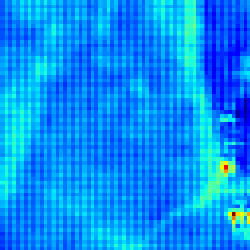}}
		\hfil
		\subfloat[\label{sfig:6_msgms_worm_9}]{\includegraphics[width=0.6in]{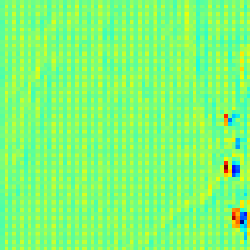}}
		\hfil
		\subfloat[\label{sfig:6_msgms_worm_10}]{\includegraphics[width=0.6in]{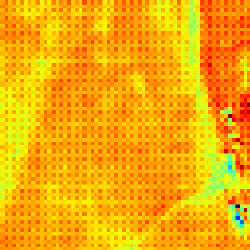}}
	\end{minipage}
	\\
	\vspace{-5mm}
	(c)
	\begin{minipage}[c]{0.95\textwidth}
		\subfloat[1st Band\label{sfig:6_l2_rm_1}]{\includegraphics[width=0.6in]{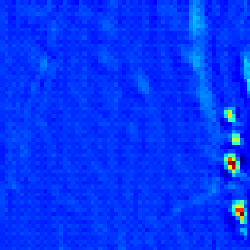}}
		\hfil
		\subfloat[2nd Band\label{sfig:6_l2_rm_2}]{\includegraphics[width=0.6in]{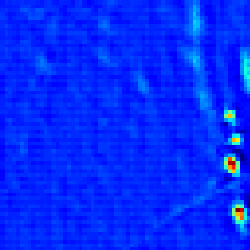}}
		\hfil
		\subfloat[3rd Band\label{sfig:6_l2_rm_3}]{\includegraphics[width=0.6in]{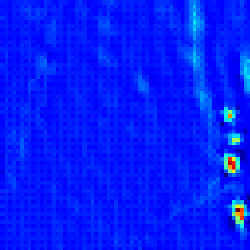}}
		\hfil
		\subfloat[4th Band\label{sfig:6_l2_rm_4}]{\includegraphics[width=0.6in]{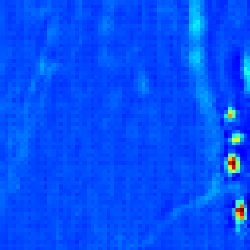}}
		\hfil
		\subfloat[5th Band\label{sfig:6_l2_rm_5}]{\includegraphics[width=0.6in]{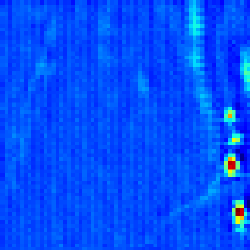}}
		\hfil
		\subfloat[6th Band\label{sfig:6_l2_rm_6}]{\includegraphics[width=0.6in]{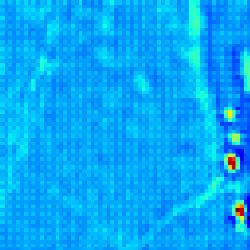}}
		\hfil
		\subfloat[7th Band\label{sfig:6_l2_rm_7}]{\includegraphics[width=0.6in]{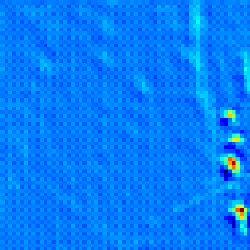}}
		\hfil
		\subfloat[8th Band\label{sfig:6_l2_rm_8}]{\includegraphics[width=0.6in]{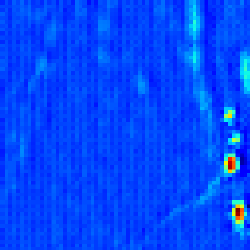}}
		\hfil
		\subfloat[9th Band\label{sfig:6_l2_rm_9}]{\includegraphics[width=0.6in]{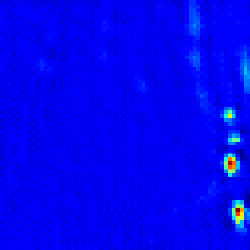}}
		\hfil
		\subfloat[10th Band\label{sfig:6_l2_rm_10}]{\includegraphics[width=0.6in]{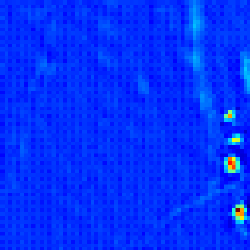}}
	\end{minipage}
	\caption{Residual output maps of AETNet achieved on the test scene 6. (a) Use MSGMS loss and Random Mask strategy. (b) Only use MSGMS loss. (c) Use L2 loss and Random Mask strategy.}
	\label{fig:loss6}
\end{figure*}	

\begin{figure*}[tb]
	\centering
\captionsetup[subfloat]{labelsep=none,format=plain,labelformat=empty}
	(a)
	\begin{minipage}[c]{0.95\textwidth}
		\subfloat[\label{sfig:7_msgms_rm_1}]{\includegraphics[width=0.6in]{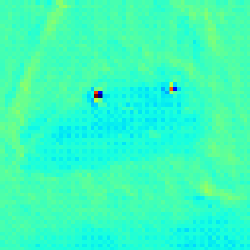}}
		\hfil
		\subfloat[\label{sfig:7_msgms_rm_2}]{\includegraphics[width=0.6in]{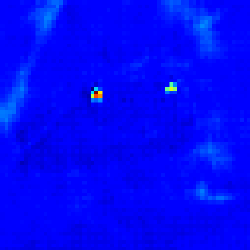}}
		\hfil
		\subfloat[\label{sfig:7_msgms_rm_3}]{\includegraphics[width=0.6in]{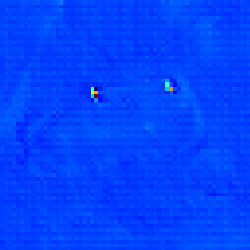}}
		\hfil
		\subfloat[\label{sfig:7_msgms_rm_4}]{\includegraphics[width=0.6in]{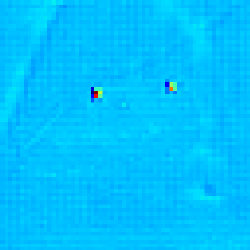}}
		\hfil
		\subfloat[\label{sfig:7_msgms_rm_5}]{\includegraphics[width=0.6in]{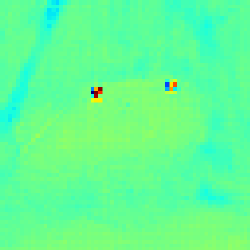}}
		\hfil
		\subfloat[\label{sfig:7_msgms_rm_6}]{\includegraphics[width=0.6in]{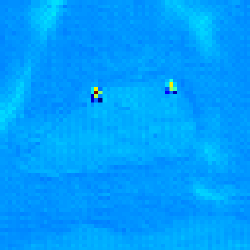}}
		\hfil
		\subfloat[\label{sfig:7_msgms_rm_7}]{\includegraphics[width=0.6in]{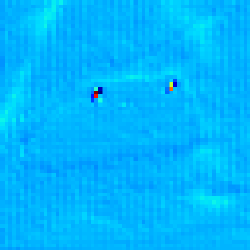}}
		\hfil
		\subfloat[\label{sfig:7_msgms_rm_8}]{\includegraphics[width=0.6in]{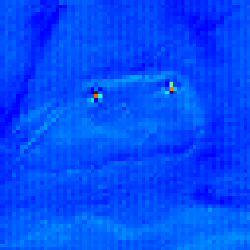}}
		\hfil
		\subfloat[\label{sfig:7_msgms_rm_9}]{\includegraphics[width=0.6in]{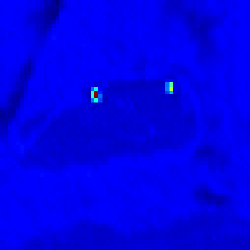}}
		\hfil
		\subfloat[\label{sfig:7_msgms_rm_10}]{\includegraphics[width=0.6in]{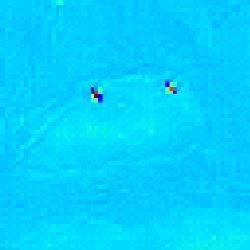}}
	\end{minipage}
	\\
	\vspace{-5mm}
	(b)
		\begin{minipage}[c]{0.95\textwidth}
	\subfloat[\label{sfig:7_msgms_worm_1}]{\includegraphics[width=0.6in]{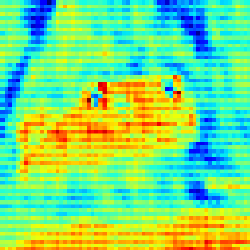}}
	\hfil
	\subfloat[\label{sfig:7_msgms_worm_2}]{\includegraphics[width=0.6in]{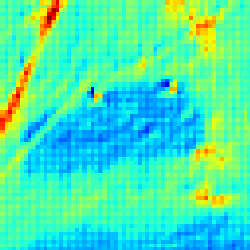}}
	\hfil
	\subfloat[\label{sfig:7_msgms_worm_3}]{\includegraphics[width=0.6in]{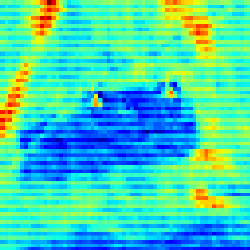}}
	\hfil
	\subfloat[\label{sfig:7_msgms_worm_4}]{\includegraphics[width=0.6in]{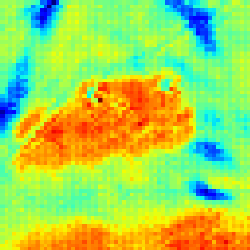}}
	\hfil
	\subfloat[\label{sfig:7_msgms_worm_5}]{\includegraphics[width=0.6in]{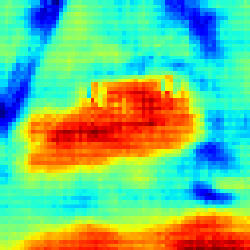}}
	\hfil
	\subfloat[\label{sfig:7_msgms_worm_6}]{\includegraphics[width=0.6in]{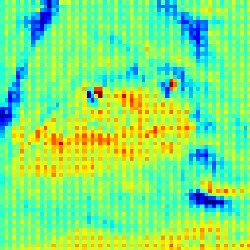}}
	\hfil
	\subfloat[\label{sfig:7_msgms_worm_7}]{\includegraphics[width=0.6in]{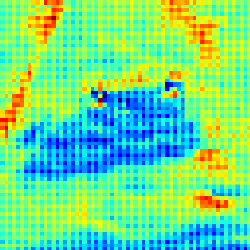}}
	\hfil
	\subfloat[\label{sfig:7_msgms_worm_8}]{\includegraphics[width=0.6in]{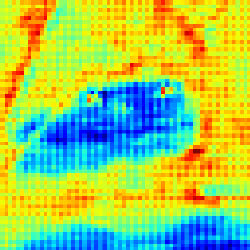}}
	\hfil
	\subfloat[\label{sfig:7_msgms_worm_9}]{\includegraphics[width=0.6in]{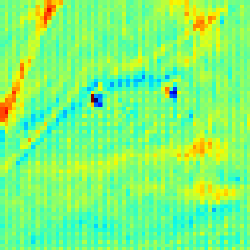}}
	\hfil
	\subfloat[\label{sfig:7_msgms_worm_10}]{\includegraphics[width=0.6in]{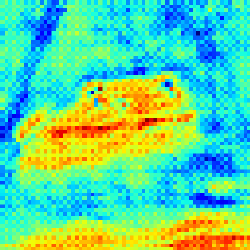}}
	\end{minipage}
	\\
	\vspace{-5mm}
	(c)
	\begin{minipage}[c]{0.95\textwidth}
	\subfloat[1st Band\label{sfig:7_l2_rm_1}]{\includegraphics[width=0.6in]{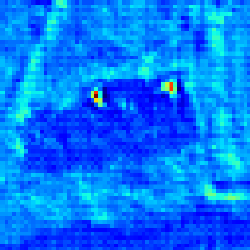}}
	\hfil
	\subfloat[2nd Band\label{sfig:7_l2_rm_2}]{\includegraphics[width=0.6in]{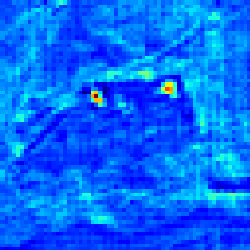}}
	\hfil
	\subfloat[3rd Band\label{sfig:7_l2_rm_3}]{\includegraphics[width=0.6in]{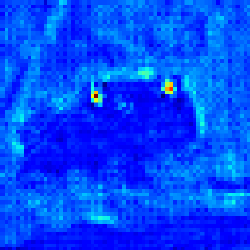}}
	\hfil
	\subfloat[4th Band\label{sfig:7_l2_rm_4}]{\includegraphics[width=0.6in]{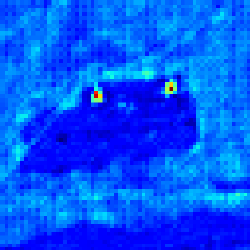}}
	\hfil
	\subfloat[5th Band\label{sfig:7_l2_rm_5}]{\includegraphics[width=0.6in]{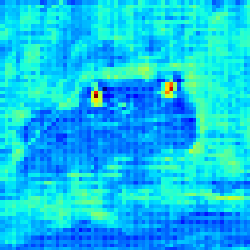}}
	\hfil
	\subfloat[6th Band\label{sfig:7_l2_rm_6}]{\includegraphics[width=0.6in]{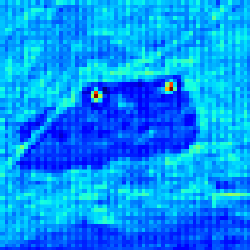}}
	\hfil
	\subfloat[7th Band\label{sfig:7_l2_rm_7}]{\includegraphics[width=0.6in]{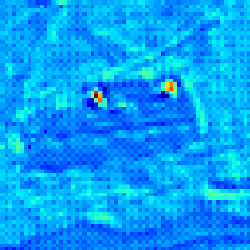}}
	\hfil
	\subfloat[8th Band\label{sfig:7_l2_rm_8}]{\includegraphics[width=0.6in]{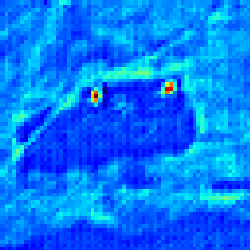}}
	\hfil
	\subfloat[9th Band\label{sfig:7_l2_rm_9}]{\includegraphics[width=0.6in]{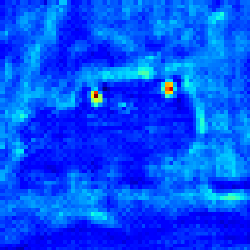}}
	\hfil
	\subfloat[10th Band\label{sfig:7_l2_rm_10}]{\includegraphics[width=0.6in]{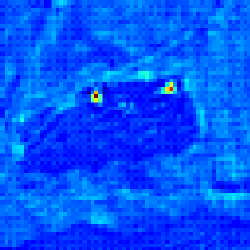}}
\end{minipage}
	\caption{Residual output maps of AETNet achieved on the test scene 7. (a) Use MSGMS loss and Random Mask strategy. (b) Only use MSGMS loss. (c) Use L2 loss and Random Mask strategy.}
	\label{fig:loss7}
\end{figure*}	

\subsection{Ablation Experiments}
\subsubsection{Loss Function}
As shown in \tabref{tab:loss}, we compare the detection accuracy of AETNet with four different loss functions.
L1 loss and L2 loss penalize pixel-wise spectral differences between input and output images.
MSGMS loss considers the spatial correlation of pixels and penalizes the structural differences between input and output images.
SSIM loss simultaneously penalizes the pixel-wise spectral differences and the structural differences.
When the Random Mask strategy is adopted, most mAUC values of AETNet with these losses are better than the baseline values achieved by GRX.
MSGMS loss outperforms the other three loss functions, while L1 loss performs the worst and fails on the first 50 bands.
SSIM loss has a slight advantage over L2 loss on the first 50 and 100 bands.
Figs. 11 and 12 show the residual output of AETNet in the inference on the test scene 6 and 7.
It can be obviously seen that anomaly targets have higher response values than background on the residual output maps when our network adopts the Random Mask strategy.
Consequently, the spectral differences between anomaly targets and background are enhanced in the transform domain generated by AETNet.
Compared with L2 loss, MSGMS loss obtains residual output maps with a cleaner background.

In contrast, when AETNet is not trained with the Random Mask strategy, the enhancement effect on anomaly targets cannot be guaranteed.
\tabref{tab:loss} shows that all four loss functions fail without the Random Mask strategy.
Figs. 11(b) and 12(b)   show that the residual output of AETNet without the Random Mask strategy does not focus on anomaly targets.
These experimental results indicate that the Random Mask strategy is the foundation for the generalization capability of AETNet.

\subsubsection{Network Structure}

\begin{table}[t]
	\centering
	\renewcommand{\arraystretch}{1.3}
	\setlength\tabcolsep{8pt}
	\caption{Detection Accuracy Performance Achieved by Different Network Architecture on the first 50 Bands of the Test Set}
	\begin{tabular}{ccccc}
		\hline
		\multirow{2}[0]{*}{CAE} & \multirow{2}[0]{*}{UNet} & {Swin Transformer} & Residual & \multirow{2}[0]{*}{mAUC} \\
		&  & Block& Connection & \\
		\hline
		$\checkmark$ &  &  &  & 0.9850 \\
		$\checkmark$ &  &  & $\checkmark$ & 0.9894 \\
		\hline
		$\checkmark$ & $\checkmark$ &  &  & 0.9814 \\
		$\checkmark$ & $\checkmark$ &  & $\checkmark$ & 0.9906 \\
		\hline
		$\checkmark$ & $\checkmark$ & $\checkmark$ &  & 0.9877 \\
		$\checkmark$ & $\checkmark$ & $\checkmark$ & $\checkmark$ & 0.9925 \\
		\hline
	\end{tabular}%
	\label{tab:network}%
\end{table}%

 \tabref{tab:network} shows the detection accuracy achieved by different network architectures on the first 50 bands. 
 A single CAE with residual connection reaches an mAUC value of 0.9894 and exceeds $\text{WeaklyAD}_{\mathcal{B}}$ with a value of 0.9887, which indicates that our image-level training paradigm is better than the traditional spectral-level training paradigm.
 Added with the UNet structure, the network reaches an mAUC value of 0.9906.
 Furthermore, the full version of AETNet, which uses Swin Transformer blocks,  can reach an mAUC value of 0.9925.
 In addition, the residual connection plays an important role in our AETNet.
 When the residual connection is removed, the detection performance of all three network structures decreases dramatically.

\section{Conclusion}
\label{sec:Cnl}

In this paper, we introduce an image-level training paradigm and  Random Mask strategy to solve the generalization problem in  hyperspectral  anomaly detection.
Our method eliminates the need for adjusting parameters or retraining on new test scenes as required by most existing methods.
 Moreover, we developed a large-scale hyperspectral anomaly detection dataset and a unified evaluation benchmark. 
Experimental results show that our method achieves a better balance between detection accuracy and inference speed than existing state-of-the-art methods. 
 We hope our work can stimulate the community toward real-world and practical hyperspectral anomaly detection.

\ifCLASSOPTIONcaptionsoff
\newpage
\fi


\bibliographystyle{IEEEtran}
\bibliography{IEEEabrv, AETNET}

\end{document}